\documentclass[%
reprint,
superscriptaddress,
showpacs,
amsmath,amssymb,
prc,
floatfix, ]%
{revtex4-1}

\usepackage{color}

\usepackage{graphicx}% Include figure files
\usepackage{dcolumn}% Align table columns on decimal point
\usepackage{bm}% bold math
\usepackage[dvipdfmx,bookmarks=true,colorlinks,%
            citecolor=blue,linkcolor=blue,anchorcolor=blue,filecolor=blue,urlcolor=blue,%
           ]{hyperref}
\allowdisplaybreaks

\begin{document}

%\begin{CJK*}{GBK}{}

\title{Multidimensionally-constrained relativistic mean field models
and potential energy surfaces of actinide nuclei}

\author{Bing-Nan Lu}% (ÂÀ±þéª)}%
%\email{bnlu@itp.ac.cn}
 \affiliation{State Key Laboratory of Theoretical Physics,
              Institute of Theoretical Physics, Chinese Academy of Sciences,
              Beijing 100190, China}
\author{Jie Zhao}% (ÕÔ½Ü)}%
%\email{zhaojie@itp.ac.cn}
 \affiliation{State Key Laboratory of Theoretical Physics,
              Institute of Theoretical Physics, Chinese Academy of Sciences,
              Beijing 100190, China}
\author{En-Guang Zhao}% (ÕÔ¶÷¹ã)}%
%\email{egzhao@mail.itp.ac.cn}
 \affiliation{State Key Laboratory of Theoretical Physics,
              Institute of Theoretical Physics, Chinese Academy of Sciences,
              Beijing 100190, China}
 \affiliation{Center of Theoretical Nuclear Physics, National Laboratory
              of Heavy Ion Accelerator, Lanzhou 730000, China}
\author{Shan-Gui Zhou}% (ÖÜÉƹó)}%
 \email{sgzhou@itp.ac.cn}
%\homepage{http://www.itp.ac.cn/~sgzhou}
 \affiliation{State Key Laboratory of Theoretical Physics,
              Institute of Theoretical Physics, Chinese Academy of Sciences,
              Beijing 100190, China}
 \affiliation{Center of Theoretical Nuclear Physics, National Laboratory
              of Heavy Ion Accelerator, Lanzhou 730000, China}

\date{\today}

\begin{abstract}
\begin{description}
\item[Background]
Many different shape degrees of freedom play crucial roles in determining
the nuclear ground state and saddle point properties and the fission path.
For the study of nuclear potential energy surfaces, it is desirable to
have microscopic and self-consistent models in which all known important
shape degrees of freedom are included.
\item[Purpose]
By breaking both the axial and the spatial reflection symmetries simultaneously,
we develop multidimensionally-constrained relativistic mean field (MDC-RMF) models.
\item[Methods]
The nuclear shape is assumed to be invariant under
the reversion of $x$ and $y$ axes, {i.e.},
the intrinsic symmetry group is $V_{4}$ and all shape degrees of freedom
$\beta_{\lambda\mu}$ with even $\mu$, such as $\beta_{20}$, $\beta_{22}$, $\beta_{30}$,
$\beta_{32}$, $\beta_{40}$, $\dots$, are included self-consistently.
The single-particle wave functions are
expanded in an axially deformed harmonic oscillator (ADHO) basis.
The RMF functional can be one of the following four forms:
the meson exchange or point-coupling nucleon interactions combined with
the nonlinear or density-dependent couplings.
The pairing effects are taken into account with the BCS approach.
\item[Results]
The one-, two, and three-dimensional potential energy surfaces of $^{240}$Pu
are illustrated for numerical checks and for the study of the effect of
the triaxiality on the fission barriers.
Potential energy curves of even-even actinide nuclei around the first and second
fission barriers are studied systematically.
Besides the first ones, the second fission barriers in these nuclei are also
lowered considerably by the triaxial deformation.
This lowering effect is independent of the effective interactions used
in the RMF functionals.
Further discussions are made about
different predictions on the effect of the
triaxiality between the macroscopic-microscopic and MDC-RMF models,
possible discontinuities on PES's from self-consistent approaches,
and the restoration of broken symmetries.
\item[Conclusions]
MDC-RMF models give a reasonably good description of fission barriers
of even-even actinide nuclei.
It is important to include both the nonaxial and the
reflection asymmetric shapes simultaneously for the study of potential
energy surfaces and fission barriers of actinide nuclei and
of those in unknown mass regions such as, {e.g.}, superheavy nuclei.
\end{description}
\end{abstract}

\pacs{21.60.Jz, 24.75.+i, 25.85.-w, 27.90.+b}
%21.60.Jz       Nuclear Density Functional Theory and extensions
%               (includes Hartree-Fock and random-phase approximations)
%24.75.+i       General properties of fission
%25.85.-w       Fission reactions
%27.90.+b       A >= 220

\maketitle

%\end{CJK*}

\section{Introduction}

The occurrence of spontaneous symmetry breaking leads to nuclear shapes
with a variety of symmetries~\cite{Bohr1998_Nucl_Structure_2, Ring1980}.
A lot of nuclear phenomena are connected with the nuclear deformation,
including small and large amplitude collective motions, {e.g.}, the rotation and
the fission~\cite{Frauendorf2001_RMP73-463, Nazarewicz2001_LNP581-102}.
%There are also many unresolved problems concerning the motion or change in
%the nuclear shape degrees of freedom.
%For example, in recent years much effort was devoted to explore
%nuclear shape phase transitions~\cite{Iachello2001_PRL87-052502,
%Li2009_PRC79-054301,*Li2009_PRC80-061301R, Cejnar2010_RMP82-2155};
%one way to investigate shape phase transitions is to examine nuclear potential
%energy surfaces (PES's) and their changes with the nucleon number~\cite{Meng2005_EPJA25-23,
%Fossion2006_PRC73-044310, Niksic2007_PRL99-092502, Guo2008_IJMPE17-539,*Guo2010_PRC82-047301,
%Zhang2010_PRC81-034302}.%,
%Nomura2012_PRL108-132501}.
The shape of a nucleus can be described by the parametrization of the nuclear surface
or the nucleon density distribution.
%There are mainly two kinds of ways to parameterize
%the nuclear shape~\cite{Brack1972_RMP44-320, Hasse1988}.
%One of them is the two-center or two-center-like parametrization~\cite{%Mustafa1972_PRL28-1536,
%Swiatecki1981_PS24-113,
%Moeller2001_Nature409-785, Diaz-Torres2008_PRL101-122501,
%Sun2013_ChinPhysC37-014102}.
%It is usually used to describe a nucleus at large deformations, such as the nuclear fission.
The multipole expansion of the nuclear surface is usually used in mean field calculations,
%The other way is of one-center and to make a multipole expansion of the nuclear surface,
\begin{equation}
 R ( \theta, \varphi ) =
 R_0 \left[  1 +
           \beta_{00} +
           \sum_{\lambda=2}^{\infty} \sum_{\mu=-\lambda}^\lambda
            \beta_{\lambda \mu}^* Y_{\lambda \mu} ( \theta, \varphi )
     \right] ,
 \label{Eq:SurfaceDeformation}
\end{equation}
where $\beta_{\lambda \mu}$'s are deformation parameters.

The majority of observed nuclear shapes is of the spheroidal form which
can be described by the axial-quadrupole deformation parameter $\beta_{20}$
though in early years the Nilsson perturbed-spheroid parameter $\epsilon_2$ was
usually adopted for numerical convenience~\cite{Nilsson1969_NPA131-1}.
%The nonaxial-quadrupole (triaxial) deformation $\beta_{22}$ (or, equivalently, $\gamma$)
%has been predicted in atomic nuclei for quite a long time and
%it may play important roles in superheavy nuclei (SHN)~\cite{Cwiok2005_Nature433-705}.
%A static triaxial shape manifests itself by the wobbling motion and
%chiral doublet bands; both have been extensively studied from experimental and
%theoretical sides~\cite{Frauendorf1997_NPA617-131, Starosta2001_PRL86-971,
%Odegard2001_PRL86-5866, Meng2010_JPG37-064025,*Meng2013_FPC8-55}.%,Meng2006_PRC73-037303,
%%Ayangeakaa2013_PRL110-172504}.
%The softness with respect to the gamma distortion also results in
%many interesting nuclear phenomena~\cite{Li2010_PRC81-034316, Zhao2011_PRL107-122501,
%Song2011_SciChinaPMA54-222, Zhang2011_SciChinaPMA54-227,
%Zhu2011_SciChinaPMA54-44, Liu2012_SciChinaPMA55-2420}.
The nonaxial-quadrupole (triaxial) deformation $\beta_{22}$ (or, equivalently, $\gamma$)
manifests itself by the wobbling motion and chiral doublet
bands~\cite{Frauendorf1997_NPA617-131, Starosta2001_PRL86-971,
Odegard2001_PRL86-5866, Meng2010_JPG37-064025,*Meng2013_FPC8-55},
and it may also play important roles in superheavy nuclei (SHN)~\cite{Cwiok2005_Nature433-705}.

The octupole shapes with $\lambda = 3$ are predicted to
exist in nuclei in several mass regions~\cite{Butler1996_RMP68-349}.
The low-lying negative parity bands observed in actinides and some
rare-earth nuclei are related to the reflection asymmetric (RA)
shapes~\cite{Shneidman2003_PRC67-014313,*Shneidman2006_PRC74-034316,
Wang2005_PRC72-024317, Yang2009_CPL26-082101, Robledo2011_PRC84-054302,
Zhu2012_PRC85-014330,*Zhu2012_NSC2012-348, Gaffney2013_Nature497-199}.
%In addition, the influence of the nonaxial octupole $\beta_{32}$ deformation
%on the low-lying spectra have been traced out theoretically
%and experimentally~\cite{Hamamoto1991_ZPD21-163, Skalski1991_PRC43-140, Li1994_PRC49-R1250,
%Takami1998_PLB431-242,*Yamagami2001_NPA693-579,
%Dudek2002_PRL88-252502,*Dudek2006_PRL97-072501,
%Olbratowski2006_IJMPE15-333,
%Zberecki2006_PRC74-051302R,
%Dudek2010_JPG37-064032}.
%For example, reflection asymmetric shell model calculations revealed
%that the observed low-energy $2^{-}$ bands in $N=150$ nuclei~\cite{Robinson2008_PRC78-034308}
%are caused by the $\beta_{32}$ deformation~\cite{Chen2008_PRC77-061305R}.
%Indeed, strong $Y_{32}$ correlations were found in some $N=150$ isotones
%from multi-dimensional constraint covariant density functional theories
%(MDC-CDFT)~\cite{Zhao2012_PRC86-057304}.
In addition, reflection asymmetric shell model calculations revealed
that the observed low-energy $2^{-}$ bands in $N=150$ nuclei~\cite{Robinson2008_PRC78-034308}
are caused by the $\beta_{32}$ deformation~\cite{Chen2008_PRC77-061305R}.
Indeed, strong $Y_{32}$ correlations were found in some $N=150$ isotones
from multidimensionally-constrained covariant density functional theories
(MDC-CDFT)~\cite{Zhao2012_PRC86-057304}.

Deformations of higher-order multipole with $\lambda>3$ are important to different extents.
The hexadecapole deformation, $\beta_{40}$ or $\epsilon_{4}$,
has been included in deformed mean field potentials since 1960s,
see, {e.g.}, Ref.~\cite{Nilsson1969_NPA131-1}.
The important effects of the higher-order deformation $\beta_{60}$ or $\epsilon_6$ on the
angular momentum alignments and dynamic moments of inertia in superheavy nuclei
were also revealed~\cite{Liu2012_PRC86-011301R, Zhang2013_PRC87-054308}.

The shape degrees of freedom are important not only for
the ground states or small amplitude collective motions, but also
for large amplitude collective motions such as fission.
Since the discovery of the nuclear fission,
the description of the fission process has been a difficult and challenging task.
The fission dynamics are mostly governed
by the barriers which prohibit the dissolving of the nucleus.
In order to study the fission problem, one should have very accurate
information about the fission barrier,
{i.e.}, the height and width, or, more precisely, the shape of the fission
barrier~\cite{Zubov2009_PPN40-847, Xia2011_SciChinaPMA54S1-109}.
Particularly, to explore the island of stability of SHN,
it is more and more desirable to have accurate predictions of
fission barriers of SHN. %~\cite{Moeller2009_PRC79-064304,
%Peter2004_EPJA22-271,*Peter2004_NPA734-192,
%Antonenko1995_PRC51-2635,*Adamian1997_NPA618-176,
%Liu2007_PRC76-034604,*Liu2013_PRC87-034616,
%Wang2008_PRC78-054607,*Wang2012_PRC85-041601R,
%Pei2009_PRL102-192501,
%Li2010_NPA834-353c,*Gan2011_SciChinaPMA54S1-61,
%Wang2011_PRC84-061601R,*Zanganeh2012_PRC85-034601,
%Walker2012_JPG39-105106}.
To now, many popular nuclear structure models have been employed to study
nuclear fission barriers, including the macroscopic-microscopic (MM)
models~\cite{Moeller2001_Nature409-785,
%Moeller2004_PRL92-072501,
%Moeller2009_PRC79-064304,
Pomorski2003_PRC67-044316, Ivanyuk2009_PRC79-054327,
Kowal2010_PRC82-014303,
Royer2012_PRC86-044326},
the extended Thomas-Fermi plus Strutinsky integral (ETFSI)
method~\cite{Mamdouh1998_NPA644-389, *Mamdouh2001_NPA679-337},
the Hartree-Fock or Hartree-Fock-Bogoliubov methods with the Skyrme
force~\cite{%
Burvenich2004_PRC69-014307,
Samyn2005_PRC72-044316,*Goriely2007_PRC75-064312,
Minato2009_NPA831-150,
Pei2009_PRL102-192501,
Kortelainen2012_PRC85-024304,
McDonnell2012_arXiv1301.7587,
Staszczak2013_PRC87-024320,
Schunck2013_arXiv1311.2616,*Schunck2013_arXiv1311.2620}
and the Gogny force~\cite{Egido2000_PRL85-1198,*Warda2012_PRC86-014322},
and the CDFTs~\cite{Blum1994_PLB323-262,
Zhang2003_CPL20-1694,
Bender2003_RMP75-121, 
Burvenich2004_PRC69-014307,
Lu2006_CPL23-2940,
Li2010_PRC81-064321,
Abusara2010_PRC82-044303,
Lu2012_PRC85-011301R,
Lu2012_PhD, Lu2012_EPJWoC38-05003,*Lu2013_arXiv1304.6830,
Abusara2012_PRC85-024314,
Prassa2012_PRC86-024317,*Prassa2013_PRC88-044324,
Afanasjev2013_arXiv1303.1206}.

Besides $\beta_{20}$ which describes the elongation of a fissile nucleus
and $\beta_{40}$ which is relevant to the size of a neck,
many other shape degrees of freedom are also crucial
for determining the shape of fission barriers and the fission path.
Let us take actinides as examples.
Due to shell effects, actinide nuclei are characterized by a two-humped
fission barrier~\cite{Brack1972_RMP44-320}.
It has long been known from the MM model calculations that the inner fission
barrier is lowered by the nonaxial-quadrupole deformation~\cite{Pashkevich1969_NPA133-400,
Moeller1970_PLB31-283, Randrup1976_PRC13-229}
and the outer one by the reflection asymmetric shape~\cite{Ledergerber1973_NPA207-1}.
Later, the important roles played by the nonaxial-quadrupole deformation and
the octupole deformation were confirmed in the nonrelativistic~\cite{Girod1983_PRC27-2317}
and relativistic~\cite{Rutz1995_NPA590-680}
density functional calculations, respectively.
Therefore what is usually done is to consider the triaxial
but reflection symmetric (RS) shapes for the inner barrier and
axially symmetric (AS) but RA shapes for the outer one~\cite{Egido2000_PRL85-1198,
Bonneau2004_EPJA21-391, Moeller2009_PRC79-064304}
though in several publications, both the nonaxial and the octupole
deformations are included~\cite{Skalski2007_PRC76-044603,*Jachimowicz2011_PRC83-054302}.

In recent years, the nuclear CDFT has been very successful in the description of
both ground states and excited states of the nuclei ranging from
light to superheavy regions.
The CDFTs have also been used to study the PES's
and the fission barriers of heavy and superheavy
nuclei~\cite{Blum1994_PLB323-262,
Zhang2003_CPL20-1694,
Bender2003_RMP75-121, 
Burvenich2004_PRC69-014307,
Lu2006_CPL23-2940,
Li2010_PRC81-064321,
Abusara2010_PRC82-044303,
Lu2012_PRC85-011301R,
Lu2012_PhD, Lu2012_EPJWoC38-05003,*Lu2013_arXiv1304.6830,
Abusara2012_PRC85-024314,
Prassa2012_PRC86-024317,*Prassa2013_PRC88-044324,
Afanasjev2013_arXiv1303.1206}.
We have developed MDC-CDFTs by breaking both the axial and reflection symmetries
simultaneously~\cite{Lu2012_PRC85-011301R,
Lu2012_PhD, Lu2012_EPJWoC38-05003,*Lu2013_arXiv1304.6830}.
In these MDC-CDFTs,
all shape degrees of freedom $\beta_{\lambda\mu}$ with even $\mu$,
{e.g.}, $\beta_{20}$, $\beta_{22}$, $\beta_{30}$, $\beta_{32}$, $\beta_{40}$, $\dots$,
are included self-consistently.
The covariant density functional can be one of the following four forms:
the meson exchange or point-coupling nucleon interactions combined with
the nonlinear or density-dependent couplings.
For the particle-particle channel, either the BCS approach or the Bogoliubov transformation
has been implemented.
%The pairing force could be $\delta$-force or the finite-range separable
%force~\cite{Tian2006_CPL23-3226,*Tian2009_PLB676-44,*Tian2009_PRC79-064301}.
For convenience, we name the MDC-CDFT with the BCS approach for the pairing
as the MDC-RMF models and those with the Bogoliubov transformation as the MDC-RHB models.
Due to the heavy computational burden, MDC-RHB models have not been used to
do multidimensionally constrained calculations yet (here ``multi'' means three or more).
The MDC-RMF models have been used to study the PES's
of actinide nuclei in the ($\beta_{20}$, $\beta_{22}$, $\beta_{30}$) deformation space
and it was found that the triaxiality also
plays an important role upon the second fission barriers~\cite{Lu2012_PRC85-011301R}.
In this paper, we will present the detailed formulae for MDC-RMF models
and some results of actinide nuclei.

The paper is organized as follows.
The formalism of MDC-RMF models will be given in Sec.~\ref{sec:formalism}.
In Sec.~\ref{sec:results} we present the numerical details and
results on the PES's of the actinide nuclei.
A summary is given in Sec.~\ref{sec:summary}.

\section{\label{sec:formalism}
Formalism of MDC-RMF Models}

In the RMF model, a nucleus is treated as a composite of nucleons interacting
through exchanges of mesons and photons~\cite{Serot1986_ANP16-1,
Reinhard1989_RPP52-439, Ring1996_PPNP37-193, Bender2003_RMP75-121,
Vretenar2005_PR409-101, Meng2006_PPNP57-470, Paar2007_RPP70-691,
Niksic2011_PPNP66-519}.
The effects of mesons are described either by mean fields or
by point-like interactions between the nucleons~\cite{Nikolaus1992_PRC46-1757,
Burvenich2002_PRC65-044308}.
Meanwhile,
the nonlinear coupling
terms~\cite{Boguta1977_NPA292-413, Brockmann1992_PRL68-3408, Sugahara1994_NPA579-557}
or
the density dependence of the coupling constants~\cite{Fuchs1995_PRC52-3043,
Niksic2002_PRC66-024306}
are introduced to give correct saturation properties of nuclear matter.
Accordingly, the form of the covariant density functional can be one of the following four:
the meson exchange or point-coupling nucleon interactions combined with
the nonlinear or density dependent couplings.
Most of the computational efforts are devoted to solving
the Dirac equation and the calculation of various densities;
this is common for all these RMF models.
In this section, we mainly present the formalism of the RMF model
with the nonlinear point-couplings (NL-PC).
%For more details of CDFTs, the readers are referred to Refs.~\cite{Serot1986_ANP16-1,
%Reinhard1989_RPP52-439, Ring1996_PPNP37-193, Bender2003_RMP75-121,
%Vretenar2005_PR409-101, Meng2006_PPNP57-470, Paar2007_RPP70-691,
%Niksic2011_PPNP66-519}.

The starting point of the RMF model with the NL-PC is the following Lagrangian,
\begin{equation}
 \mathcal{L} = \bar{\psi}(i\gamma_{\mu}\partial^{\mu}-M)\psi
              -\mathcal{L}_{{\rm lin}}
              -\mathcal{L}_{{\rm nl}}
              -\mathcal{L}_{{\rm der}}
              -\mathcal{L}_{{\rm Cou}},
\end{equation}
where
\begin{eqnarray}
 \mathcal{L}_{{\rm lin}} & = & \frac{1}{2} \alpha_{S} \rho_{S}^{2}
                              +\frac{1}{2} \alpha_{V} \rho_{V}^{2}
                              +\frac{1}{2} \alpha_{TS} \vec{\rho}_{TS}^{2}
                              +\frac{1}{2} \alpha_{TV} \vec{\rho}_{TV}^{2} ,
 \nonumber \\
 \mathcal{L}_{{\rm nl}}  & = & \frac{1}{3} \beta_{S} \rho_{S}^{3}
                              +\frac{1}{4} \gamma_{S}\rho_{S}^{4}
                              +\frac{1}{4} \gamma_{V}[\rho_{V}^{2}]^{2} ,
 \nonumber \\
 \mathcal{L}_{{\rm der}} & = & \frac{1}{2} \delta_{S}[\partial_{\nu}\rho_{S}]^{2}
                              +\frac{1}{2} \delta_{V}[\partial_{\nu}\rho_{V}]^{2}
                              +\frac{1}{2} \delta_{TS}[\partial_{\nu}\vec{\rho}_{TS}]^{2}
 \nonumber \\
 &  & \mbox{}                 +\frac{1}{2} \delta_{TV}[\partial_{\nu}\vec{\rho}_{TV}]^{2} ,
 \nonumber \\
 \mathcal{L}_{{\rm Cou}} & = & \frac{1}{4} F^{\mu\nu} F_{\mu\nu}
                             +e\frac{1-\tau_{3}}{2} A_{0} \rho_{V} ,
\label{eq:lagrangian}
\end{eqnarray}
are the linear coupling, nonlinear coupling, derivative coupling,
and the Coulomb part, respectively.
$M$ is the nucleon mass, $\alpha_{S}$, $\alpha_{V}$, $\alpha_{TS}$,
$\alpha_{TV}$, $\beta_{S}$, $\gamma_{S}$, $\gamma_{V}$, $\delta_{S}$,
$\delta_{V}$, $\delta_{TS}$, and $\delta_{TV}$ are coupling constants
for different channels, and $e$ is the electric charge.
$\rho_{S}$, $\vec{\rho}_{TS}$, $\rho_{V}$, and $\vec{\rho}_{TV}$ are the isoscalar density,
isovector density, time-like components of isoscalar current, and time-like components of
isovector current, respectively.
The densities and currents are defined as
\begin{eqnarray}
      \rho  _{S} = \bar{\psi}\psi , & \qquad &
 \vec{\rho}_{TS} = \bar{\psi}\vec{\tau}\psi ,
 \nonumber \\
      \rho_{V} = \bar{\psi} \gamma^{0} \psi , & \qquad &
 \vec{\rho}_{TV} = \bar{\psi} \vec{\tau} \gamma^{0} \psi.
 \label{eq:densities}
\end{eqnarray}

Starting from the Lagrangian, using the Slater determinants as trial wave functions,
and neglecting the Fock term as well as the contribution to the densities and currents
from the Dirac sea (the no sea approximation), we can derive the RMF equation
with the variational principle,
\begin{equation}
 \hat{h}\psi_{k}(\bm{r}) = \epsilon_{k} \psi_{k}(\bm{r}) ,
 \label{eq:Diracequation}
\end{equation}
where
\begin{equation}
 \hat{h} = \bm{\alpha} \cdot \bm{p}
         + \beta \left[ M+S(\bm{r}) \right]
         + V(\bm{r}) ,
 \label{eq:dirac}
\end{equation}
is the single-particle Hamiltonian and
\begin{eqnarray}
 S & = &
  \alpha_{S} \rho_{S}     + \alpha_{TS} \vec{\rho}_{TS} \cdot \vec{\tau}
 + \beta_{S} \rho_{S}^{2} +  \gamma_{S} \rho_{S}^{3}
 \nonumber \\
 &  & \mbox{}
 +\delta_{S} \triangle \rho_{S} + \delta_{TS} \triangle \vec{\rho}_{TS} \cdot \vec{\tau}
 ,
%\nonumber
 \\
 %V^{\mu} & = &
%   \alpha_{V} j_{V}^{\mu} + \alpha_{TV} \vec{j}_{TV}^{\mu} \cdot \vec{\tau}
% + \gamma_{V} j_{V}^{2} j_{V}^{\mu}
% \nonumber \\
% &  & \mbox{}
% + \delta_{V} \triangle j_{V}^{\mu} + \delta_{TV} \triangle \vec{j}_{TV}^{\mu} \cdot \vec{\tau}
% ,
%
V & = &
   \alpha_{V} \rho_{V} + \alpha_{TV} \vec{\rho}_{TV} \cdot \vec{\tau}
 + \gamma_{V} \rho_{V}^{2} \rho_{V}
 \nonumber \\
 &  & \mbox{}
 + \delta_{V} \triangle \rho_{V} + \delta_{TV} \triangle \vec{\rho}_{TV} \cdot \vec{\tau}
 ,
\label{eq:potential}
\end{eqnarray}
are the scalar and vector potentials, respectively.
In the present work we suppose that the nuclei in question are invariant under the time-reversion,
which means that all the time-odd or vector components of the currents
and the potentials vanish.
In this case the single-particle Hamiltonian has the time-reversal symmetry
which simplifies the calculation.

It is customary to solve the deformed RMF equations by expanding the auxiliary
single-particle wave functions in a complete basis,
{e.g.}, the harmonic oscillator (HO) basis~\cite{Gambhir1990_APNY198-132, Ring1997_CPC105-77}
or the Woods-Saxon (WS) basis~\cite{Zhou2003_PRC68-034323,*Zhou2006_AIPCP865-90,
*Zhou2010_PRC82-011301R,*Li2012_PRC85-024312, Chen2012_PRC85-067301}.
By using a basis in a two-centre harmonic oscillator potential,
a reflection asymmetric relativistic mean field (RAS-RMF) approach has been
developed~\cite{Geng2007_CPL24-1865} and
used to study PES's of even-even $^{146-156}$Sm
in which the important role of the octupole deformation on shape phase transitions
was found~\cite{Zhang2010_PRC81-034302}.
However, in our MDC-CDFTs,
the single-particle wave functions and various densities are
expanded in an axially deformed harmonic oscillator (ADHO) basis.
The ADHO basis consists of the eigenstates of the Schr\"odinger
equation~\cite{Gambhir1990_APNY198-132, Ring1997_CPC105-77, Lu2011_PRC84-014328},
\begin{eqnarray}
 \left[-\frac{\hbar^{2}}{2M}\nabla^{2}+V_{B}(z,\rho)\right]\Phi_{\alpha}(\bm{r}\sigma)
 & = & E_{\alpha}\Phi_{\alpha}(\bm{r}\sigma) ,
 \label{eq:BasSchrodinger-1}
\end{eqnarray}
where $\bm{r} = (z,\rho)$ with $\rho=\sqrt{x^2+y^2}$ and
\begin{equation}
 V_{B}(z,\rho) = \frac{1}{2} M ( \omega_{\rho}^{2} \rho^{2} + \omega_{z}^{2} z^{2})
 ,
\end{equation}
is the ADHO potential and $\omega_{z}$ and $\omega_{\rho}$ are
the oscillator frequencies along and perpendicular to the symmetry $z$ axis, respectively.
The solution of Eq.~(\ref{eq:BasSchrodinger-1}) reads
\begin{equation}
 \Phi_{\alpha}(\bm{r}\sigma)
 =
 C_{\alpha} \phi_{n_{z}}(z) R_{n_{\rho}}^{m_{l}}(\rho)
 \frac{1}{\sqrt{2\pi}} e^{im_{l}\varphi}
 \chi_{s_{z}}(\sigma),
 \label{eq:HO}
\end{equation}
where $\phi_{n_{z}}(z)$ and $R_{n_{\rho}}^{m_{l}}(\rho)$ are the harmonic oscillator wave functions,
\begin{eqnarray}
 \phi_{n_{z}}(z) & = &
 \frac{1}{\sqrt{b_{z}}} \frac{1}{\pi^{{1}/{4}} \sqrt{2^{n_{z}}n_{z}!}}
  H_{n_{z}}\left(\frac{z}{b_{z}}\right)
  e^{-\frac{z^{2}}{2b_{z}}} ,
  \\
  R_{n_{\rho}}^{m_{l}}(\rho) & = &
  \frac{1}{b_{\rho}} \sqrt{\frac{2n_{\rho}!}{(n_{\rho}+|m_{l}|)!}}
  \left( \frac{\rho}{b_{\rho}} \right)^{|m_{l}|}
  L_{n_{\rho}}^{|m_{l}|}
  \left( \frac{\rho^{2}}{b_{\rho}^{2}} \right)
  e^{ -\frac{\rho^{2}}{2b_{\rho}^{2}} }
  .
\end{eqnarray}
$\chi_{s_{z}}$ is a two-component spinor and $C_{\alpha}$ is a complex number
introduced for convenience.
Harmonic oscillator lengths $b_{z}$ and $b_{\rho}$ are related to the frequencies
by $b_{z}=1/\sqrt{M\omega_{z}}$ and $b_{\rho}=1/\sqrt{M\omega_{\rho}}$.
The corresponding eigenenergy
$E_{\alpha}=\omega_{\rho}(2n_{\rho}+|m_{l}|+1)+\omega_{z}(n_{z}+{1}/{2})$
and the major quantum number $N_{\alpha}=2n_{\rho}+|m_{l}|+n_{z}$.

These basis states
are also eigenstates of the $z$ component of the angular momentum
$\hat{j}_{z}$ with eigenvalues $K_\alpha=m_{l}+m_{s}$.
For any state $\Phi_{\alpha}(\bm{r}\sigma)$, the time-reversal state is defined as
$\Phi_{\bar{\alpha}}(\bm{r}\sigma)=\mathcal{T}\Phi_{\alpha}(\bm{r}\sigma)$,
where $\mathcal{T}=i\sigma_{y}\hat{K}$ is the time-reversal operator and
$\hat{K}$ is the complex conjugation operator.
Apparently, we have $K_{\bar{\alpha}}=-K_{\alpha}$ and $\pi_{\bar{\alpha}}=\pi_{\alpha}$,
where $\pi_\alpha = \pm 1$ is the parity.
The deformation of the basis $\beta_{{\rm basis}}$ is defined through the relations
$\omega_{z}=\omega_{0}\exp\left(-\sqrt{{5}/{4\pi}}\beta_{{\rm basis}}\right)$
and $\omega_{\rho}=\omega_{0}\exp\left(\sqrt{{5}/{16\pi}}\beta_{{\rm basis}}\right)$,
where $\omega_{0}=(\omega_{z}\omega_{\rho}^{2})^{{1}/{3}}$ is the frequency of
the corresponding spherical oscillator.

These basis states form a complete set for expanding any two-component spinors.
For a Dirac spinor with four components,
\begin{equation}
 \psi_{i}(\bm{r}\sigma) =
 \left(
  \begin{array}{c}
   \sum_{\alpha}f_{i}^{\alpha} \Phi_{\alpha}(\bm{r}\sigma) \\
   \sum_{\alpha}g_{i}^{\alpha} \Phi_{\alpha}(\bm{r}\sigma)
  \end{array}
 \right),
\label{eq:spwaveexpansion}
\end{equation}
where the sum runs over all the possible combination of the quantum numbers
$\alpha=\{n_{z},n_{\rho},m_{l},m_{s}\}$ and $f_{i}^{\alpha}$ and
$g_{i}^{\alpha}$ are the expansion coefficients.
In practical calculations the summations in Eq.~(\ref{eq:spwaveexpansion}) have to be truncated.
Following Ref.~\cite{Warda2002_PRC66-014310},
for the large component of the Dirac wave function, the states satisfying
$[ n_{z}/Q_{z}+(2n_{\rho}+|m_l|)/Q_{\rho} ] \le N_f$
are included in the expansion
where
$Q_{z}=\max(1,b_{z}/b_{0})$ and $Q_{\rho}=\max(1,b_{\rho}/b_{0})$ are
constants related to the oscillator lengths $b_{0} \equiv 1/\sqrt{M\omega_0}$,
$b_{z}$, and $b_{\rho}$.
For the expansion of the small component, the truncation is made up to
$N_g = N_f+1$ major shells
in order to avoid the spurious states~\cite{Gambhir1990_APNY198-132}.

If a nucleus is invariant under the rotation around the symmetry $z$ axis
and the spatial reflection,
the angular momentum projection on the $z$-axis and the parity are conserved and
the RMF equation (\ref{eq:Diracequation}) can be decomposed into blocks
characterized by the quantum numbers $K_\alpha$ and $\pi_\alpha$.
Usually only half of the space with $K_{\alpha}>0$ is considered due to the time-reversal symmetry.

Now let us turn to the general case that the axial symmetry as well as
the spatial reflection symmetry are broken.
Since we still expand the spinors in the ADHO basis,
components with different $K$ and $\pi$ are mixed together;
thus, we must diagonalize a larger single-particle Hamiltonian matrix with
non-zero matrix elements between two basis states with different $K$ and $\pi$.
Nevertheless, even in this case, we still have one symmetry operator
that makes the Hamiltonian matrix block-diagonal.
Due to the axial symmetry of the basis, it is convenient to introduce
the simplex operator $\hat{S}=ie^{-i\pi \hat{j}_{z}}$. % as a conserved quantity.
Note that for a fermionic system with a half-integer spin, $\hat{S}$ is a Hermitian operator
and $\hat{S}^{2}=1$.
This operator corresponds to the rotation by $\pi$ around the $z$ axis, thus
leaving the nuclear mean field invariant.
The eigenvalue of $\hat{S}$, $S$, is also a good quantum number for the basis,
$\hat{S}\Phi_{\alpha}=S\Phi_{\alpha}=(-1)^{K_{\alpha}-{1}/{2}}\Phi_{\alpha}$,
which means that the basis $\Phi_{\alpha}$ with
$K_{\alpha}=+{1}/{2}$, $-{3}/{2}$, $+{5}/{2}$, $-{7}/{2},\ \dots$
span the subspace with $S=1$, while their time-reversal states span the one with $S=-1$.
Note that now the blocks with $K=+{1}/{2}$, $-{3}/{2}$,
$+{5}/{2}$, $-{7}/{2}$, $\dots$ are mixed.
Remember that for a nucleus with the time-reversal symmetry, only the basis with $S=1$
are used in the expansions; for such a basis state, we set $C_{\alpha}=1$
[cf. Eq.~(\ref{eq:HO})] when expanding
the large component and $C_{\alpha}=i$ for the small one.
The basis states with $S_{\alpha}=-1$ are obtained by simply applying $\mathcal{T}$ on those
with $S_{\alpha}=1$.
Furthermore, for systems with the time-reversal symmetry, it is only necessary
to diagonalize the matrix with $S=1$ and the other half is obtained
by a time-reversal operation on the obtained single-particle wave functions.

For deformed nuclei with the $V_{4}$ symmetry, we expand the potentials $V(\bm{r})$,
$S(\bm{r})$, and the densities in Eq.~(\ref{eq:densities}) in terms of the Fourier series,
\begin{equation}
 f(\rho,\varphi,z)
 = \sum_{\mu=-\infty}^{\infty} f_{\mu}(\rho,z)
   \frac{1}{\sqrt{2\pi}} \exp(i\mu\varphi)
 .
 \label{eq:potentialexpansion}
\end{equation}
Applying the symmetry conditions,
it is easy to see that $f_{\mu}=f_{\mu}^{*}=f_{\bar{\mu}}$ and $f_{n}=0$ for odd $n$.
Thus the expansion (\ref{eq:potentialexpansion}) can be simplified as
\begin{equation}
 f(\rho,\varphi,z) =
  f_{0}(\rho,z) \frac{1}{\sqrt{2\pi}}
+ \sum_{n=1}^{\infty} f_{n}(\rho,z) \frac{1}{\sqrt{\pi}} \cos(2n\varphi),
\end{equation}
where
\begin{eqnarray}
 f_{0}(\rho_{,}z) & = & \frac{1}{\sqrt{2\pi}}\int_{0}^{2\pi}
                         d\varphi f(\rho,\varphi,z) ,
 \label{eq:potentialexpansion_0}
 \\
 f_{n}(\rho,z)    & = & \frac{1}{\sqrt{ \pi}}\int_{0}^{2\pi}
                         d\varphi f(\rho,\varphi,z)\cos(2n\varphi) ,
 \label{eq:potentialexpansion_n}
\end{eqnarray}
are real functions of $\rho$ and $z$.
The details for calculating the matrix elements of the Dirac Hamiltonian and
various densities and their derivatives are given in the Appendixes.

For open shell nuclei the pairing interaction becomes crucial and must be included.
It has been shown that fission barriers depend very much on the form and
strength of the effective pairing interactions~\cite{Karatzikos2010_PLB689-72}.
Several methods have been developed to treat the pairing effects,
{e.g.}, the BCS approach, the Bogoliubov transformation, and the particle number
conserving method~\cite{Zeng1983_NPA405-1, Molique1997_PRC56-1795, Meng2006_FPC1-38,
Pillet2002_NPA697-141,*Hao2012_PRC86-064307,
Zhang2011_PRC83-011304R,*Zhang2012_PRC85-014324};
all of them have been used in the study of PES's and fission barriers.
Since we use the BCS approach in our MDC-RMF calculations,
we only show the formulas for the BCS approximation.
In the particle-particle channel, the gap equation reads~\cite{Ring1980},
\begin{equation}
 \Delta_{k} =
 \sum_{k^{\prime}>0}
  \frac{1}{2} V_{k\bar{k}k^{\prime}\bar{k^{\prime}}}^{{\rm pp}}
  \frac{\Delta_{k^{\prime}}}
       {\sqrt{\tilde{\epsilon}_{k^{\prime}}^{2} + \Delta_{k^{\prime}}^{2}}},
\end{equation}
where $\tilde{\epsilon}_{k} = \epsilon_{k}-\lambda$ and $\lambda$ is the Fermi energy.
The total pairing energy is
\begin{equation}
 E_{{\rm pair}} = \frac{1}{4} V_{0} \int d^{3}\bm{r} \kappa^{*}(\bm{r}) \kappa(\bm{r}) .
\end{equation}
In our models, either the $\delta$-force or the finite-range separable
force~\cite{Tian2006_CPL23-3226,*Tian2009_PLB676-44,*Tian2009_PRC79-064301}
is implemented.

To obtain a PES one can perform a constraint calculation which is equivalent
to adding an external potential during the iteration~\cite{Ring1980}.
The quadratic constraint method is usually used,
\begin{equation}
 E^{\prime} = E_{{\rm RMF}}
            + \sum_{\lambda\mu} \frac{1}{2}
               C_{\lambda\mu} \left( Q_{\lambda\mu}-m_{\lambda\mu} \right)^{2},
\end{equation}
where $C_{\lambda\mu}$ is the spring constant and $m_{\lambda\mu}$'s are desired moments.
With this method the calculation always converges to a deformation point on the PES
other than the desired one.
To overcome this shortcoming and to get a PES with equally distributed points,
we use a modified linear constraint method. The Routhian reads,
\begin{equation}
 E^{\prime} = E_{{\rm RMF}} +
              \sum_{\lambda\mu} \frac{1}{2} C_{\lambda\mu}Q_{\lambda\mu} ,
\end{equation}
where the variables $C_{\lambda\mu}$'s change their values during the iteration
through the following relation,
\begin{equation}
 C_{\lambda\mu}^{(n+1)} =
 C_{\lambda\mu}^{(n)} +
  k_{\lambda\mu} \left( \beta_{\lambda\mu}^{(n)} - \beta_{\lambda\mu} \right),
\end{equation}
where $\beta_{\lambda\mu}$ is the desired deformation,
$k_{\lambda\mu}$ is a constant, and $C_{\lambda\mu}^{(n)}$ is the value at the $n$th step.
This constraint method works well in our multidimensionally-constrained calculations.

The total energy of the nucleus reads
\begin{eqnarray}
 E_{{\rm total}} & = &
 \int d^{3}\bm{r}
  \left\{ \sum_{k}
           v_{k}^{2} \psi_{k}^{\dagger}
           \left( \bm{\alpha}\cdot\bm{p}+\beta M \right) \psi_{k} \right.
 \nonumber \\
 &  & \mbox{}
         +\frac{1}{2} \alpha_{ S} \rho_{ S}^{2} + \frac{1}{2} \alpha_{ V} \rho_{ V}^{2}
         +\frac{1}{2} \alpha_{TS} \rho_{TS}^{2} + \frac{1}{2} \alpha_{TV} \rho_{TV}^{2}
 \nonumber \\
 &  & \mbox{}
         +\frac{1}{3} \beta_{S} \rho_{S}^{3} + \frac{1}{4} \gamma_{S} \rho_{S}^{4}
                                             + \frac{1}{4} \gamma_{V} \rho_{V}^{4}
 \nonumber \\
 &  & \mbox{}
         +\frac{1}{2} \delta_{S} \rho_{S} \Delta\rho_{S}
         +\frac{1}{2} \delta_{V} \rho_{V} \Delta\rho_{V}
 \nonumber \\
 &  & \left. \mbox{}
         +\frac{1}{2} \delta_{TS} \rho_{TS} \Delta\rho_{TS}
         +\frac{1}{2} \delta_{TV} \rho_{TV} \Delta\rho_{TV}
         +\frac{1}{2}e\rho_{C}A
  \right\}
 \nonumber \\
 &  & \mbox{}
         + E_{{\rm pair}} + E_{{\rm c.m.}}
  ,
\end{eqnarray}
where $E_{{\rm c.m.}}$ is the center of mass correction.
Depending on the effective interactions used in the RMF functional,
$E_{{\rm c.m.}}$ can be calculated either in the oscillator approximation,
\begin{eqnarray}
 E_{\rm c.m.} = -\dfrac{3}{4} \times 41 A^{1/3}\ \mathrm{MeV},
\end{eqnarray}
or
from the quasi-particle vacuum,
\begin{equation}
 E_{{\rm c.m.}} = -\frac{\langle \hat{P}^{2}\rangle}{2MA} ,
 \label{eq:Ecm_mic}
\end{equation}
where $\hat{P}$ is the total linear momentum and $A$ is the nuclear mass number.

The intrinsic multipole moments are calculated from the density by
\begin{equation}
 Q_{\lambda\mu} = \int d^{3}\bm{r} \rho_{V}(\bm{r}) r^{\lambda} Y_{\lambda\mu}(\Omega),
\end{equation}
where $Y_{\lambda\mu}(\Omega)$ is the spherical harmonics.
The deformation parameter $\beta_{\lambda\mu}$ is obtained from
the corresponding multipole moment by
\begin{equation}
 \beta_{\lambda\mu} = \frac{4\pi} {3NR^{\lambda}} Q_{\lambda\mu},
\end{equation}
where $R=1.2\times A^{{1}/{3}}$~fm is the nuclear radius
and $N$ is the number of protons, neutrons, or nucleons.

\section{~\label{sec:results}Results and discussions}

\subsection{\label{sec:numerical}Numerical details}

%\begin{figure}
%\begin{centering}
%%\includegraphics[width=0.45\textwidth]{pic/PU240_convg_check}
%\includegraphics[width=0.80\columnwidth]{PU240_TNsh}
%\par\end{centering}
%\caption{(Color online)~\label{fig:truncation}
%The potential energy curve of $^{240}$Pu calculated with different truncations
%of the ADHO basis from MDC-RMF calculations.
%The axial symmetry is imposed.
%The results calculated with $N_f=16$, 18, 20, and 22 are depicted by dashed, dotted,
%dot-dashed, and solid curves, respectively.
%The four sub-figures show the detailed structure of the potential energy curve
%near the ground state, the inner barrier, the isomeric state, and the outer barrier.
%The width of each sub-figure is 0.1 and the height is 1 MeV.
%}
%\end{figure}
In this section we present the numerical details and some illustrative calculations
of MDC-RMF models.
The potentials and densities are calculated in a spatial lattice
[cf. Eqs.~(\ref{eq:potentialexpansion}), (\ref{eq:potentialexpansion_0}),
and (\ref{eq:potentialexpansion_n}) and Appendix~\ref{appendix:density}] in which
mesh points in the $\rho$ and $z$ directions are designed in a way that
the Gaussian quadrature can be made and
those for the azimuthal angle $\phi$ are equally distributed.
Since we keep the mirror reflection symmetry with respect to the $x=0$ or $y=0$ planes,
only mesh points with positive $x$ and $y$ are considered.
For the azimuthal angle $\phi$, more than 10 mesh points for light nuclei and
about 20 mesh points for heavy nuclei are used, which are enough for most of
practical applications.
In two special cases the number of mesh points can be further reduced:
(1) for axially symmetric nuclei the azimuthal degree of freedom vanishes and
(2) for reflection symmetric nuclei the mesh points with $z<0$ can be omitted.
With these choices the values of the localized fields and potentials
in the full lattice space can be simply obtained by symmetry transformations
such as rotations or the spatial reflection.
The Coulomb field must be treated carefully due to its long-range nature.
In the pairing channel, we use a density-independent $\delta$ force with a smooth cutoff.
The pairing strength parameters are $V_{\rm n}=-349.5$ MeV fm$^{3}$ and
$V_{\rm p}=-330.0$ MeV fm$^{3}$ which are obtained by fitting the average pairing
gaps~\cite{Bender2000_EPJA8-59,Zhao2010_PRC82-054319}.

\begin{figure}
\begin{centering}
\includegraphics[width=0.95\columnwidth]{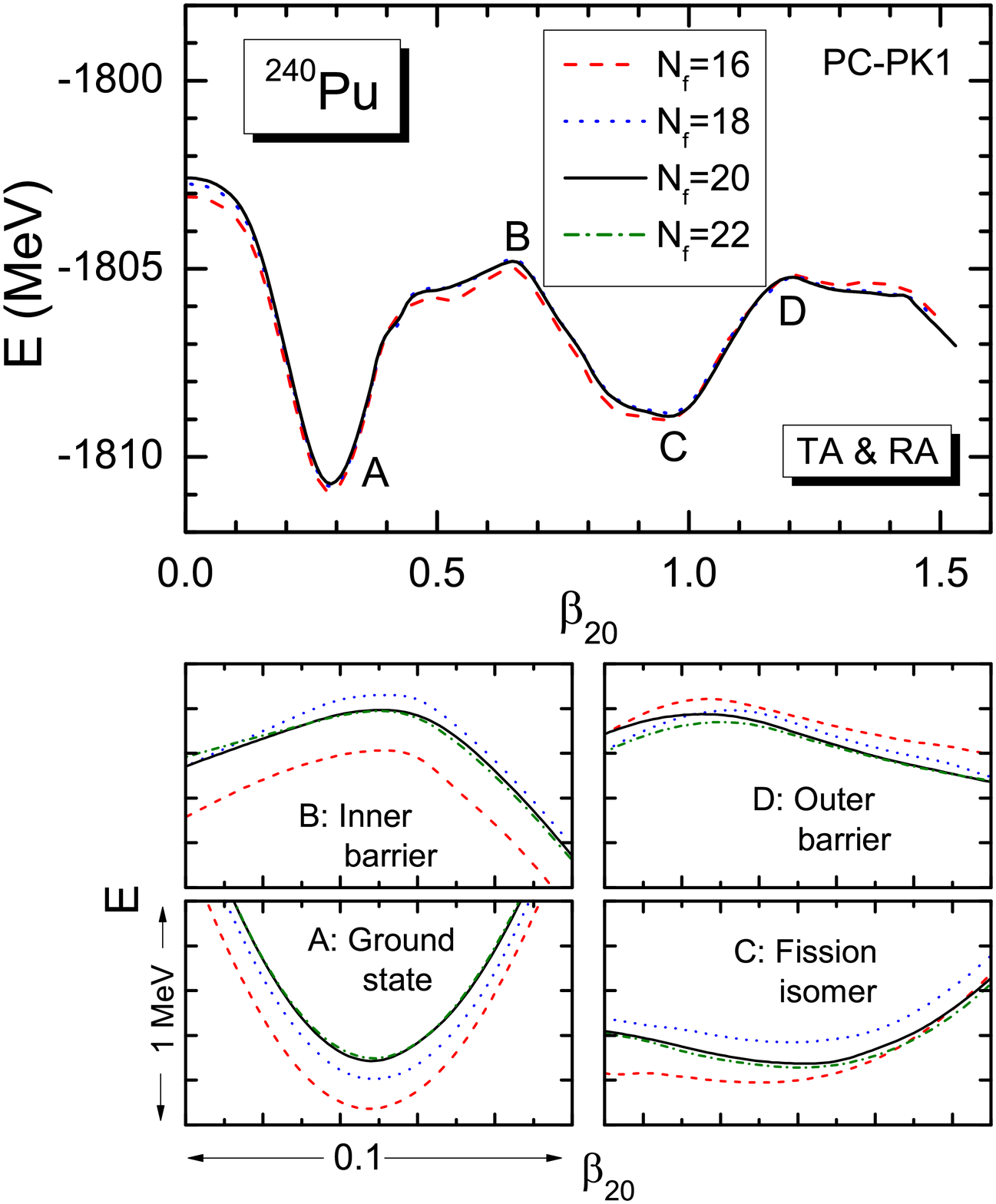}
\par\end{centering}
\caption{(Color online)~\label{fig:truncation}
%The same as Fig.~\ref{fig:truncation}, but for the triaxial+octupole calculations.
The potential energy curve of $^{240}$Pu from MDC-RMF calculations
with different truncations of the ADHO basis.
Both triaxial (TA) and reflection asymmetric (RA) deformations are allowed.
%The axial symmetry is imposed.
The results calculated with $N_f=16$, 18, and 20 are depicted by dashed, dotted,
and solid curves, respectively.
The four sub-figures show the detailed structure of the potential energy curve
near the ground state (A), the inner barrier (B), the isomeric state (C),
and the outer barrier (D).
The results calculated with $N_f=22$ (dot-dashed curves) are also included in the sub-figures.
The width of each sub-figure is 0.1 and the height is 1 MeV.
}
\end{figure}

The calculated physical observables should converge as the truncation
$N_f \rightarrow \infty$.
In Fig.~\ref{fig:truncation} we show the potential energy curve of $^{240}$Pu
calculated with different truncations, $N_f=$ 16, 18, and 20.
The effective interaction PC-PK1~\cite{Zhao2010_PRC82-054319,*Zhao2012_PRC86-064324} is used;
accordingly, $E_\mathrm{c.m.}$ is calculated with Eq.~(\ref{eq:Ecm_mic}).
%For simplicity these calculations are restricted to the axially symmetric case.
As is found in most of earlier calculations, the results show a typical two-humped
structure.
To see more clearly the truncation errors we have amplified the figure near
four important points, {i.e.}, the ground state,
the top of the inner barrier, the isomeric state, and the top of the outer barrier.
In these four sub-figures, the results with $N_f=$ 22 are also shown
and we can investigate in more details the convergence properties of our model.
When $N_f$ increases from 16 to 20, the binding energy changes
differently for different points in the potential energy curve;
the largest changes are near the ground state and about 0.3 MeV.
%(around the
%ground state and the inner barrier) or less than (around the fission isomer and the outer barrier)
%about 0.2 MeV.
Around the ground state and the second minimum, $^{240}$Pu is axially deformed;
the energy obtained from calculations with $N_f = $ 20 and 22 are almost the same.
This means a good convergence; in the present work,
we use $N_f=20$ in axially symmetric calculations.
Around the two fission barriers, the triaxial deformation is very important.
One finds that for the inner barrier, the results from $N_f = $ 20 and 22 are
also almost identical; for the outer one, the difference between the barrier heights
from calculations with $N_f = $ 20 and 22 is about several tens keV.
It is very time consuming to make calculations with both axial and reflection
symmetries broken.
%In our systematic calculations, $N_f=16$ is used when the nonaxial shapes
%are allowed.
%For the energy of the saddle points, this may result in a truncation error
%around 0.2 MeV. %(cf. sub-figure B in Fig.~\ref{fig:truncation})
%%or even less (cf. sub-figure D in Fig.~\ref{fig:truncation}).
%\label{modification:convergence}
%To summarize, the following strategy is adopted in the systematic calculations:
%Around the local minima which are axially deformed, $N_f=20$ is used and
%the truncation error is tiny (cf. sub-figures A and C in Fig.~\ref{fig:truncation});
%around the saddle points which are both nonaxial and reflection asymmetric,
%$N_f=16$ is used and the truncation error is around 0.2 MeV
%(cf. sub-figures B and D in Fig.~\ref{fig:truncation}).
%Therefore, in our calculations, the inaccuracy for the barrier height
%should be within 0.3 MeV.
%Thus, in our systematic calculations,
%around the ground states and the first saddle points, $N_f=20$ is used;
%around the second saddle points, $N_f=16$ is used 
%since they are both nonaxial and reflection asymmetric.
%As a result, inner fission barriers are described with an accuracy of $\sim 0.15$ MeV 
%and outer fission barriers with an accuracy of $\sim 0.4$ MeV.
%However, the truncation error for the outer barriers shoud be also within 0.2 MeV
%if $N_f=20$ is used.
In our systematic calculations presented in this paper, $N_f=20$ is used around 
the ground state and the first saddle point, $N_f=16$ is used around the fission isomer 
and the second saddle point. As a result, the inner barrier height is described with 
an accuracy of $\sim 0.15$ MeV and the outer one with an accuracy of $\sim 0.4$ MeV.
If in future calculations, $N_f=20$ can also be used around the second barrier,
the accuracy for its height should be within 0.2 MeV.

\begin{figure}
\begin{centering}
\includegraphics[width=0.90\columnwidth]{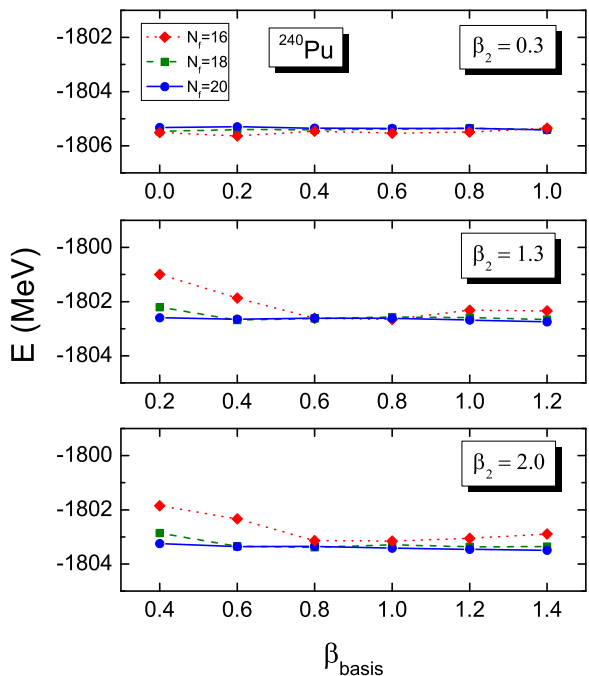}
\caption{(Color online)~\label{fig:basisdeformation}
Mean field energy ({i.e.}, $E_\mathrm{c.m.}$ is not included) of $^{240}$Pu calculated
by using basis with different
basis deformation $\beta_\mathrm{basis}$ and different $N_f$.
The results calculated with $N_f=16$, 18, and 20 shells are denoted by red diamonds, green squares,
and blue dots, respectively.
The deformation is constrained to $\beta_{20}=0.3$, 1.3, and 2.0, respectively.
}
\par\end{centering}
\end{figure}

Next we show how the results depend on the deformation of the ADHO potential which is
used to generate the ADHO basis; for brevity, we will call it the basis deformation and
label it with $\beta_\mathrm{basis}$.
In Fig.~\ref{fig:basisdeformation} we depict the calculated mean field energy of $^{240}$Pu
as a function of the basis deformation.
Note that $E_\mathrm{c.m.}$ is not included in Fig.~\ref{fig:basisdeformation}.
In principle, if the basis space is complete,
the results should not change when the basis deformation changes.
Near the ground state, $\beta_{20}=0.3$, the calculated energies are rather stable
against the basis deformation.
Furthermore, the results with $N_f=16$, 18, and 20 almost coincide with each other.
This conclusion holds also for the second barrier, $\beta_{20}=1.3$, except that two points
with small basis deformations and $N_f=16$ are very high.
For even larger deformations, $\beta_{20}=2.0$, the results with $N_f=16$
deviate a bit from those with $N_f=18$ and 20; the differences between
the results with $N_f=18$ and 20 are still very small.
In the present work, the basis deformation is chosen in the following way:
$\beta_{\rm basis} = \beta_{20}$ for $\beta_{20}<0.3$ and
$\beta_{\rm basis} = \beta_{20}/2$ for $0.3<\beta_{20}<2.0$.

\begin{figure}
\begin{centering}
\includegraphics[width=0.90\columnwidth]{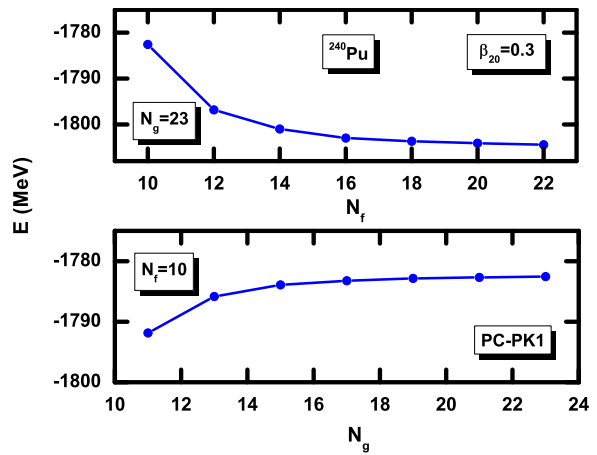}
\caption{(Color online)~\label{fig:Dirac}
Mean field energy ({i.e.}, $E_\mathrm{c.m.}$ is not included) of $^{240}$Pu calculated
with different truncations on large
(upper panel) and small component (lower panel).
In the calculations for the upper panel, $N_g$ are fixed to 23, while the calculations
for the lower panel are preformed with $N_f=10$.
The deformation is constrained to $\beta_{20}=0.3$.
}
\par\end{centering}
\end{figure}

When the basis is not complete, the calculated single-particle energies contain
the contributions from the other levels in both the Fermi sea and the Dirac sea.
This causes the ``variational collapse'' problem~\cite{Kutzelnigg1984_IJQChem25-107,
Fillion-Gourdeau2012_PRA85-022506}.
As shown in Fig.~\ref{fig:Dirac}, if we fix the truncation for the small component $N_g$
and increase $N_f$ alone, the nucleus becomes more and more bound.
Note that we have to set $N_f < N_g$ to prohibit the occurrence of the spurious states,
as we mentioned earlier.
However, if we fix the truncation for the large component $N_f$
and increase $N_g$ alone, the nucleus becomes less and less bound.
Thus, as more basis states are included in the expansion, the binding energy may not
change monotonically;
this is different from the nonrelativistic calculations.
%where the calculated binding energy always increases when the basis size is increased.

\begin{figure}[h]
\begin{centering}
\includegraphics[width=0.95\columnwidth]{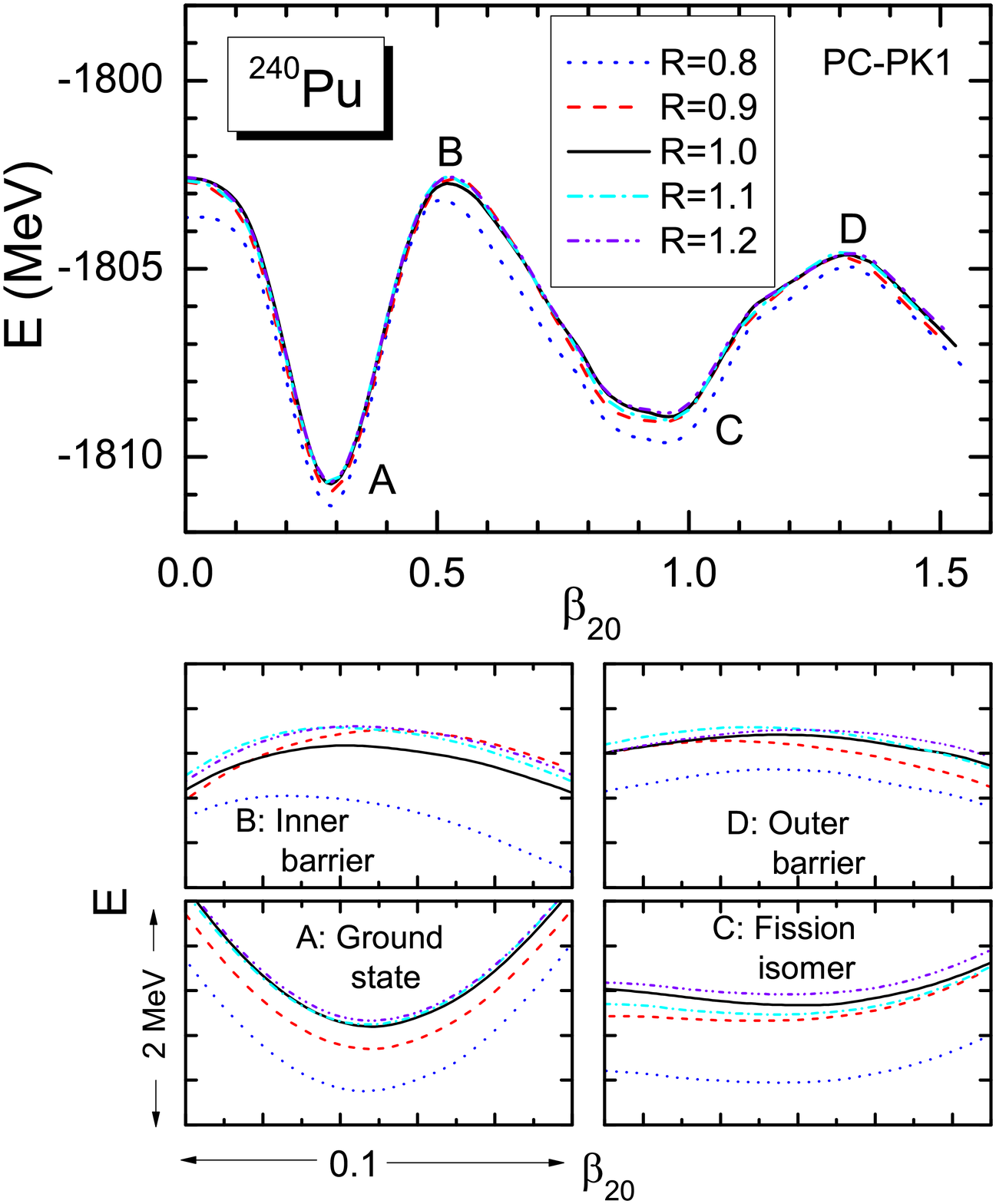}
\par\end{centering}
\caption{(Color online)~\label{fig:osilength}
The potential energy curve of $^{240}$Pu calculated by using basis
with different oscillator frequency $\hbar\omega_0$.
The axial symmetry is imposed and $N_f=20$ is used.
The results calculated with
$R\equiv\hbar\omega_0 / ( 41A^{-1/3}\ \mathrm{MeV}) = 0.8$, 0.9, 1.0, 1.1,
and 1.2 are depicted by dotted, dashed, solid, dot-dashed, and dash-dot-dotted curves, respectively.
The four sub-figures show the detailed structure of the potential energy curve
near the ground state (A), the inner barrier (B), the isomeric state (C),
and the outer barrier (D).
The width of each sub-figure is 0.1 and the height is 2 MeV.
}
\end{figure}

In the calculations, the oscillator length or, equivalently, the frequency of the
oscillator potential $\hbar\omega_0$ for the ADHO basis should also be chosen carefully.
The dependence of the binding energy of $^{240}$Pu on $\hbar\omega_0$ has been
investigated in detail.
As shown in Fig.~\ref{fig:osilength},
when $R\equiv \hbar\omega_0 / ( 41A^{-1/3}\ \mathrm{MeV})$ increases from 0.8 to 1.2,
the binding energy of $^{240}$Pu in the whole potential energy curve
varies by less than 1 MeV (0.05\% of the absolute value).\label{modification:osilength}
Moreover, when $\hbar\omega_0$ is around $41A^{-1/3}$ MeV, the binding energy
changes very slightly.
So, we set $\hbar\omega_0=41A^{-1/3}$ MeV in all calculations.

\subsection{\label{sec:240Pu}Three-dimensional PES of $^{240}$Pu}

\begin{figure*}%[h]
\begin{centering}
\includegraphics[width=0.9\textwidth]{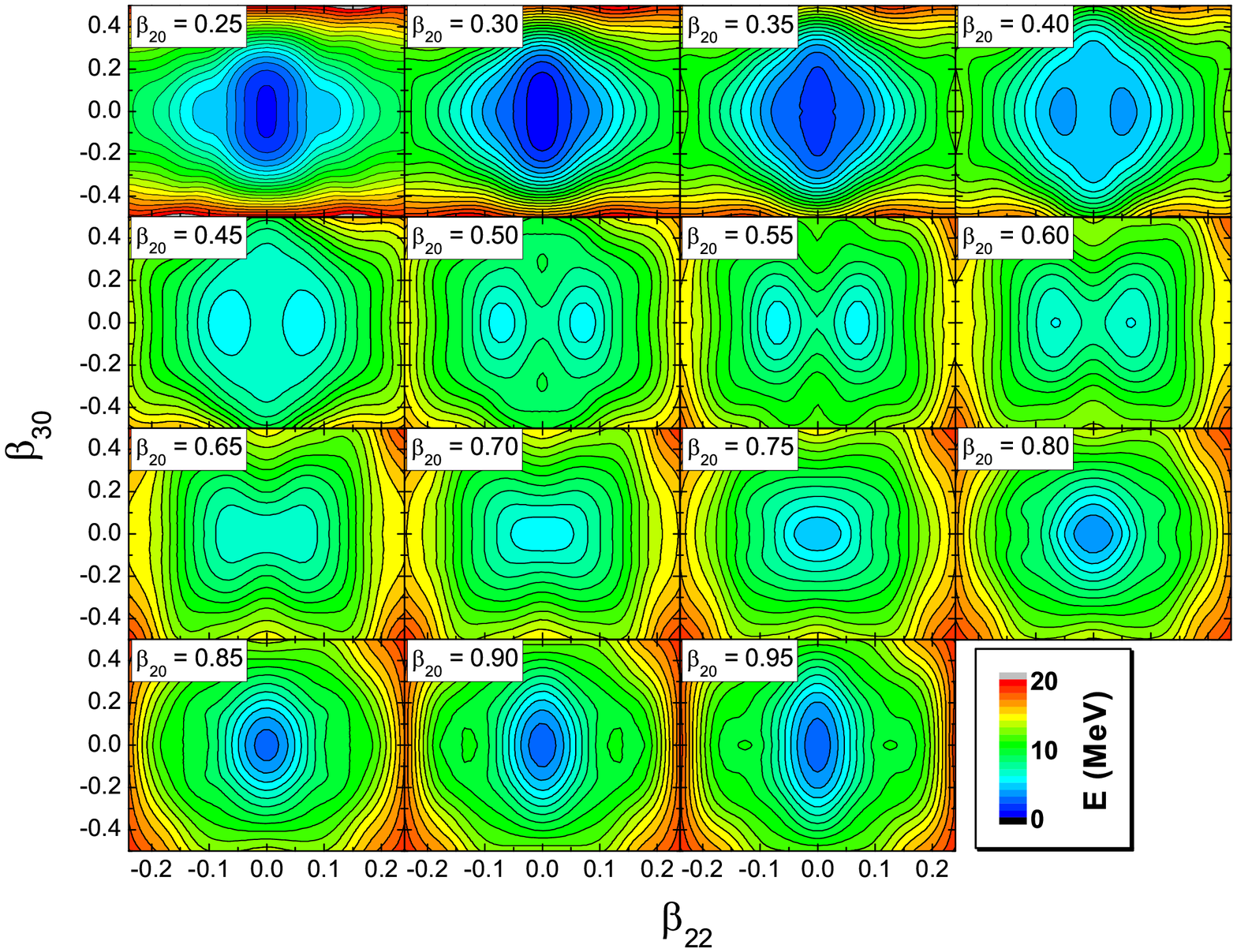}
\par\end{centering}
\caption{(Color online)~\label{fig:d3plotin}
Sections of the three-dimensional potential energy surface,
$E = E(\beta_{20},\beta_{22},\beta_{30})$,
of $^{240}$Pu around the ground state, the inner barrier, and the fission isomer
from MDC-RMF calculations.
In each sub-figure the energy is shown as a function of the deformation parameters
$\beta_{22}$ and $\beta_{30}$ when $\beta_{20}$ is fixed at a certain value.
The energies are normalized with respect to the binding energy of the ground state.
The contour interval is 1 MeV.}
\end{figure*}

\begin{figure*}%[h]
\begin{centering}
\includegraphics[width=0.9\textwidth]{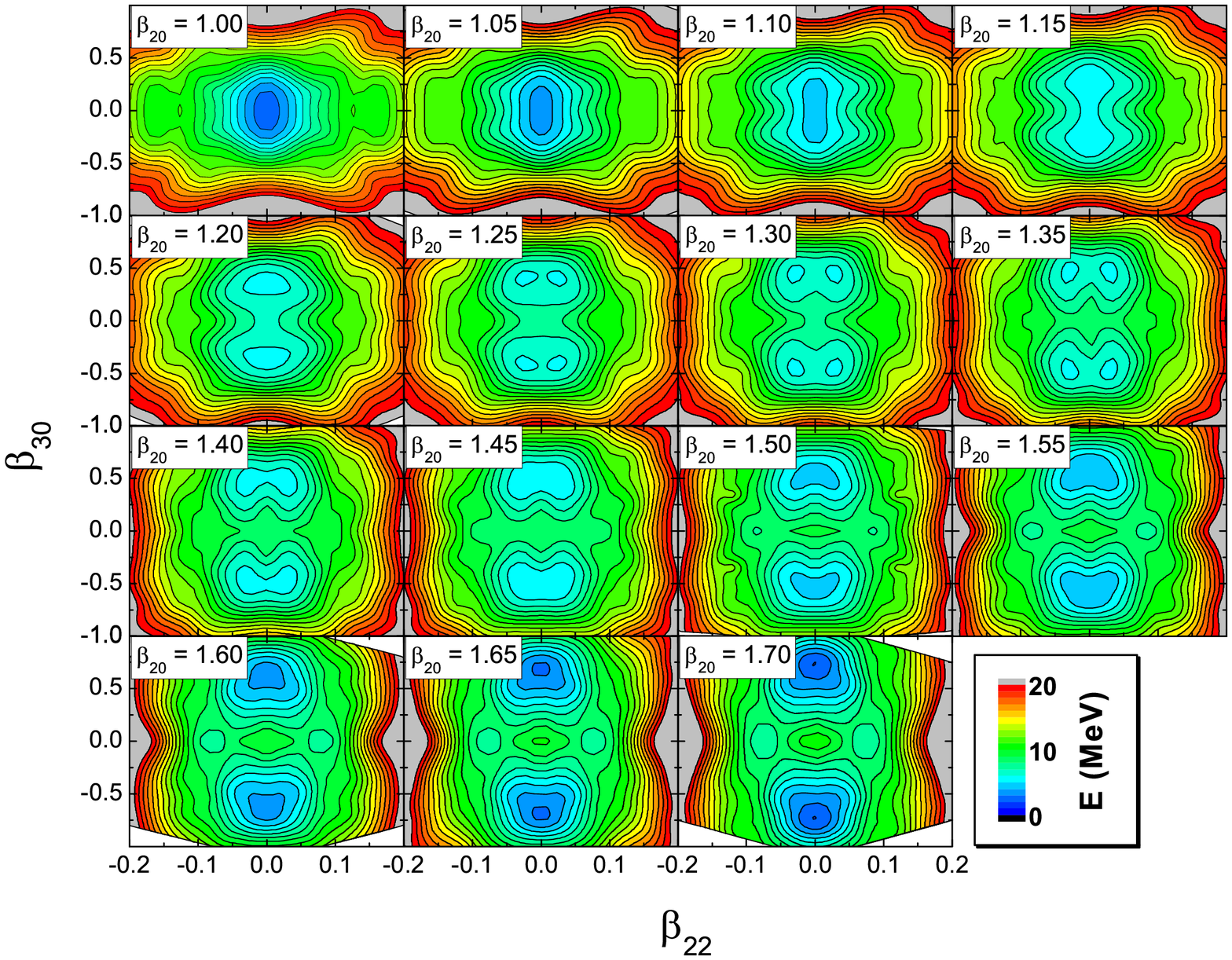}
\par\end{centering}
\caption{(Color online)~\label{fig:d3plotout}
Sections of the three-dimensional potential energy surface,
$E = E(\beta_{20},\beta_{22},\beta_{30})$,
of $^{240}$Pu around the fission isomer and the outer barrier from MDC-RMF calculations.
In each sub-figure the energy is shown as a function of the deformation parameters
$\beta_{22}$ and $\beta_{30}$ when $\beta_{20}$ is fixed at a certain value.
The energies are normalized with respect to the binding energy of the ground state.
The contour interval is 1 MeV.}
\end{figure*}

The double-humped fission barriers of actinide nuclei are usually used as
a benchmark for theoretical models, see, e.g., Refs.~\cite{Reichstein1976_AoP98-322,
Pomorska1979_NPA327-1,
Girod1983_PRC27-2317,
Tondeur1985_NPA442-460,
Blum1994_PLB323-262,
Rutz1995_NPA590-680,
Bender2004_PRC70-054304,
Andreev2005_EPJA26-327,
Lu2006_CPL23-2940,
Robledo2008_PRC77-051301,
Younes2009_PRC80-054313,
Li2010_PRC81-064321,
Kortelainen2012_PRC85-024304,
Hao2012_PRC86-064307}.
As a nucleus evolves from the ground state to the fission configurations,
many shape degrees of freedom play important and different roles in determining
the heights of the inner and outer barriers.
It has long been known that the inner barrier is lowered when
the triaxial deformation is allowed~\cite{Pashkevich1969_NPA133-400,
Moeller1970_PLB31-283, Randrup1976_PRC13-229},
while for the outer barrier,
the reflection asymmetric shape is favored~\cite{Ledergerber1973_NPA207-1}.
We have shown that the reflection asymmetric outer barrier may be further lowered by including the
nonaxial shape degrees of freedom~\cite{Lu2012_PRC85-011301R}.
In this section we show some results of multi-dimensional PES's for actinides.
We will first present detailed results about the three-dimensional PES of $^{240}$Pu,
then display systematic results of the actinide nuclei.
In these calculations, the effective interaction PC-PK1~\cite{Zhao2010_PRC82-054319,*Zhao2012_PRC86-064324} is used.

In Figs.~\ref{fig:d3plotin} and \ref{fig:d3plotout} is shown the calculated
three-dimensional PES of $^{240}$Pu.
In these calculations, both non-axial and reflection asymmetric shapes are
allowed and $N_f=16$ is used.
In each sub-figure we fix the value of $\beta_{20}$ and display the energy as a function of
$\beta_{22}$ and $\beta_{30}$.
In other words, we have made three-dimensional constraints on the corresponding multipole moments.
Note that other shape degrees of freedom $\beta_{\lambda\mu}$ with even $\mu$,
{e.g.}, $\beta_{32}$, $\beta_{40}$, $\beta_{42}$, $\beta_{44}$, $\beta_{50}$, $\dots$,
are also included in the calculations self-consistently.
The MDC-RMF equations are solved for each point on the deformation lattice
$(\beta_{20},\beta_{22},\beta_{30})$ in which
$\beta_{20}$ runs from 0.25 to 1.70 with a step size of 0.05,
$\beta_{22}$      from 0    to 0.25 with a step size of 0.01, and
$\beta_{30}$      from 0    to 0.50 with a step size of 0.05.
The points with $\beta_{22}(\beta_{30})<0$ are obtained through the relation
$E(\beta_{20},\beta_{22},\beta_{30})=E(\beta_{20},|\beta_{22}|,|\beta_{30}|)$.
That is, for each sub-figure, 26 (for $\beta_{22}$) $\times$ 11
(for $\beta_{30}$) = 286 points are calculated, and there are totally 286
$\times$ 30 (for $\beta_{20}$) = 8580 points
in Figs.~\ref{fig:d3plotin} and \ref{fig:d3plotout}.
This deformation lattice covers the shape space of the most interest for $^{240}$Pu,
from the ground state to the isomeric state and fission configurations.

Now we examine the first three sub-figures with $\beta_{20}=0.25$, 0.30, and 0.35.
It is clear that the ground state with $\beta_{20}\sim 0.3$ is
both axially symmetric and reflection symmetric,
though it is a little soft against the octupole distortion.
When the nucleus is stretched by the quadrupole constraining potential,
it becomes softer against the triaxial as well as the octupole distortions.
From the sub-figures with $\beta_{20}=0.40\sim0.65$ one finds that
two symmetric minima with non-zero triaxial deformation $\beta_{22}$ appear and
the corresponding fission paths are much more favored than the axially symmetric one,
which is consistent with earlier calculations~\cite{Abusara2010_PRC82-044303}.
Although $\beta_{30}=0$ for all minima in these sub-figures, the softness against
the octupole distortion changes as the nucleus is elongated.
The inner fission barrier locates near the deformation $\beta_{20}\sim0.60$.
For the last six sub-figures in Fig.~\ref{fig:d3plotin} with $\beta_{20}=0.70\sim0.95$,
the situation becomes much simpler, where the nucleus becomes axially symmetric
and reflection symmetric again.
Nevertheless, the trend to become softer against the octupole distortion can be seen from
these sub-figures, which indicates that the reflection asymmetric shape becomes more
relevant.

As the deformation $\beta_{20}$ becomes larger, the second minimum
and the second saddle point of the PES appear.
From the last three sub-figures of Fig.~\ref{fig:d3plotin} and
the first three sub-figures of Fig.~\ref{fig:d3plotout} we see that,
the nucleus keeps reflection symmetric near the fission isomeric state with $\beta_{20}\sim 0.95$,
but it becomes softer against the $\beta_{30}$ distortion.
Note that the scales of the $\beta_{30}$ axes are different
in Figs.~\ref{fig:d3plotin} and \ref{fig:d3plotout}.
At $\beta_{20}=1.15$, two minima corresponding to reflection asymmetric shapes appear.
Here the effect of the nonaxial deformation is not apparent,
but along the $\beta_{22}$ direction the PES becomes softer.
At $\beta_{20}=1.2$ the energy of the reflection asymmetric shape is lower
by about 1 MeV than that of the reflection symmetric one.
Interestingly, when $\beta_{20}$ increases further, around the top of the second
barrier,
each reflection asymmetric minimum splits into two minima with non-zero $\beta_{22}$.
This is the lowering effects of triaxiality on the outer barrier found in
Ref.~\cite{Lu2012_PRC85-011301R}.
Around the second barrier, the largest energy gain due to the triaxial distortion is about 1 MeV.
The nucleus becomes axially symmetric again when $\beta_{20}>1.6$.

From the above discussions, we can draw the following conclusions for $^{240}$Pu:
(1) Both the ground state and the fission isomeric state are axial and reflection symmetric;
(2) Around the first fission barrier it assumes triaxial and reflection symmetric shapes;
(3) Around the second fission barrier both triaxial and octupole deformations are important.

\subsection{PES's of even-even actinide nuclei around two fission barriers}
\label{subsec:barriers}

\begin{figure*}
\begin{centering}
\includegraphics[width=0.9\textwidth]{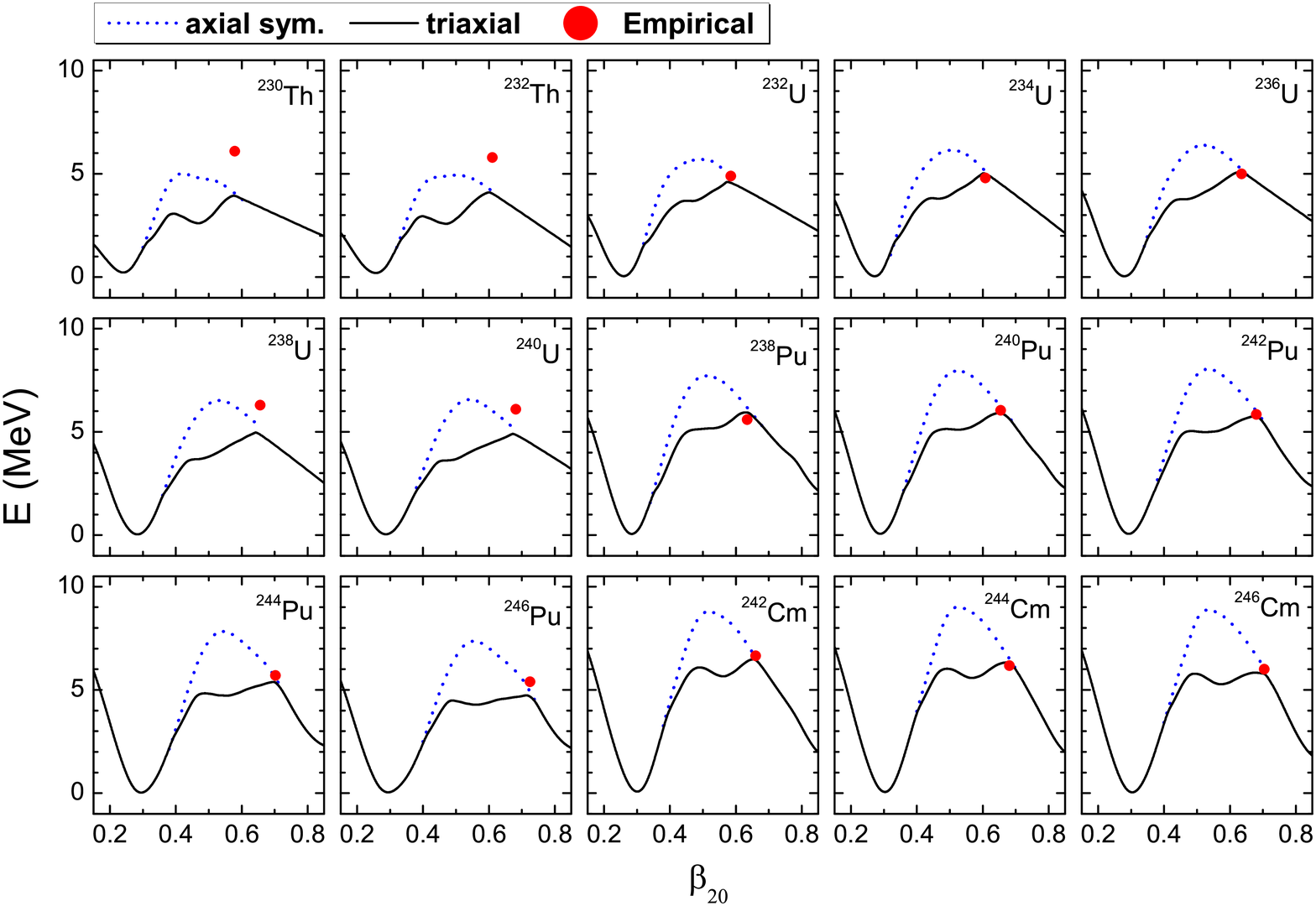}
\par\end{centering}
\caption{(Color online)~\label{fig:actinidePESinner}
Potential energy curves of even-even actinide nuclei around the
ground states and the first fission barriers
from MDC-RMF calculations.
The axially symmetric results are displayed by dotted curves,
while those from the triaxial calculations are shown by solid curves.
The binding energy is normalized with respect to that of
the ground state of each nucleus.
The empirical values of fission barriers are taken from Ref.~\cite{Capote2009_NDS110-3107}
and shown by red dots.
}
\end{figure*}

The self-consistent three-dimensionally-constrained calculations are very time-consuming,
we have only performed such calculations for $^{240}$Pu.
From this benchmark study we learned many experiences about the important roles
played by various shape degrees of freedom in different regions of the deformation space,
including the conclusions listed in the end of Sec.~\ref{sec:240Pu}.
These experiences are used in a systematic study of even-even actinide nuclei.

Since around the inner barrier an actinide nucleus assumes triaxial
but reflection symmetric shapes,
we can make a one-dimensional constraint calculation with
the triaxial deformation allowed and the reflection symmetry imposed.
In Fig.~\ref{fig:actinidePESinner} we present
potential energy curves of even-even actinide nuclei around the ground state and
the first fission barrier from MDC-RMF calculations.
The axially symmetric results are displayed by dotted curves
and those from triaxial calculations are shown by solid curves.
The empirical values of fission barrier heights
are taken from Ref.~\cite{Capote2009_NDS110-3107}
and shown with red dots.
We also list in Table~\ref{tab:barriers} the heights of the first and second fission barriers of
actinide nuclei from MDC-RMF calculations compared with the empirical values.
It is clearly seen that the triaxial deformation lowers the inner barrier
of these actinide nuclei by about $1 \sim 4$ MeV and
this is consistent with Ref.~\cite{Abusara2010_PRC82-044303}.
The agreement of our calculation results with the empirical ones
is good for most of the nuclei studied here with exceptions in the two thorium
isotopes and $^{238,240}$U.
We have discussed possible reasons for these discrepancies in Ref.~\cite{Lu2012_PRC85-011301R}.

\begin{figure*}
\begin{centering}
\includegraphics[width=0.9\textwidth]{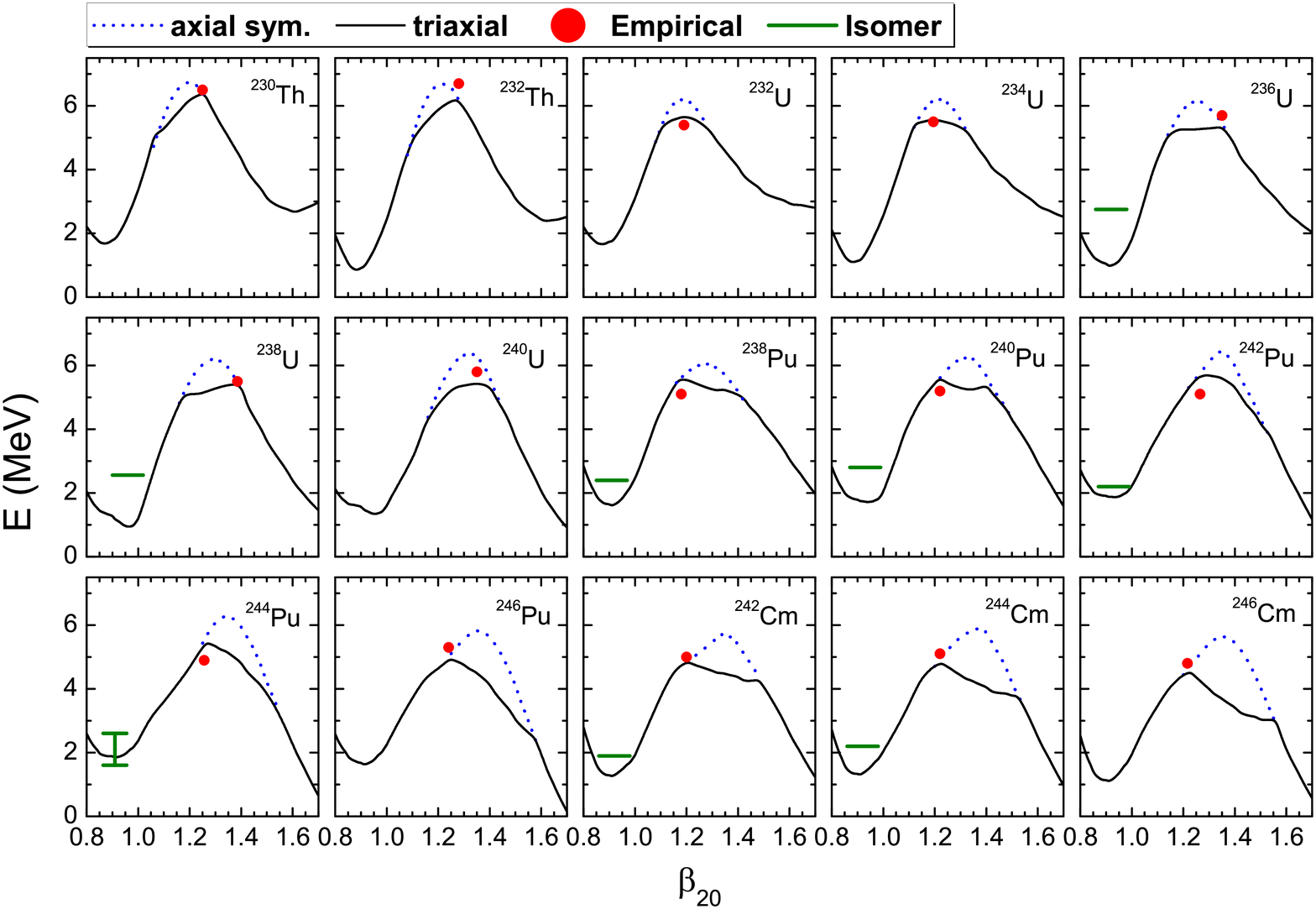}
\par\end{centering}
\caption{(Color online)~\label{fig:actinidePESouter}
Potential energy curves of even-even actinide nuclei around the fission isomers and
the second fission barriers from MDC-RMF calculations.
The reflection asymmetric shapes are allowed in the calculations.
The axially symmetric results are displayed by dotted curves,
while those from the triaxial calculations are shown by solid curves.
The binding energy is normalized with respect to that of
the ground state of each nucleus.
The empirical values of fission barriers are taken from Ref.~\cite{Capote2009_NDS110-3107}
and shown by red dots.
Experimental values of energies of fission isomers are taken from Ref.~\cite{Singh2002_NDS97-241}
and shown by green horizontal lines.
For $^{244}$Pu, only an energy range was given for the isomer.
}
\end{figure*}

It is more complicated to determine the second fission barrier height
because more shape degrees of freedom become
important and there are often two or more fission paths.
What we have done is the following~\cite{Lu2012_PRC85-011301R}:
(1) The axial symmetry is assumed and a two-dimensional constraint calculation
in the $(\beta_{20},\beta_{30})$ plane is made;
(2) From the two-dimensional calculation the lowest fission path
$\beta_{30}^\mathrm{lowest}(\beta_{20})$ is approximately identified;
(3) Along this fission path, a one-dimensional $\beta_{20}$-constrained
calculation is performed with the triaxial and octupole deformations allowed
and at each point with $\beta_{20}$ fixed, the initial deformations are taken as
$\beta_{22}^\mathrm{ini.} = 0$ and
$\beta_{30}^\mathrm{ini.} = \beta_{30}^\mathrm{lowest}(\beta_{20})$;
(4) In this one-dimensional potential energy curve, the second saddle point
is located and the second barrier height is extracted.

Potential energy curves of even-even actinide nuclei around the fission isomer and
the second fission barrier
from MDC-RMF calculations are shown in Fig.~\ref{fig:actinidePESouter}.
In addition, the experimental values of the energies for fission isomers~\cite{Singh2002_NDS97-241}
are also shown by short horizontal lines.
For $^{244}$Pu, a suggested isomeric energy range (1.6 MeV to 2.6 MeV) from systematics was shown.
\label{modification:isomer}
From Fig.~\ref{fig:actinidePESouter} and Table~\ref{tab:barriers} one finds that
for most of the nuclei investigated here, the triaxiality lowers the second barrier by
about 0.5 $\sim$ 1 MeV (about 10 $\sim$ 20\% of the barrier height)~\cite{Lu2012_PRC85-011301R}.
The calculation results with the triaxiality included agrees well with
the empirical values for all actinide nuclei shown in Fig.~\ref{fig:actinidePESouter}.
In Ref.~\cite{Lu2012_PRC85-011301R}, we have found that for $^{248}$Cm,
the height of the second
barrier without the triaxial deformation included is already
smaller than the empirical value.
From Table~\ref{tab:barriers} it is seen that this is also the case
for $^{250}$Cm and $^{250,252}$Cf.
For these four nuclei, it is difficult to find the second saddle point
when the triaxial deformation is included.
Therefore the heights of the second barriers with the triaxial deformation
included are not listed for them.
The reason for these discrepancies may be related to
a strong competition between the two or among more fission paths and
the assumption on the barrier width made when the empirical value of
the barrier height is evaluated, as discussed in Ref.~\cite{Lu2012_PRC85-011301R}.
%For $^{244}$Pu only an energy range was given
%Due to the zero-point vibration, the energy of an isomer is always higher than
%the second minimum in the potential energy curve.
%They are always higher than our calculations
%except that for $^{244}$Pu only an energy range was given.

\begin{table}
\begin{centering}
\caption{\label{tab:barriers}%
Heights of the first and second fission barriers of some even-even actinide nuclei
from MDC-RMF calculations
compared with the empirical values (``Emp'') which are taken from
Ref.~\cite{Capote2009_NDS110-3107}.
The calculation results with and without the axial symmetry imposed are denoted by
``AS'' (axially symmetric) and ``TA'' (triaxial), respectively.
Note that around the first barrier, the reflection symmetry is assumed and
for the second barrier, the octupole deformation is included.
The height is in MeV.
}
\begin{ruledtabular}
\begin{tabular}{rcccccccccccc}
    Nucleus && $Z$& $N$ & $A$ &  &\multicolumn{3}{c}{First barrier}
                                                      &  & \multicolumn{3}{c}{Second barrier} \\
            &&    &     &     &  & AS   &   TA &  Emp &  &   AS &   TA & Emp \tabularnewline
\hline
 $^{230}$Th && 90 & 140 & 230 &  & 5.03 & 3.96 & 6.10 &  & 6.80 & 6.37 & 6.50\tabularnewline
 $^{232}$Th && 90 & 142 & 232 &  & 4.94 & 4.12 & 5.80 &  & 6.70 & 6.18 & 6.70\tabularnewline
  $^{232}$U && 92 & 140 & 232 &  & 5.71 & 4.81 & 4.90 &  & 6.20 & 5.64 & 5.40\tabularnewline
  $^{234}$U && 92 & 142 & 234 &  & 6.15 & 5.09 & 4.80 &  & 6.20 & 5.55 & 5.50\tabularnewline
  $^{236}$U && 92 & 144 & 236 &  & 6.40 & 5.11 & 5.00 &  & 6.15 & 5.31 & 5.70\tabularnewline
  $^{238}$U && 92 & 146 & 238 &  & 6.54 & 5.03 & 6.30 &  & 6.20 & 5.42 & 5.50\tabularnewline
  $^{240}$U && 92 & 148 & 240 &  & 6.58 & 4.96 & 6.10 &  & 6.38 & 5.43 & 5.80\tabularnewline
 $^{238}$Pu && 94 & 144 & 238 &  & 7.72 & 5.96 & 5.60 &  & 6.05 & 5.56 & 5.10\tabularnewline
 $^{240}$Pu && 94 & 146 & 240 &  & 7.98 & 5.92 & 6.05 &  & 6.24 & 5.60 & 5.20\tabularnewline
 $^{242}$Pu && 94 & 148 & 242 &  & 8.05 & 5.77 & 5.85 &  & 6.43 & 5.74 & 5.10\tabularnewline
 $^{244}$Pu && 94 & 150 & 244 &  & 7.85 & 5.40 & 5.70 &  & 6.26 & 5.49 & 4.90\tabularnewline
 $^{246}$Pu && 94 & 152 & 246 &  & 7.37 & 4.76 & 5.40 &  & 5.84 & 4.96 & 5.30\tabularnewline
 $^{242}$Cm && 96 & 146 & 242 &  & 8.80 & 6.49 & 6.65 &  & 5.72 & 4.85 & 5.00\tabularnewline
 $^{244}$Cm && 96 & 148 & 244 &  & 9.04 & 6.34 & 6.18 &  & 5.90 & 4.88 & 5.10\tabularnewline
 $^{246}$Cm && 96 & 150 & 246 &  & 8.89 & 5.84 & 6.00 &  & 5.40 & 4.62 & 4.80\tabularnewline
 $^{248}$Cm && 96 & 152 & 248 &  & 8.43 & 5.35 & 5.80 &  & 4.10 & ---  & 4.80\tabularnewline
 $^{250}$Cm && 96 & 154 & 250 &  & 7.77 & 4.79 & 5.40 &  & 2.60 & ---  & 4.40\tabularnewline
 $^{250}$Cf && 98 & 152 & 250 &  & 8.87 & 5.70 & 5.60 &  & 2.40 & ---  & 3.80\tabularnewline
 $^{252}$Cf && 98 & 154 & 252 &  & 8.41 & 5.26 & 5.30 &  & 1.20 & ---  & 3.50\tabularnewline
\end{tabular}
\end{ruledtabular}
\par\end{centering}
\end{table}

\subsection{Parameter dependence of the effect of triaxiality around the second barrier}

\begin{table}
\caption{\label{tab:para_dep}%
Height of the second barrier of $^{234}$U calculated from MDC-RMF models
with different parameter sets.
The results with and without nonaxial deformations included and their difference
are denoted by $B_{\rm AS}$, $B_{\rm TA}$, and $\Delta B$,
respectively.
}
\begin{ruledtabular}
\begin{tabular}{l | c c c}
 Parameter set & $B_{\rm AS}$ (MeV)& $B_{\rm TA}$ (MeV)& $\Delta B$ (MeV)\\
\hline
 NL3*          & 7.54              & 6.85              & 0.69            \\
 NL-Z2         & 4.83              & 3.91              & 0.92            \\
 PC-PK1        & 6.20              & 5.55              & 0.65            \\
 DD-ME2        & 8.19              & 7.51              & 0.68            \\
 DD-PC1        & 6.13              & 5.64              & 0.49            \\
\end{tabular}
\end{ruledtabular}
\end{table}

From the above detailed studies of PES's of actinide nuclei
in the ($\beta_{20}$, $\beta_{22}$, $\beta_{30}$) space,
it is clear that the triaxiality plays an important role
upon the second fission barriers of actinide nuclei.
We have studied the parameter dependence of the effect of triaxiality on the second barrier
height and some results can be found in Refs.~\cite{Lu2012_PRC85-011301R, Lu2012_PhD, Lu2013_arXiv1304.6830}.
In Table~\ref{tab:para_dep} are listed the second barrier heights of $^{234}$U calculated with
different parameter sets, including meson-exchange ones
NL3{*}~\cite{Lalazissis1997_PRC55-540,*Lalazissis2009_PLB671-36},
NL-Z2~\cite{Bender1999_PRC60-034304},
DD-ME2~\cite{Lalazissis2005_PRC71-024312}, and
point-coupling ones PC-PK1~\cite{Zhao2010_PRC82-054319,*Zhao2012_PRC86-064324} and
DD-PC1~\cite{Niksic2008_PRC78-034318}.
One finds that the second barrier height may differ by a few MeV with different
effective interactions, but in all cases the barriers are lowered considerably
by the nonaxial deformations.
For this specific nucleus, the lowering effect with different effective interactions
is about 0.5 MeV to 0.9 MeV
which are clearly larger than the possible errors from the basis truncations discussed
in Sec.~\ref{sec:numerical}.

\subsection{Further discussions}

A systematic study of fission barriers has been performed for even-even superheavy nuclei
with $Z=112\sim120$ by using CDFTs and the outer fission
barriers are found to be considerably affected by the triaxiality~\cite{Abusara2012_PRC85-024314}.
However, in a recent work within the MM model,
it was found that the influence of the triaxiality
on the second fission barriers is negligible~\cite{Jachimowicz2012_PRC85-034305}.
This raises an open problem: What are the main reasons for the different effects
of the triaxiality on the second fission barriers predicted by the MM model and CDFTs?
It would be interesting to make more investigations with different models, especially,
with the nonrelativistic DFTs.

One of the problems concerning the PES calculated from self-consistent approaches
is that there may exist unexpected discontinuities on PES's~\cite{Moeller2009_PRC79-064304}.
This is mainly due to the complexity of multi-dimensional PES's.
When a high-dimensional PES is projected onto a low-dimensional one,
the minimizing procedure incorporated implicitly in the self-consistent approaches
may cause a sudden change in the total energy at some points in
the low-dimensional deformation space.
This may result in spurious saddle points.
In our MDC-RMF calculations, we have tried some ways to avoid discontinuities
on the PES's and to exclude spurious saddle points~\cite{Lu2012_PRC85-011301R}.
To be completely free of discontinuities or spurious saddle points, one certainly
should carry out multidimensionally-constrained calculations with higher-order multipole
deformations included.
In Ref.~\cite{Dubray2012_CPC183-2035} the authors analyzed the origin of
the discontinuities and proposed a numerical method to identify them.
It will be useful to implement this method in the present MDC-RMF calculations;
this will be one of our topics in the future.

Since many symmetries are broken in mean field calculations, quantum numbers
related to these symmetries are no longer conserved~\cite{Ring1980}.
For example, the total angular momentum or nuclear spin $J$ is not a good quantum number
when $\beta_{20}$ is not zero; similarly, the projection of the total angular momentum
on the symmetry axis $K$ is also not conserved if a nucleus is triaxially
deformed~\cite{Yao2010_PRC81-044311,*Yao2011_PRC83-014308,*Yao2011_SCG54-198}.
When the reflection symmetry is broken, the parity is not a good quantum number.
In our three-dimensional constraint calculations, all these quantum numbers are
not good ones.
One may introduce techniques of projection to restore the broken symmetries~\cite{Ring1980}.
In recent years, the importance of various projections on the PES's and fission
barriers have been investigated~\cite{Egido2000_PRL85-1198, Liu2011_EPJA47-135,
Hao2012_IJMPE21-1250051, Hao2012_PRC86-064307%
}.

\section{~\label{sec:summary}Summary}

We developed multidimensionally-constrained relativistic mean field (MDC-RMF) models.
In these models, the nuclear shape is assumed to be invariant under
the reversion of $x$ and $y$ axes, {i.e.},
the intrinsic symmetry group is $V_{4}$ and all shape degrees of freedom
$\beta_{\lambda\mu}$ with even $\mu$ ($\beta_{20}$, $\beta_{22}$, $\beta_{30}$,
$\beta_{32}$, $\beta_{40}$, $\dots$) are included self-consistently.
The RMF functional can be one of the following four forms:
the meson exchange or point-coupling nucleon interactions combined with
the nonlinear or density-dependent couplings.
We solve the Dirac equation in an axially deformed harmonic oscillator (ADHO) basis.
The convergence of the calculated results against the basis truncation,
the oscillator length,
and the basis deformation is studied and it is shown that reasonably
large ADHO basis is able to provide a desired accuracy.

Three-dimensional potential energy surface in the ($\beta_{20}$, $\beta_{22}$, $\beta_{30}$)
plane is obtained for $^{240}$Pu and
potential energy curves of actinide nuclei around the first and second
fission barriers are studied systematically.
It is found that the triaxiality is crucial in determining the height of both the
first and the second fission barriers.
Taking $^{234}$U as an example, we have studied the parameter dependence of
the effects of triaxiality on the second barrier height and found that
the height of the second barrier may differ by a few MeV with different
effective interactions, but in all cases the barriers are lowered considerably
by the nonaxial deformations.

We conclude that it is important to include both the nonaxial and
the spatial reflection asymmetric
shapes simultaneously for the study of potential energy surfaces and fission barriers
of actinide nuclei and of those in unknown mass regions such as, {e.g.}, superheavy nuclei.

\begin{acknowledgments}
Helpful discussions with
Haozhao Liang, Jie Meng, P. Ring, N. Schunck, D. Vretenar, Xi-Zhen Wu, and Zhen-Hua Zhang
are acknowledged.
This work has been supported by Major State Basic Research Development
Program of China (Grant No. 2013CB834400),
National Natural Science Foundation of China (Grants No. 11121403, No. 11175252,
No. 11120101005, No. 11211120152, and No. 11275248),
Knowledge Innovation Project of Chinese Academy of Sciences (Grant No. KJCX2-EW-N01).
The results described in this paper are obtained on 
the High-performance Computing Clusters of ITP/CAS and 
the ScGrid of the Supercomputing Center, Computer Network Information Center of 
the Chinese Academy of Sciences.
\end{acknowledgments}

\appendix

\section{Matrix elements of the Dirac Hamiltonian}

In the ADHO basis, the Dirac equation (\ref{eq:Diracequation}) is transformed into
a matrix eigenvalue problem,
\begin{eqnarray}
  \sum_{\alpha'} \langle \alpha|M_{+}(\bm{\bm{r}})    |\alpha' \rangle f_{i}^{\alpha'}
 +\sum_{\alpha'} \langle \alpha|\bm{\sigma}\cdot\bm{p}|\alpha' \rangle g_{i}^{\alpha'}
 & = &
  \epsilon_{i} f_{i}^{\alpha} ,
 \nonumber \\
  \sum_{\alpha'} \langle \alpha|\bm{\sigma}\cdot\bm{p}|\alpha' \rangle f_{i}^{\alpha'}
 +\sum_{\alpha'} \langle \alpha|M_{-}(\bm{r})         |\alpha' \rangle g_{i}^{\alpha'}
 & = &
  \epsilon_{i} g_{i}^{\alpha}.
 \nonumber
\end{eqnarray}
with $M_\pm = \pm M + V \pm S$.

In the cylindrical coordinate system the kinetic energy operator
$\bm{\sigma}\cdot\bm{p}$ reads,
\begin{equation}
 \bm{\sigma}\cdot\bm{p} =
 -i \left( \begin{array}{cc}
            \partial_{z} & e^{-i\varphi}(\partial_{\rho}-i\rho^{-1}\partial_{\varphi}) \\
            e^{i\varphi}(\partial_{\rho}+i\rho^{-1}\partial_{\varphi}) & -\partial_{z}
           \end{array}
    \right).
 \nonumber
\end{equation}
The matrix elements of the kinetic energy operator $\bm{\sigma}\cdot\bm{p}$ can be calculated as
\begin{eqnarray}
 \langle \alpha | \bm{\sigma}\cdot\bm{p} | \alpha' \rangle
 & = &
 \langle n_{z} ,n_{\rho} ,m_{l} ,m_{s} | \bm{\sigma}\cdot\bm{p} |
         n_{z}',n_{\rho}',m_{l}',m_{s}'
 \rangle
 \nonumber \\
 & = &
 -i C_{\alpha}^{*} C_{\alpha'} \delta_{K,K'}
 \left[ \delta_{m_{l},m_{l}'} \delta_{n_{\rho},n_{\rho}'} \delta_{m_{s},m_{s}'} A_{\alpha\alpha'}
 \right.
 \nonumber \\
 &  &
% \left.
 \mbox{}
 +\delta_{m_{s},-\frac{1}{2}} \delta_{m_{s}',\frac{1}{2}}
  \delta_{n_{z},n_{z}'}
  \left( B_{\alpha\alpha'} - m_{l}'C_{\alpha\alpha'} \right)
% \right.
 \nonumber \\
 &  &
 \left. \mbox{}
 +\delta_{m_{s},\frac{1}{2}} \delta_{m_{s}',-\frac{1}{2}}
  \delta_{n_{z},n_{z}'}
  \left( B_{\alpha\alpha'} + m_{l}'C_{\alpha\alpha'} \right)
 \right],
 \nonumber \\
\end{eqnarray}
where the integrals read,
\begin{eqnarray}
 A_{\alpha\alpha'} & = & \int dz    \phi_{n_{z}} \partial_{z} \phi_{n_{z}'} , \\
 B_{\alpha\alpha'} & = & \int d\rho R_{n_{\rho}}^{m_{l}} \rho \partial_{\rho}R_{n_{\rho}'}^{m_{l}'} , \\
 C_{\alpha\alpha'} & = & \int d\rho R_{n_{\rho}}^{m_{l}} R_{n_{\rho}'}^{m_{l}'} .
\end{eqnarray}

The matrix elements of the potentials $U(\bm{r}) \equiv M_{\pm}(\bm{r})$ are calculated
by expanding $U(\bm{r})$ into a Fourier series,
\begin{equation}
 U(\bm{r}) =
 U(z,\rho,\varphi) = \frac{1}{2\pi} \sum_{\mu=-\infty}^{\infty} U^{(\mu)}(z,\rho) e^{i\mu\varphi},
\end{equation}
where the components $U^{(\mu)}(z,\rho)$ are calculated by the inverse Fourier transformation,
\begin{equation}
 U^{(\mu)}(z,\rho) = \int_{0}^{2\pi} d\varphi U(z,\rho,\varphi) e^{-i\mu\varphi},
\end{equation}
where the integral over $\varphi$ is performed on a uniformly distributed mesh points.
From the symmetry conditions one finds,
\begin{equation}
 U^{(\mu)} = U^{(-\mu)} = (-1)^{\mu} U^{(-\mu)}.
\end{equation}
Thus the Fourier series is abbreviated as
\begin{equation}
 U(\bm{r}) = \frac{1}{2\pi} \left(                      U^{(0)} (z,\rho)
                                 + 2\sum_{n=1}^{\infty} U^{(2n)}(z,\rho) \cos(2n\varphi)
                            \right).
\end{equation}
Finally the matrix element of $U(\bm{r})$ reads,
\begin{eqnarray}
 \langle \alpha | U |\alpha' \rangle
 & = &
 \langle n_{z} ,n_{r} ,m_{l} ,m_{s} | U |
         n_{z}',n_{r}',m_{l}',m_{s}'
 \rangle
 \nonumber \\
 & = &
 \delta_{m_{s},m_{s}'} C_{\alpha}^{*} C_{\alpha'}
 \frac{1}{2\pi} \left[ \delta_{K,K'} D_{\alpha\alpha'}^{(0)}
                \right.
 \nonumber \\
 &  &           \left. \mbox{}
                      +\sum_{n=1}^{\infty} \left( \delta_{K'-K,2n} + \delta_{K-K',2n} \right)
                                           D_{\alpha\alpha'}^{(2n)}
                \right] ,
\end{eqnarray}
where we have performed the integrals over $\varphi$ and
\begin{equation}
 D_{\alpha\alpha'}^{(\mu)}
 =
 \int_{-\infty}^{\infty} dz \int_{0}^{\infty} \rho d\rho
  U^{(\mu)}(z,\rho) \phi_{n_{z}} \phi_{n_{z}'}
  R_{n_{\rho}}^{m_{l}} R_{n_{\rho}'}^{m_{l}'}
 ,
\end{equation}
are performed numerically by using the Gaussian quadrature. For axially symmetric potentials
only the first term survives, while for triaxially
deformed nucleus we must calculate additional terms with $n\neq0$.

\section{~\label{appendix:density}Densities and their derivatives}

After solving the Dirac equation, we can calculate the densities
from the wave functions.
All the densities are linear combinations of the quantities
$\rho_{f}(\bm{r},\tau)$ and $\rho_{g}(\bm{r},\tau)$ which are the contributions
to the vector density from the large and small components, respectively.
$\tau$ represents the isospin. For vector potential we have
\begin{eqnarray}
 \rho_{V}(\bm{r})
 & = &
 \sum_{i=1}^{N} v_{i}^{2} \sum_{p=f,g} \psi_{i}^{p\dagger} \psi_{i}^{p}
 \nonumber \\
 & = &
 \sum_{p=f,g} \sum_{\alpha\alpha'}
  \left( \sum_{i=1}^{N} v_{i}^{2} p_{i}^{\alpha} p_{i}^{\alpha'} \right)
  \Phi_{\alpha}^{\dagger} \Phi_{\alpha'} .
\end{eqnarray}
Note that the factors with $\varphi$ in $\Phi_{\alpha}$ and $\Phi_{\alpha'}$
are in an exponential form, rearranging the terms,
$\rho_{V}(\bm{r})$ can be written as,
\begin{equation}
 \rho_{V}(\bm{r})
 =
 \frac{1}{2\pi} \left(  \rho_{V}^{(0) }(z,\rho)
 + 2\sum_{n=1}^{\infty} \rho_{V}^{(2n)}(z,\rho) \cos(2n\varphi)
                \right),
 \label{eq:rhovexpansion}
\end{equation}
where the components
\begin{eqnarray}
 \rho_{V}^{(\mu)} & = &
 \sum_{p=f,g} \sum_{\alpha\alpha'}
  \left[ \left( \sum_{i=1}^{N} v_{i}^{2} p_{i}^{\alpha} p_{i}^{\alpha'} \right)
         \delta_{K'-K,\mu} \delta_{m_{s},m_{s}'}
  \right.
  \nonumber \\
 &  &
  \left.
   \times C_{\alpha}^{*} C_{\alpha'} \phi_{n_{z}} \phi_{n_{z}'}
          R_{n_{\rho}}^{m_{l}} R_{n_{\rho}'}^{m_{l}'}
  \right] ,
\end{eqnarray}
are calculated with the wave functions $f_{i}^{\alpha}$ and $g_{i}^{\alpha}$.

The scalar density $\rho_{S}(\bm{r})$ can be calculated similarly.
The isovector densities are just the linear combinations of the corresponding densities
of protons and neutrons.

The derivatives of the vector density reads
\begin{equation}
 \nabla^{2} \rho_{V}(\bm{r}) =
 \sum_{p=f,g} \sum_{\alpha\alpha'}
  \left( \sum_{i=1}^{N} v_{i}^{2} p_{i}^{\alpha} p_{i}^{\alpha'} \right)
  \nabla^{2}(\Phi_{\alpha}^{\dagger}\Phi_{\alpha'}),
\end{equation}
where the derivatives of the basis can be calculated as,
\begin{equation}
 \nabla^{2}(\Phi_{\alpha}^{\dagger}\Phi_{\alpha'})
 =
 I_{\alpha\alpha'} + 2J_{\alpha\alpha'} + I_{\alpha'\alpha}^{*}
 ,
 \label{eq:Derivatieofdensity}
\end{equation}
with $I_{\alpha\alpha'}$ and $J_{\alpha\alpha'}$ defined later.
First, from Eq.~(\ref{eq:BasSchrodinger-1}) we get
\begin{equation}
 \nabla^{2} \Phi_{\alpha} = -2M \left[ E_{\alpha}-V_{B}(z,\rho) \right] \Phi_{\alpha},
\end{equation}
thus
\begin{eqnarray}
 I_{\alpha\alpha'}
 & \equiv &
 \Phi_{\alpha}^{\dagger} \nabla^{2} \Phi_{\alpha'}
 \nonumber \\
 & = &
 -2M \left[ E_{\alpha}-V_{B}(z,\rho) \right]
 \delta_{m_{s},m_{s}'} C_{\alpha}^{*} C_{\alpha'}
 \nonumber \\
 &  &
 \times \phi_{n_{z}} \phi_{n_{z}'} R_{n_{\rho}}^{m_{l}} R_{n_{\rho}'}^{m_{l}'}
 \times \frac{1}{2\pi} e^{i(m_{l}'-m_{l}) \varphi} .
\end{eqnarray}
Second,
\begin{eqnarray}
 J_{\alpha\alpha^{\prime}}
 & \equiv &
 \bm{\nabla} \Phi_{\alpha}^{\dagger} \cdot \bm{\nabla} \Phi_{\alpha'} \nonumber \\
 & = &
 \delta_{m_{s},m_{s}'} C_{\alpha}^{*} C_{\alpha'} \nonumber \\
 &   &
 \times \left[\phi_{n_{z}} \phi_{n_{z}'}
              \left( \partial_{\rho} R_{n_{\rho} }^{m_{l} }
                     \partial_{\rho} R_{n_{\rho}'}^{m_{l}'} +
                     m_{l}m_{l}' \frac{1}{\rho^{2}} R_{n_{\rho}}^{m_{l}} R_{n_{\rho}'}^{m_{l}'}
              \right)
        \right. \nonumber \\
 &  &   \left. \mbox{}
             +R_{n_{\rho}}^{m_{l}} R_{n_{\rho}'}^{m_{l}'}
              \partial_{z} \phi_{n_{z}} \partial_{z} \phi_{n_{z}'}
        \right] \times \frac{1}{2\pi} e^{i(m_{l}'-m_{l}) \varphi}.
\end{eqnarray}
Substituting them into Eq.~(\ref{eq:Derivatieofdensity}) and combining
the same exponentials of $\varphi$, we get
\begin{eqnarray}
 \nabla^{2} \rho_{V}(\bm{r}) & = &
 \frac{1}{2\pi} \left(  \tilde{\rho}_{V}^{(0) }(z,\rho) +
   2\sum_{n=1}^{\infty} \tilde{\rho}_{V}^{(2n)}(z,\rho) \cos\left(2n\varphi\right)
                \right), \nonumber \\
\end{eqnarray}
in which
\begin{eqnarray}
 \tilde{\rho}_{V}^{(\mu)}
 & = &
 \sum_{p=f,g} \sum_{\alpha\alpha'}
  \left( \sum_{i=1}^{N}v_{i}^{2} p_{i}^{\alpha} p_{i}^{\alpha'} \right)
  \delta_{K'-K,\mu} \delta_{m_{s},m_{s}'}
 \nonumber \\
 &  &
  \times C_{\alpha}^{*} C_{\alpha'}
   \left\{ D_{\alpha\alpha'}(z,\rho) \phi_{n_{z}} \phi_{n_{z}'}
           R_{n_{\rho}}^{m_{l}} R_{n_{\rho}'}^{m_{l}'}
   \right. \nonumber \\
 &  &
   \left. \mbox{}
  + 2\phi_{n_{z}} \phi_{n_{z}'} \partial_{\rho}R_{n_{\rho} }^{m_{l}}
                                \partial_{\rho}R_{n_{\rho}'}^{m_{l}'}
  + 2R_{n_{\rho}}^{m_{l}} R_{n_{\rho}'}^{m_{l}'} \partial_{z}\phi_{n_{z}}
                                                 \partial_{z}\phi_{n_{z}'}
   \right\} ,
 \nonumber \\
\end{eqnarray}
are the Fourier components and
\begin{equation}
 D_{\alpha\alpha'}(z,\rho) =
 \frac{2m_{l}m_{l}'}{\rho^{2}} - 2M \left[ E_{\alpha'} + E_{\alpha}-2V_{B}(z,\rho) \right].
\end{equation}

%\bibliographystyle{unsrt}
%\bibliography{fission,reference,/home/bnlu/Works/worksphd/Documents/Mypapers/oct-tri/prc-v0.1/nuclear,Tetrahedra}
%\bibliography{../../../information/refs/JabRef/sgzhou}

\begin{thebibliography}{133}%
\makeatletter
\providecommand \@ifxundefined [1]{%
 \@ifx{#1\undefined}
}%
\providecommand \@ifnum [1]{%
 \ifnum #1\expandafter \@firstoftwo
 \else \expandafter \@secondoftwo
 \fi
}%
\providecommand \@ifx [1]{%
 \ifx #1\expandafter \@firstoftwo
 \else \expandafter \@secondoftwo
 \fi
}%
\providecommand \natexlab [1]{#1}%
\providecommand \enquote  [1]{``#1''}%
\providecommand \bibnamefont  [1]{#1}%
\providecommand \bibfnamefont [1]{#1}%
\providecommand \citenamefont [1]{#1}%
\providecommand \href@noop [0]{\@secondoftwo}%
\providecommand \href [0]{\begingroup \@sanitize@url \@href}%
\providecommand \@href[1]{\@@startlink{#1}\@@href}%
\providecommand \@@href[1]{\endgroup#1\@@endlink}%
\providecommand \@sanitize@url [0]{\catcode `\\12\catcode `\$12\catcode
  `\&12\catcode `\#12\catcode `\^12\catcode `\_12\catcode `\%12\relax}%
\providecommand \@@startlink[1]{}%
\providecommand \@@endlink[0]{}%
\providecommand \url  [0]{\begingroup\@sanitize@url \@url }%
\providecommand \@url [1]{\endgroup\@href {#1}{\urlprefix }}%
\providecommand \urlprefix  [0]{URL }%
\providecommand \Eprint [0]{\href }%
\providecommand \doibase [0]{http://dx.doi.org/}%
\providecommand \selectlanguage [0]{\@gobble}%
\providecommand \bibinfo  [0]{\@secondoftwo}%
\providecommand \bibfield  [0]{\@secondoftwo}%
\providecommand \translation [1]{[#1]}%
\providecommand \BibitemOpen [0]{}%
\providecommand \bibitemStop [0]{}%
\providecommand \bibitemNoStop [0]{.\EOS\space}%
\providecommand \EOS [0]{\spacefactor3000\relax}%
\providecommand \BibitemShut  [1]{\csname bibitem#1\endcsname}%
\let\auto@bib@innerbib\@empty
%</preamble>
\bibitem [{\citenamefont {Bohr}\ and\ \citenamefont
  {Mottelson}(1998)}]{Bohr1998_Nucl_Structure_2}%
  \BibitemOpen
  \bibfield  {author} {\bibinfo {author} {\bibfnamefont {A.}~\bibnamefont
  {Bohr}}\ and\ \bibinfo {author} {\bibfnamefont {B.~R.}\ \bibnamefont
  {Mottelson}},\ }\href@noop {} {\emph {\bibinfo {title} {Nuclear
  Structure}}},\ Vol.~\bibinfo {volume} {II}\ (\bibinfo  {publisher} {World
  Scientific},\ \bibinfo {year} {1998})\BibitemShut {NoStop}%
\bibitem [{\citenamefont {Ring}\ and\ \citenamefont {Schuck}(1980)}]{Ring1980}%
  \BibitemOpen
  \bibfield  {author} {\bibinfo {author} {\bibfnamefont {P.}~\bibnamefont
  {Ring}}\ and\ \bibinfo {author} {\bibfnamefont {P.}~\bibnamefont {Schuck}},\
  }\href@noop {} {\emph {\bibinfo {title} {The Nuclear Many-Body Problem}}}\
  (\bibinfo  {publisher} {Springer-Verlag Berlin, Heidelberg, and New York},\
  \bibinfo {year} {1980})\BibitemShut {NoStop}%
\bibitem [{\citenamefont {Frauendorf}(2001)}]{Frauendorf2001_RMP73-463}%
  \BibitemOpen
  \bibfield  {author} {\bibinfo {author} {\bibfnamefont {S.}~\bibnamefont
  {Frauendorf}},\ }\href {\doibase 10.1103/RevModPhys.73.463} {\bibfield
  {journal} {\bibinfo  {journal} {Rev. Mod. Phys.}\ }\textbf {\bibinfo {volume}
  {73}},\ \bibinfo {pages} {463} (\bibinfo {year} {2001})}\BibitemShut
  {NoStop}%
\bibitem [{\citenamefont {Nazarewicz}(2001)}]{Nazarewicz2001_LNP581-102}%
  \BibitemOpen
  \bibfield  {author} {\bibinfo {author} {\bibfnamefont {W.}~\bibnamefont
  {Nazarewicz}},\ }in\ \href {\doibase 10.1007/3-540-44620-6_4} {\emph
  {\bibinfo {booktitle} {Lecture Notes in Physics}}},\ Vol.\ \bibinfo {volume}
  {581},\ \bibinfo {editor} {edited by\ \bibinfo {editor} {\bibfnamefont
  {J.}~\bibnamefont {Arias}}\ and\ \bibinfo {editor} {\bibfnamefont
  {M.}~\bibnamefont {Lozano}}}\ (\bibinfo  {publisher} {Springer Berlin
  Heidelberg},\ \bibinfo {year} {2001})\ pp.\ \bibinfo {pages}
  {102--140}\BibitemShut {NoStop}%
\bibitem [{\citenamefont {Nilsson}\ \emph {et~al.}(1969)\citenamefont
  {Nilsson}, \citenamefont {Tsang}, \citenamefont {Sobiczewski}, \citenamefont
  {Szymanski}, \citenamefont {Wycech}, \citenamefont {Gustafson}, \citenamefont
  {Lamm}, \citenamefont {M\"oller},\ and\ \citenamefont
  {Nilsson}}]{Nilsson1969_NPA131-1}%
  \BibitemOpen
  \bibfield  {author} {\bibinfo {author} {\bibfnamefont {S.~G.}\ \bibnamefont
  {Nilsson}}, \bibinfo {author} {\bibfnamefont {C.~F.}\ \bibnamefont {Tsang}},
  \bibinfo {author} {\bibfnamefont {A.}~\bibnamefont {Sobiczewski}}, \bibinfo
  {author} {\bibfnamefont {Z.}~\bibnamefont {Szymanski}}, \bibinfo {author}
  {\bibfnamefont {S.}~\bibnamefont {Wycech}}, \bibinfo {author} {\bibfnamefont
  {C.}~\bibnamefont {Gustafson}}, \bibinfo {author} {\bibfnamefont {I.-L.}\
  \bibnamefont {Lamm}}, \bibinfo {author} {\bibfnamefont {P.}~\bibnamefont
  {M\"oller}}, \ and\ \bibinfo {author} {\bibfnamefont {B.}~\bibnamefont
  {Nilsson}},\ }\href {\doibase 10.1016/0375-9474(69)90809-4} {\bibfield
  {journal} {\bibinfo  {journal} {Nucl. Phys. A}\ }\textbf {\bibinfo {volume}
  {131}},\ \bibinfo {pages} {1} (\bibinfo {year} {1969})}\BibitemShut {NoStop}%
\bibitem [{\citenamefont {Frauendorf}\ and\ \citenamefont
  {Meng}(1997)}]{Frauendorf1997_NPA617-131}%
  \BibitemOpen
  \bibfield  {author} {\bibinfo {author} {\bibfnamefont {S.}~\bibnamefont
  {Frauendorf}}\ and\ \bibinfo {author} {\bibfnamefont {J.}~\bibnamefont
  {Meng}},\ }\href {\doibase 10.1016/S0375-9474(97)00004-3} {\bibfield
  {journal} {\bibinfo  {journal} {Nucl. Phys. A}\ }\textbf {\bibinfo {volume}
  {617}},\ \bibinfo {pages} {131} (\bibinfo {year} {1997})}\BibitemShut
  {NoStop}%
\bibitem [{\citenamefont {Starosta}\ \emph {et~al.}(2001)\citenamefont
  {Starosta}, \citenamefont {Koike}, \citenamefont {Chiara}, \citenamefont
  {Fossan}, \citenamefont {LaFosse}, \citenamefont {Hecht}, \citenamefont
  {Beausang}, \citenamefont {Caprio}, \citenamefont {Cooper}, \citenamefont
  {Krucken}, \citenamefont {Novak}, \citenamefont {Zamfir}, \citenamefont
  {Zyromski}, \citenamefont {Hartley}, \citenamefont {Balabanski},
  \citenamefont {Zhang}, \citenamefont {Frauendorf},\ and\ \citenamefont
  {Dimitrov}}]{Starosta2001_PRL86-971}%
  \BibitemOpen
  \bibfield  {author} {\bibinfo {author} {\bibfnamefont {K.}~\bibnamefont
  {Starosta}}, \bibinfo {author} {\bibfnamefont {T.}~\bibnamefont {Koike}},
  \bibinfo {author} {\bibfnamefont {C.~J.}\ \bibnamefont {Chiara}}, \bibinfo
  {author} {\bibfnamefont {D.~B.}\ \bibnamefont {Fossan}}, \bibinfo {author}
  {\bibfnamefont {D.~R.}\ \bibnamefont {LaFosse}}, \bibinfo {author}
  {\bibfnamefont {A.~A.}\ \bibnamefont {Hecht}}, \bibinfo {author}
  {\bibfnamefont {C.~W.}\ \bibnamefont {Beausang}}, \bibinfo {author}
  {\bibfnamefont {M.~A.}\ \bibnamefont {Caprio}}, \bibinfo {author}
  {\bibfnamefont {J.~R.}\ \bibnamefont {Cooper}}, \bibinfo {author}
  {\bibfnamefont {R.}~\bibnamefont {Krucken}}, \bibinfo {author} {\bibfnamefont
  {J.~R.}\ \bibnamefont {Novak}}, \bibinfo {author} {\bibfnamefont {N.~V.}\
  \bibnamefont {Zamfir}}, \bibinfo {author} {\bibfnamefont {K.~E.}\
  \bibnamefont {Zyromski}}, \bibinfo {author} {\bibfnamefont {D.~J.}\
  \bibnamefont {Hartley}}, \bibinfo {author} {\bibfnamefont {D.~L.}\
  \bibnamefont {Balabanski}}, \bibinfo {author} {\bibfnamefont {J.-y.}\
  \bibnamefont {Zhang}}, \bibinfo {author} {\bibfnamefont {S.}~\bibnamefont
  {Frauendorf}}, \ and\ \bibinfo {author} {\bibfnamefont {V.~I.}\ \bibnamefont
  {Dimitrov}},\ }\href {\doibase 10.1103/PhysRevLett.86.971} {\bibfield
  {journal} {\bibinfo  {journal} {Phys. Rev. Lett.}\ }\textbf {\bibinfo
  {volume} {86}},\ \bibinfo {pages} {971} (\bibinfo {year} {2001})}\BibitemShut
  {NoStop}%
\bibitem [{\citenamefont {Odegard}\ \emph {et~al.}(2001)\citenamefont
  {Odegard}, \citenamefont {Hagemann}, \citenamefont {Jensen}, \citenamefont
  {Bergstroem}, \citenamefont {Herskind}, \citenamefont {Sletten},
  \citenamefont {Toermaenen}, \citenamefont {Wilson}, \citenamefont {Tjom},
  \citenamefont {Hamamoto}, \citenamefont {Spohr}, \citenamefont {Huebel},
  \citenamefont {Goergen}, \citenamefont {Schoenwasser}, \citenamefont
  {Bracco}, \citenamefont {Leoni}, \citenamefont {Maj}, \citenamefont
  {Petrache}, \citenamefont {Bednarczyk},\ and\ \citenamefont
  {Curien}}]{Odegard2001_PRL86-5866}%
  \BibitemOpen
  \bibfield  {author} {\bibinfo {author} {\bibfnamefont {S.~W.}\ \bibnamefont
  {Odegard}}, \bibinfo {author} {\bibfnamefont {G.~B.}\ \bibnamefont
  {Hagemann}}, \bibinfo {author} {\bibfnamefont {D.~R.}\ \bibnamefont
  {Jensen}}, \bibinfo {author} {\bibfnamefont {M.}~\bibnamefont {Bergstroem}},
  \bibinfo {author} {\bibfnamefont {B.}~\bibnamefont {Herskind}}, \bibinfo
  {author} {\bibfnamefont {G.}~\bibnamefont {Sletten}}, \bibinfo {author}
  {\bibfnamefont {S.}~\bibnamefont {Toermaenen}}, \bibinfo {author}
  {\bibfnamefont {J.~N.}\ \bibnamefont {Wilson}}, \bibinfo {author}
  {\bibfnamefont {P.~O.}\ \bibnamefont {Tjom}}, \bibinfo {author}
  {\bibfnamefont {I.}~\bibnamefont {Hamamoto}}, \bibinfo {author}
  {\bibfnamefont {K.}~\bibnamefont {Spohr}}, \bibinfo {author} {\bibfnamefont
  {H.}~\bibnamefont {Huebel}}, \bibinfo {author} {\bibfnamefont
  {A.}~\bibnamefont {Goergen}}, \bibinfo {author} {\bibfnamefont
  {G.}~\bibnamefont {Schoenwasser}}, \bibinfo {author} {\bibfnamefont
  {A.}~\bibnamefont {Bracco}}, \bibinfo {author} {\bibfnamefont
  {S.}~\bibnamefont {Leoni}}, \bibinfo {author} {\bibfnamefont
  {A.}~\bibnamefont {Maj}}, \bibinfo {author} {\bibfnamefont {C.~M.}\
  \bibnamefont {Petrache}}, \bibinfo {author} {\bibfnamefont {P.}~\bibnamefont
  {Bednarczyk}}, \ and\ \bibinfo {author} {\bibfnamefont {D.}~\bibnamefont
  {Curien}},\ }\href {\doibase 10.1103/PhysRevLett.86.5866} {\bibfield
  {journal} {\bibinfo  {journal} {Phys. Rev. Lett.}\ }\textbf {\bibinfo
  {volume} {86}},\ \bibinfo {pages} {5866} (\bibinfo {year}
  {2001})}\BibitemShut {NoStop}%
\bibitem [{\citenamefont {Meng}\ and\ \citenamefont
  {Zhang}(2010)}]{Meng2010_JPG37-064025}%
  \BibitemOpen
  \bibfield  {author} {\bibinfo {author} {\bibfnamefont {J.}~\bibnamefont
  {Meng}}\ and\ \bibinfo {author} {\bibfnamefont {S.~Q.}\ \bibnamefont
  {Zhang}},\ }\href {\doibase 10.1088/0954-3899/37/6/064025} {\bibfield
  {journal} {\bibinfo  {journal} {J. Phys. G: Nucl. Phys.}\ }\textbf {\bibinfo
  {volume} {37}},\ \bibinfo {pages} {064025} (\bibinfo {year}
  {2010})}\BibitemShut {NoStop}%
\bibitem [{\citenamefont {Meng}\ \emph {et~al.}(2013)\citenamefont {Meng},
  \citenamefont {Peng}, \citenamefont {Zhang},\ and\ \citenamefont
  {Zhao}}]{Meng2013_FPC8-55}%
  \BibitemOpen
  \bibfield  {author} {\bibinfo {author} {\bibfnamefont {J.}~\bibnamefont
  {Meng}}, \bibinfo {author} {\bibfnamefont {J.}~\bibnamefont {Peng}}, \bibinfo
  {author} {\bibfnamefont {S.-Q.}\ \bibnamefont {Zhang}}, \ and\ \bibinfo
  {author} {\bibfnamefont {P.-W.}\ \bibnamefont {Zhao}},\ }\href {\doibase
  10.1007/s11467-013-0287-y} {\bibfield  {journal} {\bibinfo  {journal} {Front.
  Phys.}\ }\textbf {\bibinfo {volume} {8}},\ \bibinfo {pages} {55} (\bibinfo
  {year} {2013})}\BibitemShut {NoStop}%
\bibitem [{\citenamefont {\'{C}wiok}\ \emph {et~al.}(2005)\citenamefont
  {\'{C}wiok}, \citenamefont {Heenen},\ and\ \citenamefont
  {Nazarewicz}}]{Cwiok2005_Nature433-705}%
  \BibitemOpen
  \bibfield  {author} {\bibinfo {author} {\bibfnamefont {S.}~\bibnamefont
  {\'{C}wiok}}, \bibinfo {author} {\bibfnamefont {P.-H.}\ \bibnamefont
  {Heenen}}, \ and\ \bibinfo {author} {\bibfnamefont {W.}~\bibnamefont
  {Nazarewicz}},\ }\href {\doibase 10.1038/nature03336} {\bibfield  {journal}
  {\bibinfo  {journal} {Nature}\ }\textbf {\bibinfo {volume} {433}},\ \bibinfo
  {pages} {705} (\bibinfo {year} {2005})}\BibitemShut {NoStop}%
\bibitem [{\citenamefont {Butler}\ and\ \citenamefont
  {Nazarewicz}(1996)}]{Butler1996_RMP68-349}%
  \BibitemOpen
  \bibfield  {author} {\bibinfo {author} {\bibfnamefont {P.~A.}\ \bibnamefont
  {Butler}}\ and\ \bibinfo {author} {\bibfnamefont {W.}~\bibnamefont
  {Nazarewicz}},\ }\href {\doibase 10.1103/RevModPhys.68.349} {\bibfield
  {journal} {\bibinfo  {journal} {Rev. Mod. Phys.}\ }\textbf {\bibinfo {volume}
  {68}},\ \bibinfo {pages} {349} (\bibinfo {year} {1996})}\BibitemShut
  {NoStop}%
\bibitem [{\citenamefont {Shneidman}\ \emph {et~al.}(2003)\citenamefont
  {Shneidman}, \citenamefont {Adamian}, \citenamefont {Antonenko},
  \citenamefont {Jolos},\ and\ \citenamefont
  {Scheid}}]{Shneidman2003_PRC67-014313}%
  \BibitemOpen
  \bibfield  {author} {\bibinfo {author} {\bibfnamefont {T.~M.}\ \bibnamefont
  {Shneidman}}, \bibinfo {author} {\bibfnamefont {G.~G.}\ \bibnamefont
  {Adamian}}, \bibinfo {author} {\bibfnamefont {N.~V.}\ \bibnamefont
  {Antonenko}}, \bibinfo {author} {\bibfnamefont {R.~V.}\ \bibnamefont
  {Jolos}}, \ and\ \bibinfo {author} {\bibfnamefont {W.}~\bibnamefont
  {Scheid}},\ }\href {\doibase 10.1103/PhysRevC.67.014313} {\bibfield
  {journal} {\bibinfo  {journal} {Phys. Rev. C}\ }\textbf {\bibinfo {volume}
  {67}},\ \bibinfo {pages} {014313} (\bibinfo {year} {2003})}\BibitemShut
  {NoStop}%
\bibitem [{\citenamefont {Shneidman}\ \emph {et~al.}(2006)\citenamefont
  {Shneidman}, \citenamefont {Adamian}, \citenamefont {Antonenko},\ and\
  \citenamefont {Jolos}}]{Shneidman2006_PRC74-034316}%
  \BibitemOpen
  \bibfield  {author} {\bibinfo {author} {\bibfnamefont {T.~M.}\ \bibnamefont
  {Shneidman}}, \bibinfo {author} {\bibfnamefont {G.~G.}\ \bibnamefont
  {Adamian}}, \bibinfo {author} {\bibfnamefont {N.~V.}\ \bibnamefont
  {Antonenko}}, \ and\ \bibinfo {author} {\bibfnamefont {R.~V.}\ \bibnamefont
  {Jolos}},\ }\href {\doibase 10.1103/PhysRevC.74.034316} {\bibfield  {journal}
  {\bibinfo  {journal} {Phys. Rev. C}\ }\textbf {\bibinfo {volume} {74}},\
  \bibinfo {pages} {034316} (\bibinfo {year} {2006})}\BibitemShut {NoStop}%
\bibitem [{\citenamefont {Wang}\ \emph {et~al.}(2005)\citenamefont {Wang},
  \citenamefont {Hua}, \citenamefont {Meng}, \citenamefont {Li}, \citenamefont
  {Zhang}, \citenamefont {Xu}, \citenamefont {Liu}, \citenamefont {Ye},
  \citenamefont {Jiang}, \citenamefont {Zheng}, \citenamefont {Wang},
  \citenamefont {Chen}, \citenamefont {Wu}, \citenamefont {Zhang},
  \citenamefont {Pang}, \citenamefont {Wang}, \citenamefont {Lou},
  \citenamefont {Guo}, \citenamefont {Jin}, \citenamefont {Zhou}, \citenamefont
  {Zhu}, \citenamefont {Wu}, \citenamefont {Li}, \citenamefont {Wen},
  \citenamefont {He}, \citenamefont {Cui},\ and\ \citenamefont
  {Liu}}]{Wang2005_PRC72-024317}%
  \BibitemOpen
  \bibfield  {author} {\bibinfo {author} {\bibfnamefont {S.}~\bibnamefont
  {Wang}}, \bibinfo {author} {\bibfnamefont {H.}~\bibnamefont {Hua}}, \bibinfo
  {author} {\bibfnamefont {J.}~\bibnamefont {Meng}}, \bibinfo {author}
  {\bibfnamefont {Z.~H.}\ \bibnamefont {Li}}, \bibinfo {author} {\bibfnamefont
  {S.~Q.}\ \bibnamefont {Zhang}}, \bibinfo {author} {\bibfnamefont {F.~R.}\
  \bibnamefont {Xu}}, \bibinfo {author} {\bibfnamefont {H.~L.}\ \bibnamefont
  {Liu}}, \bibinfo {author} {\bibfnamefont {Y.~L.}\ \bibnamefont {Ye}},
  \bibinfo {author} {\bibfnamefont {D.~X.}\ \bibnamefont {Jiang}}, \bibinfo
  {author} {\bibfnamefont {T.}~\bibnamefont {Zheng}}, \bibinfo {author}
  {\bibfnamefont {Q.~J.}\ \bibnamefont {Wang}}, \bibinfo {author}
  {\bibfnamefont {Z.~Q.}\ \bibnamefont {Chen}}, \bibinfo {author}
  {\bibfnamefont {C.~E.}\ \bibnamefont {Wu}}, \bibinfo {author} {\bibfnamefont
  {G.~L.}\ \bibnamefont {Zhang}}, \bibinfo {author} {\bibfnamefont {D.~Y.}\
  \bibnamefont {Pang}}, \bibinfo {author} {\bibfnamefont {J.}~\bibnamefont
  {Wang}}, \bibinfo {author} {\bibfnamefont {J.~L.}\ \bibnamefont {Lou}},
  \bibinfo {author} {\bibfnamefont {B.}~\bibnamefont {Guo}}, \bibinfo {author}
  {\bibfnamefont {G.}~\bibnamefont {Jin}}, \bibinfo {author} {\bibfnamefont
  {S.~G.}\ \bibnamefont {Zhou}}, \bibinfo {author} {\bibfnamefont {L.~H.}\
  \bibnamefont {Zhu}}, \bibinfo {author} {\bibfnamefont {X.~G.}\ \bibnamefont
  {Wu}}, \bibinfo {author} {\bibfnamefont {G.~S.}\ \bibnamefont {Li}}, \bibinfo
  {author} {\bibfnamefont {S.~X.}\ \bibnamefont {Wen}}, \bibinfo {author}
  {\bibfnamefont {C.~Y.}\ \bibnamefont {He}}, \bibinfo {author} {\bibfnamefont
  {X.~Z.}\ \bibnamefont {Cui}}, \ and\ \bibinfo {author} {\bibfnamefont
  {Y.}~\bibnamefont {Liu}},\ }\href {\doibase 10.1103/PhysRevC.72.024317}
  {\bibfield  {journal} {\bibinfo  {journal} {Phys. Rev. C}\ }\textbf {\bibinfo
  {volume} {72}},\ \bibinfo {pages} {024317} (\bibinfo {year}
  {2005})}\BibitemShut {NoStop}%
\bibitem [{\citenamefont {Yang}\ \emph {et~al.}(2009)\citenamefont {Yang},
  \citenamefont {Lu}, \citenamefont {Liu}, \citenamefont {Wang}, \citenamefont
  {Ma}, \citenamefont {Yang}, \citenamefont {Han}, \citenamefont {Zhao},
  \citenamefont {Ma}, \citenamefont {Zhu}, \citenamefont {Wu},\ and\
  \citenamefont {Li}}]{Yang2009_CPL26-082101}%
  \BibitemOpen
  \bibfield  {author} {\bibinfo {author} {\bibfnamefont {D.}~\bibnamefont
  {Yang}}, \bibinfo {author} {\bibfnamefont {J.-B.}\ \bibnamefont {Lu}},
  \bibinfo {author} {\bibfnamefont {Y.-Z.}\ \bibnamefont {Liu}}, \bibinfo
  {author} {\bibfnamefont {L.-L.}\ \bibnamefont {Wang}}, \bibinfo {author}
  {\bibfnamefont {K.-Y.}\ \bibnamefont {Ma}}, \bibinfo {author} {\bibfnamefont
  {C.-D.}\ \bibnamefont {Yang}}, \bibinfo {author} {\bibfnamefont {D.-K.}\
  \bibnamefont {Han}}, \bibinfo {author} {\bibfnamefont {Y.-X.}\ \bibnamefont
  {Zhao}}, \bibinfo {author} {\bibfnamefont {Y.-J.}\ \bibnamefont {Ma}},
  \bibinfo {author} {\bibfnamefont {L.-H.}\ \bibnamefont {Zhu}}, \bibinfo
  {author} {\bibfnamefont {X.-G.}\ \bibnamefont {Wu}}, \ and\ \bibinfo {author}
  {\bibfnamefont {G.-S.}\ \bibnamefont {Li}},\ }\href {\doibase
  10.1088/0256-307X/26/8/082101} {\bibfield  {journal} {\bibinfo  {journal}
  {Chin. Phys. Lett.}\ }\textbf {\bibinfo {volume} {26}},\ \bibinfo {pages}
  {082101} (\bibinfo {year} {2009})}\BibitemShut {NoStop}%
\bibitem [{\citenamefont {Robledo}\ and\ \citenamefont
  {Bertsch}(2011)}]{Robledo2011_PRC84-054302}%
  \BibitemOpen
  \bibfield  {author} {\bibinfo {author} {\bibfnamefont {L.~M.}\ \bibnamefont
  {Robledo}}\ and\ \bibinfo {author} {\bibfnamefont {G.~F.}\ \bibnamefont
  {Bertsch}},\ }\href {\doibase 10.1103/PhysRevC.84.054302} {\bibfield
  {journal} {\bibinfo  {journal} {Phys. Rev. C}\ }\textbf {\bibinfo {volume}
  {84}},\ \bibinfo {pages} {054302} (\bibinfo {year} {2011})}\BibitemShut
  {NoStop}%
\bibitem [{\citenamefont {Zhu}\ \emph {et~al.}(2012{\natexlab{a}})\citenamefont
  {Zhu}, \citenamefont {Sakhaee}, \citenamefont {Hamilton}, \citenamefont
  {Ramayya}, \citenamefont {Brewer}, \citenamefont {Hwang}, \citenamefont
  {Liu}, \citenamefont {Yeoh}, \citenamefont {Xiao}, \citenamefont {Xu},
  \citenamefont {Zhang}, \citenamefont {Luo}, \citenamefont {Rasmussen},
  \citenamefont {Lee}, \citenamefont {Li},\ and\ \citenamefont
  {Ma}}]{Zhu2012_PRC85-014330}%
  \BibitemOpen
  \bibfield  {author} {\bibinfo {author} {\bibfnamefont {S.~J.}\ \bibnamefont
  {Zhu}}, \bibinfo {author} {\bibfnamefont {M.}~\bibnamefont {Sakhaee}},
  \bibinfo {author} {\bibfnamefont {J.~H.}\ \bibnamefont {Hamilton}}, \bibinfo
  {author} {\bibfnamefont {A.~V.}\ \bibnamefont {Ramayya}}, \bibinfo {author}
  {\bibfnamefont {N.~T.}\ \bibnamefont {Brewer}}, \bibinfo {author}
  {\bibfnamefont {J.~K.}\ \bibnamefont {Hwang}}, \bibinfo {author}
  {\bibfnamefont {S.~H.}\ \bibnamefont {Liu}}, \bibinfo {author} {\bibfnamefont
  {E.~Y.}\ \bibnamefont {Yeoh}}, \bibinfo {author} {\bibfnamefont {Z.~G.}\
  \bibnamefont {Xiao}}, \bibinfo {author} {\bibfnamefont {Q.}~\bibnamefont
  {Xu}}, \bibinfo {author} {\bibfnamefont {Z.}~\bibnamefont {Zhang}}, \bibinfo
  {author} {\bibfnamefont {Y.~X.}\ \bibnamefont {Luo}}, \bibinfo {author}
  {\bibfnamefont {J.~O.}\ \bibnamefont {Rasmussen}}, \bibinfo {author}
  {\bibfnamefont {I.~Y.}\ \bibnamefont {Lee}}, \bibinfo {author} {\bibfnamefont
  {K.}~\bibnamefont {Li}}, \ and\ \bibinfo {author} {\bibfnamefont {W.~C.}\
  \bibnamefont {Ma}},\ }\href {\doibase 10.1103/PhysRevC.85.014330} {\bibfield
  {journal} {\bibinfo  {journal} {Phys. Rev. C}\ }\textbf {\bibinfo {volume}
  {85}},\ \bibinfo {pages} {014330} (\bibinfo {year}
  {2012}{\natexlab{a}})}\BibitemShut {NoStop}%
\bibitem [{\citenamefont {Zhu}\ \emph {et~al.}(2012{\natexlab{b}})\citenamefont
  {Zhu}, \citenamefont {Hamilton}, \citenamefont {Ramayya}, \citenamefont
  {Hwang}, \citenamefont {Chen}, \citenamefont {Zhu}, \citenamefont {Li},
  \citenamefont {Xiao}, \citenamefont {Yeoh}, \citenamefont {Wang},
  \citenamefont {Luo}, \citenamefont {Liu}, \citenamefont {Rasmussen},\ and\
  \citenamefont {Lee}}]{Zhu2012_NSC2012-348}%
  \BibitemOpen
  \bibfield  {author} {\bibinfo {author} {\bibfnamefont {S.~J.}\ \bibnamefont
  {Zhu}}, \bibinfo {author} {\bibfnamefont {J.~H.}\ \bibnamefont {Hamilton}},
  \bibinfo {author} {\bibfnamefont {A.~V.}\ \bibnamefont {Ramayya}}, \bibinfo
  {author} {\bibfnamefont {J.~K.}\ \bibnamefont {Hwang}}, \bibinfo {author}
  {\bibfnamefont {Y.~J.}\ \bibnamefont {Chen}}, \bibinfo {author}
  {\bibfnamefont {L.~Y.}\ \bibnamefont {Zhu}}, \bibinfo {author} {\bibfnamefont
  {H.~J.}\ \bibnamefont {Li}}, \bibinfo {author} {\bibfnamefont {Z.~G.}\
  \bibnamefont {Xiao}}, \bibinfo {author} {\bibfnamefont {E.~Y.}\ \bibnamefont
  {Yeoh}}, \bibinfo {author} {\bibfnamefont {J.~G.}\ \bibnamefont {Wang}},
  \bibinfo {author} {\bibfnamefont {Y.~X.}\ \bibnamefont {Luo}}, \bibinfo
  {author} {\bibfnamefont {S.~H.}\ \bibnamefont {Liu}}, \bibinfo {author}
  {\bibfnamefont {J.~O.}\ \bibnamefont {Rasmussen}}, \ and\ \bibinfo {author}
  {\bibfnamefont {I.~Y.}\ \bibnamefont {Lee}},\ }in\ \href
  {http://www.nuclearstructure.org/resource/NSC2012-Book/NSC2012.html} {\emph
  {\bibinfo {booktitle} {Nuclear Structure in China 2012: Proceedings of the
  14th National Conference on Nuclear Structure in China (NSC2012)}}},\
  \bibinfo {editor} {edited by\ \bibinfo {editor} {\bibfnamefont
  {J.}~\bibnamefont {Meng}}, \bibinfo {editor} {\bibfnamefont {C.-W.}\
  \bibnamefont {Shen}}, \bibinfo {editor} {\bibfnamefont {E.-G.}\ \bibnamefont
  {Zhao}}, \ and\ \bibinfo {editor} {\bibfnamefont {S.-G.}\ \bibnamefont
  {Zhou}}}\ (\bibinfo  {publisher} {World Scientific},\ \bibinfo {year}
  {2012})\ pp.\ \bibinfo {pages} {348--356}\BibitemShut {NoStop}%
\bibitem [{\citenamefont {Gaffney}\ \emph {et~al.}(2013)\citenamefont
  {Gaffney}, \citenamefont {Butler}, \citenamefont {Scheck}, \citenamefont
  {Hayes}, \citenamefont {Wenander}, \citenamefont {Albers}, \citenamefont
  {Bastin}, \citenamefont {Bauer}, \citenamefont {Blazhev}, \citenamefont
  {Bonig}, \citenamefont {Bree}, \citenamefont {Cederkall}, \citenamefont
  {Chupp}, \citenamefont {Cline}, \citenamefont {Cocolios}, \citenamefont
  {Davinson}, \citenamefont {De~Witte}, \citenamefont {Diriken}, \citenamefont
  {Grahn}, \citenamefont {Herzan}, \citenamefont {Huyse}, \citenamefont
  {Jenkins}, \citenamefont {Joss}, \citenamefont {Kesteloot}, \citenamefont
  {Konki}, \citenamefont {Kowalczyk}, \citenamefont {Kroll}, \citenamefont
  {Kwan}, \citenamefont {Lutter}, \citenamefont {Moschner}, \citenamefont
  {Napiorkowski}, \citenamefont {Pakarinen}, \citenamefont {Pfeiffer},
  \citenamefont {Radeck}, \citenamefont {Reiter}, \citenamefont {Reynders},
  \citenamefont {Rigby}, \citenamefont {Robledo}, \citenamefont {Rudigier},
  \citenamefont {Sambi}, \citenamefont {Seidlitz}, \citenamefont {Siebeck},
  \citenamefont {Stora}, \citenamefont {Thoele}, \citenamefont {Van~Duppen},
  \citenamefont {Vermeulen}, \citenamefont {von Schmid}, \citenamefont
  {Voulot}, \citenamefont {Warr}, \citenamefont {Wimmer}, \citenamefont
  {Wrzosek-Lipska}, \citenamefont {Wu},\ and\ \citenamefont
  {Zielinska}}]{Gaffney2013_Nature497-199}%
  \BibitemOpen
  \bibfield  {author} {\bibinfo {author} {\bibfnamefont {L.~P.}\ \bibnamefont
  {Gaffney}}, \bibinfo {author} {\bibfnamefont {P.~A.}\ \bibnamefont {Butler}},
  \bibinfo {author} {\bibfnamefont {M.}~\bibnamefont {Scheck}}, \bibinfo
  {author} {\bibfnamefont {A.~B.}\ \bibnamefont {Hayes}}, \bibinfo {author}
  {\bibfnamefont {F.}~\bibnamefont {Wenander}}, \bibinfo {author}
  {\bibfnamefont {M.}~\bibnamefont {Albers}}, \bibinfo {author} {\bibfnamefont
  {B.}~\bibnamefont {Bastin}}, \bibinfo {author} {\bibfnamefont
  {C.}~\bibnamefont {Bauer}}, \bibinfo {author} {\bibfnamefont
  {A.}~\bibnamefont {Blazhev}}, \bibinfo {author} {\bibfnamefont
  {S.}~\bibnamefont {Bonig}}, \bibinfo {author} {\bibfnamefont
  {N.}~\bibnamefont {Bree}}, \bibinfo {author} {\bibfnamefont {J.}~\bibnamefont
  {Cederkall}}, \bibinfo {author} {\bibfnamefont {T.}~\bibnamefont {Chupp}},
  \bibinfo {author} {\bibfnamefont {D.}~\bibnamefont {Cline}}, \bibinfo
  {author} {\bibfnamefont {T.~E.}\ \bibnamefont {Cocolios}}, \bibinfo {author}
  {\bibfnamefont {T.}~\bibnamefont {Davinson}}, \bibinfo {author}
  {\bibfnamefont {H.}~\bibnamefont {De~Witte}}, \bibinfo {author}
  {\bibfnamefont {J.}~\bibnamefont {Diriken}}, \bibinfo {author} {\bibfnamefont
  {T.}~\bibnamefont {Grahn}}, \bibinfo {author} {\bibfnamefont
  {A.}~\bibnamefont {Herzan}}, \bibinfo {author} {\bibfnamefont
  {M.}~\bibnamefont {Huyse}}, \bibinfo {author} {\bibfnamefont {D.~G.}\
  \bibnamefont {Jenkins}}, \bibinfo {author} {\bibfnamefont {D.~T.}\
  \bibnamefont {Joss}}, \bibinfo {author} {\bibfnamefont {N.}~\bibnamefont
  {Kesteloot}}, \bibinfo {author} {\bibfnamefont {J.}~\bibnamefont {Konki}},
  \bibinfo {author} {\bibfnamefont {M.}~\bibnamefont {Kowalczyk}}, \bibinfo
  {author} {\bibfnamefont {T.}~\bibnamefont {Kroll}}, \bibinfo {author}
  {\bibfnamefont {E.}~\bibnamefont {Kwan}}, \bibinfo {author} {\bibfnamefont
  {R.}~\bibnamefont {Lutter}}, \bibinfo {author} {\bibfnamefont
  {K.}~\bibnamefont {Moschner}}, \bibinfo {author} {\bibfnamefont
  {P.}~\bibnamefont {Napiorkowski}}, \bibinfo {author} {\bibfnamefont
  {J.}~\bibnamefont {Pakarinen}}, \bibinfo {author} {\bibfnamefont
  {M.}~\bibnamefont {Pfeiffer}}, \bibinfo {author} {\bibfnamefont
  {D.}~\bibnamefont {Radeck}}, \bibinfo {author} {\bibfnamefont
  {P.}~\bibnamefont {Reiter}}, \bibinfo {author} {\bibfnamefont
  {K.}~\bibnamefont {Reynders}}, \bibinfo {author} {\bibfnamefont {S.~V.}\
  \bibnamefont {Rigby}}, \bibinfo {author} {\bibfnamefont {L.~M.}\ \bibnamefont
  {Robledo}}, \bibinfo {author} {\bibfnamefont {M.}~\bibnamefont {Rudigier}},
  \bibinfo {author} {\bibfnamefont {S.}~\bibnamefont {Sambi}}, \bibinfo
  {author} {\bibfnamefont {M.}~\bibnamefont {Seidlitz}}, \bibinfo {author}
  {\bibfnamefont {B.}~\bibnamefont {Siebeck}}, \bibinfo {author} {\bibfnamefont
  {T.}~\bibnamefont {Stora}}, \bibinfo {author} {\bibfnamefont
  {P.}~\bibnamefont {Thoele}}, \bibinfo {author} {\bibfnamefont
  {P.}~\bibnamefont {Van~Duppen}}, \bibinfo {author} {\bibfnamefont {M.~J.}\
  \bibnamefont {Vermeulen}}, \bibinfo {author} {\bibfnamefont {M.}~\bibnamefont
  {von Schmid}}, \bibinfo {author} {\bibfnamefont {D.}~\bibnamefont {Voulot}},
  \bibinfo {author} {\bibfnamefont {N.}~\bibnamefont {Warr}}, \bibinfo {author}
  {\bibfnamefont {K.}~\bibnamefont {Wimmer}}, \bibinfo {author} {\bibfnamefont
  {K.}~\bibnamefont {Wrzosek-Lipska}}, \bibinfo {author} {\bibfnamefont
  {C.~Y.}\ \bibnamefont {Wu}}, \ and\ \bibinfo {author} {\bibfnamefont
  {M.}~\bibnamefont {Zielinska}},\ }\href {\doibase 10.1038/nature12073}
  {\bibfield  {journal} {\bibinfo  {journal} {Nature}\ }\textbf {\bibinfo
  {volume} {497}},\ \bibinfo {pages} {199} (\bibinfo {year}
  {2013})}\BibitemShut {NoStop}%
\bibitem [{\citenamefont {Robinson}\ \emph {et~al.}(2008)\citenamefont
  {Robinson}, \citenamefont {Khoo}, \citenamefont {Ahmad}, \citenamefont
  {Tandel}, \citenamefont {Kondev}, \citenamefont {Nakatsukasa}, \citenamefont
  {Seweryniak}, \citenamefont {Asai}, \citenamefont {Back}, \citenamefont
  {Carpenter}, \citenamefont {Chowdhury}, \citenamefont {Davids}, \citenamefont
  {Eeckhaudt}, \citenamefont {Greene}, \citenamefont {Greenlees}, \citenamefont
  {Gros}, \citenamefont {Heinz}, \citenamefont {Herzberg}, \citenamefont
  {Janssens}, \citenamefont {Jones}, \citenamefont {Lauritsen}, \citenamefont
  {Lister}, \citenamefont {Peterson}, \citenamefont {Qian}, \citenamefont
  {Tandel}, \citenamefont {Wang},\ and\ \citenamefont
  {Zhu}}]{Robinson2008_PRC78-034308}%
  \BibitemOpen
  \bibfield  {author} {\bibinfo {author} {\bibfnamefont {A.~P.}\ \bibnamefont
  {Robinson}}, \bibinfo {author} {\bibfnamefont {T.~L.}\ \bibnamefont {Khoo}},
  \bibinfo {author} {\bibfnamefont {I.}~\bibnamefont {Ahmad}}, \bibinfo
  {author} {\bibfnamefont {S.~K.}\ \bibnamefont {Tandel}}, \bibinfo {author}
  {\bibfnamefont {F.~G.}\ \bibnamefont {Kondev}}, \bibinfo {author}
  {\bibfnamefont {T.}~\bibnamefont {Nakatsukasa}}, \bibinfo {author}
  {\bibfnamefont {D.}~\bibnamefont {Seweryniak}}, \bibinfo {author}
  {\bibfnamefont {M.}~\bibnamefont {Asai}}, \bibinfo {author} {\bibfnamefont
  {B.~B.}\ \bibnamefont {Back}}, \bibinfo {author} {\bibfnamefont {M.~P.}\
  \bibnamefont {Carpenter}}, \bibinfo {author} {\bibfnamefont {P.}~\bibnamefont
  {Chowdhury}}, \bibinfo {author} {\bibfnamefont {C.~N.}\ \bibnamefont
  {Davids}}, \bibinfo {author} {\bibfnamefont {S.}~\bibnamefont {Eeckhaudt}},
  \bibinfo {author} {\bibfnamefont {J.~P.}\ \bibnamefont {Greene}}, \bibinfo
  {author} {\bibfnamefont {P.~T.}\ \bibnamefont {Greenlees}}, \bibinfo {author}
  {\bibfnamefont {S.}~\bibnamefont {Gros}}, \bibinfo {author} {\bibfnamefont
  {A.}~\bibnamefont {Heinz}}, \bibinfo {author} {\bibfnamefont {R.-D.}\
  \bibnamefont {Herzberg}}, \bibinfo {author} {\bibfnamefont {R.~V.~F.}\
  \bibnamefont {Janssens}}, \bibinfo {author} {\bibfnamefont {G.~D.}\
  \bibnamefont {Jones}}, \bibinfo {author} {\bibfnamefont {T.}~\bibnamefont
  {Lauritsen}}, \bibinfo {author} {\bibfnamefont {C.~J.}\ \bibnamefont
  {Lister}}, \bibinfo {author} {\bibfnamefont {D.}~\bibnamefont {Peterson}},
  \bibinfo {author} {\bibfnamefont {J.}~\bibnamefont {Qian}}, \bibinfo {author}
  {\bibfnamefont {U.~S.}\ \bibnamefont {Tandel}}, \bibinfo {author}
  {\bibfnamefont {X.}~\bibnamefont {Wang}}, \ and\ \bibinfo {author}
  {\bibfnamefont {S.}~\bibnamefont {Zhu}},\ }\href {\doibase
  10.1103/PhysRevC.78.034308} {\bibfield  {journal} {\bibinfo  {journal} {Phys.
  Rev. C}\ }\textbf {\bibinfo {volume} {78}},\ \bibinfo {pages} {034308}
  (\bibinfo {year} {2008})}\BibitemShut {NoStop}%
\bibitem [{\citenamefont {Chen}\ \emph {et~al.}(2008)\citenamefont {Chen},
  \citenamefont {Sun},\ and\ \citenamefont {Gao}}]{Chen2008_PRC77-061305R}%
  \BibitemOpen
  \bibfield  {author} {\bibinfo {author} {\bibfnamefont {Y.-S.}\ \bibnamefont
  {Chen}}, \bibinfo {author} {\bibfnamefont {Y.}~\bibnamefont {Sun}}, \ and\
  \bibinfo {author} {\bibfnamefont {Z.-C.}\ \bibnamefont {Gao}},\ }\href
  {\doibase 10.1103/PhysRevC.77.061305} {\bibfield  {journal} {\bibinfo
  {journal} {Phys. Rev. C}\ }\textbf {\bibinfo {volume} {77}},\ \bibinfo
  {pages} {061305(R)} (\bibinfo {year} {2008})}\BibitemShut {NoStop}%
\bibitem [{\citenamefont {Zhao}\ \emph
  {et~al.}(2012{\natexlab{a}})\citenamefont {Zhao}, \citenamefont {Lu},
  \citenamefont {Zhao},\ and\ \citenamefont {Zhou}}]{Zhao2012_PRC86-057304}%
  \BibitemOpen
  \bibfield  {author} {\bibinfo {author} {\bibfnamefont {J.}~\bibnamefont
  {Zhao}}, \bibinfo {author} {\bibfnamefont {B.-N.}\ \bibnamefont {Lu}},
  \bibinfo {author} {\bibfnamefont {E.-G.}\ \bibnamefont {Zhao}}, \ and\
  \bibinfo {author} {\bibfnamefont {S.-G.}\ \bibnamefont {Zhou}},\ }\href
  {\doibase 10.1103/PhysRevC.86.057304} {\bibfield  {journal} {\bibinfo
  {journal} {Phys. Rev. C}\ }\textbf {\bibinfo {volume} {86}},\ \bibinfo
  {pages} {057304} (\bibinfo {year} {2012}{\natexlab{a}})} \BibitemShut
  {NoStop}%
\bibitem [{\citenamefont {Liu}\ \emph {et~al.}(2012)\citenamefont {Liu},
  \citenamefont {Xu},\ and\ \citenamefont {Walker}}]{Liu2012_PRC86-011301R}%
  \BibitemOpen
  \bibfield  {author} {\bibinfo {author} {\bibfnamefont {H.~L.}\ \bibnamefont
  {Liu}}, \bibinfo {author} {\bibfnamefont {F.~R.}\ \bibnamefont {Xu}}, \ and\
  \bibinfo {author} {\bibfnamefont {P.~M.}\ \bibnamefont {Walker}},\ }\href
  {\doibase 10.1103/PhysRevC.86.011301} {\bibfield  {journal} {\bibinfo
  {journal} {Phys. Rev. C}\ }\textbf {\bibinfo {volume} {86}},\ \bibinfo
  {pages} {011301(R)} (\bibinfo {year} {2012})}\BibitemShut {NoStop}%
\bibitem [{\citenamefont {Zhang}\ \emph {et~al.}(2013)\citenamefont {Zhang},
  \citenamefont {Meng}, \citenamefont {Zhao},\ and\ \citenamefont
  {Zhou}}]{Zhang2013_PRC87-054308}%
  \BibitemOpen
  \bibfield  {author} {\bibinfo {author} {\bibfnamefont {Z.-H.}\ \bibnamefont
  {Zhang}}, \bibinfo {author} {\bibfnamefont {J.}~\bibnamefont {Meng}},
  \bibinfo {author} {\bibfnamefont {E.-G.}\ \bibnamefont {Zhao}}, \ and\
  \bibinfo {author} {\bibfnamefont {S.-G.}\ \bibnamefont {Zhou}},\ }\href
  {\doibase 10.1103/PhysRevC.87.054308} {\bibfield  {journal} {\bibinfo
  {journal} {Phys. Rev. C}\ }\textbf {\bibinfo {volume} {87}},\ \bibinfo
  {pages} {054308} (\bibinfo {year} {2013})}
%,\ \Eprint {http://arxiv.org/abs/1208.1156} {arXiv:1208.1156 [nucl-th]}
  \BibitemShut {NoStop}%
\bibitem [{\citenamefont {Zubov}\ \emph {et~al.}(2009)\citenamefont {Zubov},
  \citenamefont {Adamian},\ and\ \citenamefont
  {Antonenko}}]{Zubov2009_PPN40-847}%
  \BibitemOpen
  \bibfield  {author} {\bibinfo {author} {\bibfnamefont {A.}~\bibnamefont
  {Zubov}}, \bibinfo {author} {\bibfnamefont {G.}~\bibnamefont {Adamian}}, \
  and\ \bibinfo {author} {\bibfnamefont {N.}~\bibnamefont {Antonenko}},\ }\href
  {\doibase 10.1134/S1063779609060057} {\bibfield  {journal} {\bibinfo
  {journal} {Phys. Part. Nucl.}\ }\textbf {\bibinfo {volume} {40}},\ \bibinfo
  {pages} {847} (\bibinfo {year} {2009})}\BibitemShut {NoStop}%
\bibitem [{\citenamefont {Xia}\ \emph {et~al.}(2011)\citenamefont {Xia},
  \citenamefont {Sun}, \citenamefont {Zhao},\ and\ \citenamefont
  {Zhou}}]{Xia2011_SciChinaPMA54S1-109}%
  \BibitemOpen
  \bibfield  {author} {\bibinfo {author} {\bibfnamefont {C.-J.}\ \bibnamefont
  {Xia}}, \bibinfo {author} {\bibfnamefont {B.-X.}\ \bibnamefont {Sun}},
  \bibinfo {author} {\bibfnamefont {E.-G.}\ \bibnamefont {Zhao}}, \ and\
  \bibinfo {author} {\bibfnamefont {S.-G.}\ \bibnamefont {Zhou}},\ }\href
  {\doibase 10.1007/s11433-011-4438-2} {\bibfield  {journal} {\bibinfo
  {journal} {Sci. China-Phys. Mech. Astron.}\ }\textbf {\bibinfo {volume} {54
  (Suppl. 1)}},\ \bibinfo {pages} {s109} (\bibinfo {year} {2011})}
%,\ \Eprint {http://arxiv.org/abs/1101.2725} {arXiv:1101.2725 [nucl-th]}
  \BibitemShut {NoStop}%
\bibitem [{\citenamefont {M\"oller}\ \emph {et~al.}(2001)\citenamefont
  {M\"oller}, \citenamefont {Madland}, \citenamefont {Sierk},\ and\
  \citenamefont {Iwamoto}}]{Moeller2001_Nature409-785}%
  \BibitemOpen
  \bibfield  {author} {\bibinfo {author} {\bibfnamefont {P.}~\bibnamefont
  {M\"oller}}, \bibinfo {author} {\bibfnamefont {D.~G.}\ \bibnamefont
  {Madland}}, \bibinfo {author} {\bibfnamefont {A.~J.}\ \bibnamefont {Sierk}},
  \ and\ \bibinfo {author} {\bibfnamefont {A.}~\bibnamefont {Iwamoto}},\ }\href
  {\doibase 10.1038/35057204} {\bibfield  {journal} {\bibinfo  {journal}
  {Nature}\ }\textbf {\bibinfo {volume} {409}},\ \bibinfo {pages} {785}
  (\bibinfo {year} {2001})}\BibitemShut {NoStop}%
\bibitem [{\citenamefont {Pomorski}\ and\ \citenamefont
  {Dudek}(2003)}]{Pomorski2003_PRC67-044316}%
  \BibitemOpen
  \bibfield  {author} {\bibinfo {author} {\bibfnamefont {K.}~\bibnamefont
  {Pomorski}}\ and\ \bibinfo {author} {\bibfnamefont {J.}~\bibnamefont
  {Dudek}},\ }\href {\doibase 10.1103/PhysRevC.67.044316} {\bibfield  {journal}
  {\bibinfo  {journal} {Phys. Rev. C}\ }\textbf {\bibinfo {volume} {67}},\
  \bibinfo {pages} {044316} (\bibinfo {year} {2003})}\BibitemShut {NoStop}%
\bibitem [{\citenamefont {Ivanyuk}\ and\ \citenamefont
  {Pomorski}(2009)}]{Ivanyuk2009_PRC79-054327}%
  \BibitemOpen
  \bibfield  {author} {\bibinfo {author} {\bibfnamefont {F.~A.}\ \bibnamefont
  {Ivanyuk}}\ and\ \bibinfo {author} {\bibfnamefont {K.}~\bibnamefont
  {Pomorski}},\ }\href {\doibase 10.1103/PhysRevC.79.054327} {\bibfield
  {journal} {\bibinfo  {journal} {Phys. Rev. C}\ }\textbf {\bibinfo {volume}
  {79}},\ \bibinfo {pages} {054327} (\bibinfo {year} {2009})}\BibitemShut
  {NoStop}%
\bibitem [{\citenamefont {Kowal}\ \emph {et~al.}(2010)\citenamefont {Kowal},
  \citenamefont {Jachimowicz},\ and\ \citenamefont
  {Sobiczewski}}]{Kowal2010_PRC82-014303}%
  \BibitemOpen
  \bibfield  {author} {\bibinfo {author} {\bibfnamefont {M.}~\bibnamefont
  {Kowal}}, \bibinfo {author} {\bibfnamefont {P.}~\bibnamefont {Jachimowicz}},
  \ and\ \bibinfo {author} {\bibfnamefont {A.}~\bibnamefont {Sobiczewski}},\
  }\href {\doibase 10.1103/PhysRevC.82.014303} {\bibfield  {journal} {\bibinfo
  {journal} {Phys. Rev. C}\ }\textbf {\bibinfo {volume} {82}},\ \bibinfo
  {pages} {014303} (\bibinfo {year} {2010})}\BibitemShut {NoStop}%
\bibitem [{\citenamefont {Royer}\ \emph {et~al.}(2012)\citenamefont {Royer},
  \citenamefont {Jaffre},\ and\ \citenamefont
  {Moreau}}]{Royer2012_PRC86-044326}%
  \BibitemOpen
  \bibfield  {author} {\bibinfo {author} {\bibfnamefont {G.}~\bibnamefont
  {Royer}}, \bibinfo {author} {\bibfnamefont {M.}~\bibnamefont {Jaffre}}, \
  and\ \bibinfo {author} {\bibfnamefont {D.}~\bibnamefont {Moreau}},\ }\href
  {\doibase 10.1103/PhysRevC.86.044326} {\bibfield  {journal} {\bibinfo
  {journal} {Phys. Rev. C}\ }\textbf {\bibinfo {volume} {86}},\ \bibinfo
  {pages} {044326} (\bibinfo {year} {2012})}\BibitemShut {NoStop}%
\bibitem [{\citenamefont {Mamdouh}\ \emph {et~al.}(1998)\citenamefont
  {Mamdouh}, \citenamefont {Pearson}, \citenamefont {Rayet},\ and\
  \citenamefont {Tondeur}}]{Mamdouh1998_NPA644-389}%
  \BibitemOpen
  \bibfield  {author} {\bibinfo {author} {\bibfnamefont {A.}~\bibnamefont
  {Mamdouh}}, \bibinfo {author} {\bibfnamefont {J.~M.}\ \bibnamefont
  {Pearson}}, \bibinfo {author} {\bibfnamefont {M.}~\bibnamefont {Rayet}}, \
  and\ \bibinfo {author} {\bibfnamefont {F.}~\bibnamefont {Tondeur}},\ }\href
  {\doibase 10.1016/S0375-9474(98)00576-4} {\bibfield  {journal} {\bibinfo
  {journal} {Nucl. Phys. A}\ }\textbf {\bibinfo {volume} {644}},\ \bibinfo
  {pages} {389} (\bibinfo {year} {1998})}\BibitemShut {NoStop}%
\bibitem [{\citenamefont {Mamdouh}\ \emph {et~al.}(2001)\citenamefont
  {Mamdouh}, \citenamefont {Pearson}, \citenamefont {Rayet},\ and\
  \citenamefont {Tondeur}}]{Mamdouh2001_NPA679-337}%
  \BibitemOpen
  \bibfield  {author} {\bibinfo {author} {\bibfnamefont {A.}~\bibnamefont
  {Mamdouh}}, \bibinfo {author} {\bibfnamefont {J.~M.}\ \bibnamefont
  {Pearson}}, \bibinfo {author} {\bibfnamefont {M.}~\bibnamefont {Rayet}}, \
  and\ \bibinfo {author} {\bibfnamefont {F.}~\bibnamefont {Tondeur}},\ }\href
  {\doibase 10.1016/S0375-9474(00)00358-4} {\bibfield  {journal} {\bibinfo
  {journal} {Nucl. Phys. A}\ }\textbf {\bibinfo {volume} {679}},\ \bibinfo
  {pages} {337} (\bibinfo {year} {2001})}\BibitemShut {NoStop}%
\bibitem [{\citenamefont {Burvenich}\ \emph {et~al.}(2004)\citenamefont
  {Burvenich}, \citenamefont {Bender}, \citenamefont {Maruhn},\ and\
  \citenamefont {Reinhard}}]{Burvenich2004_PRC69-014307}%
  \BibitemOpen
  \bibfield  {author} {\bibinfo {author} {\bibfnamefont {T.}~\bibnamefont
  {Burvenich}}, \bibinfo {author} {\bibfnamefont {M.}~\bibnamefont {Bender}},
  \bibinfo {author} {\bibfnamefont {J.~A.}\ \bibnamefont {Maruhn}}, \ and\
  \bibinfo {author} {\bibfnamefont {P.-G.}\ \bibnamefont {Reinhard}},\ }\href
  {\doibase 10.1103/PhysRevC.69.014307} {\bibfield  {journal} {\bibinfo
  {journal} {Phys. Rev. C}\ }\textbf {\bibinfo {volume} {69}},\ \bibinfo
  {pages} {014307} (\bibinfo {year} {2004})}\BibitemShut {NoStop}%
\bibitem [{\citenamefont {Samyn}\ \emph {et~al.}(2005)\citenamefont {Samyn},
  \citenamefont {Goriely},\ and\ \citenamefont
  {Pearson}}]{Samyn2005_PRC72-044316}%
  \BibitemOpen
  \bibfield  {author} {\bibinfo {author} {\bibfnamefont {M.}~\bibnamefont
  {Samyn}}, \bibinfo {author} {\bibfnamefont {S.}~\bibnamefont {Goriely}}, \
  and\ \bibinfo {author} {\bibfnamefont {J.~M.}\ \bibnamefont {Pearson}},\
  }\href {\doibase 10.1103/PhysRevC.72.044316} {\bibfield  {journal} {\bibinfo
  {journal} {Phys. Rev. C}\ }\textbf {\bibinfo {volume} {72}},\ \bibinfo
  {pages} {044316} (\bibinfo {year} {2005})}\BibitemShut {NoStop}%
\bibitem [{\citenamefont {Goriely}\ \emph {et~al.}(2007)\citenamefont
  {Goriely}, \citenamefont {Samyn},\ and\ \citenamefont
  {Pearson}}]{Goriely2007_PRC75-064312}%
  \BibitemOpen
  \bibfield  {author} {\bibinfo {author} {\bibfnamefont {S.}~\bibnamefont
  {Goriely}}, \bibinfo {author} {\bibfnamefont {M.}~\bibnamefont {Samyn}}, \
  and\ \bibinfo {author} {\bibfnamefont {J.~M.}\ \bibnamefont {Pearson}},\
  }\href {\doibase 10.1103/PhysRevC.75.064312} {\bibfield  {journal} {\bibinfo
  {journal} {Phys. Rev. C}\ }\textbf {\bibinfo {volume} {75}},\ \bibinfo
  {pages} {064312} (\bibinfo {year} {2007})}\BibitemShut {NoStop}%
\bibitem [{\citenamefont {Minato}\ \emph {et~al.}(2009)\citenamefont {Minato},
  \citenamefont {Chiba},\ and\ \citenamefont {Hagino}}]{Minato2009_NPA831-150}%
  \BibitemOpen
  \bibfield  {author} {\bibinfo {author} {\bibfnamefont {F.}~\bibnamefont
  {Minato}}, \bibinfo {author} {\bibfnamefont {S.}~\bibnamefont {Chiba}}, \
  and\ \bibinfo {author} {\bibfnamefont {K.}~\bibnamefont {Hagino}},\ }\href
  {\doibase 10.1016/j.nuclphysa.2009.09.063} {\bibfield  {journal} {\bibinfo
  {journal} {Nucl. Phys. A}\ }\textbf {\bibinfo {volume} {831}},\ \bibinfo
  {pages} {150} (\bibinfo {year} {2009})}\BibitemShut {NoStop}%
\bibitem [{\citenamefont {Pei}\ \emph {et~al.}(2009)\citenamefont {Pei},
  \citenamefont {Nazarewicz}, \citenamefont {Sheikh},\ and\ \citenamefont
  {Kerman}}]{Pei2009_PRL102-192501}%
  \BibitemOpen
  \bibfield  {author} {\bibinfo {author} {\bibfnamefont {J.~C.}\ \bibnamefont
  {Pei}}, \bibinfo {author} {\bibfnamefont {W.}~\bibnamefont {Nazarewicz}},
  \bibinfo {author} {\bibfnamefont {J.~A.}\ \bibnamefont {Sheikh}}, \ and\
  \bibinfo {author} {\bibfnamefont {A.~K.}\ \bibnamefont {Kerman}},\ }\href
  {\doibase 10.1103/PhysRevLett.102.192501} {\bibfield  {journal} {\bibinfo
  {journal} {Phys. Rev. Lett.}\ }\textbf {\bibinfo {volume} {102}},\ \bibinfo
  {pages} {192501} (\bibinfo {year} {2009})}\BibitemShut {NoStop}%
\bibitem [{\citenamefont {Kortelainen}\ \emph {et~al.}(2012)\citenamefont
  {Kortelainen}, \citenamefont {McDonnell}, \citenamefont {Nazarewicz},
  \citenamefont {Reinhard}, \citenamefont {Sarich}, \citenamefont {Schunck},
  \citenamefont {Stoitsov},\ and\ \citenamefont
  {Wild}}]{Kortelainen2012_PRC85-024304}%
  \BibitemOpen
  \bibfield  {author} {\bibinfo {author} {\bibfnamefont {M.}~\bibnamefont
  {Kortelainen}}, \bibinfo {author} {\bibfnamefont {J.}~\bibnamefont
  {McDonnell}}, \bibinfo {author} {\bibfnamefont {W.}~\bibnamefont
  {Nazarewicz}}, \bibinfo {author} {\bibfnamefont {P.-G.}\ \bibnamefont
  {Reinhard}}, \bibinfo {author} {\bibfnamefont {J.}~\bibnamefont {Sarich}},
  \bibinfo {author} {\bibfnamefont {N.}~\bibnamefont {Schunck}}, \bibinfo
  {author} {\bibfnamefont {M.~V.}\ \bibnamefont {Stoitsov}}, \ and\ \bibinfo
  {author} {\bibfnamefont {S.~M.}\ \bibnamefont {Wild}},\ }\href {\doibase
  10.1103/PhysRevC.85.024304} {\bibfield  {journal} {\bibinfo  {journal} {Phys.
  Rev. C}\ }\textbf {\bibinfo {volume} {85}},\ \bibinfo {pages} {024304}
  (\bibinfo {year} {2012})}\BibitemShut {NoStop}%
\bibitem [{\citenamefont {McDonnell}\ \emph {et~al.}(2013)\citenamefont
  {McDonnell}, \citenamefont {Schunck},\ and\ \citenamefont
  {Nazarewicz}}]{McDonnell2012_arXiv1301.7587}%
  \BibitemOpen
  \bibfield  {author} {\bibinfo {author} {\bibfnamefont {J.}~\bibnamefont
  {McDonnell}}, \bibinfo {author} {\bibfnamefont {N.}~\bibnamefont {Schunck}},
  \ and\ \bibinfo {author} {\bibfnamefont {W.}~\bibnamefont {Nazarewicz}},\
  } in \textit{Fission and Properties of Neutron-Rich Nuclei: Proceedings of 
  the Fifth International Conference on Fission and Properties of Neutron-Rich
  Nuclei (ICFN5)}, Nov. 4-10, 2012, Sanibel Island, Florida, USA, Edited by 
  J. H. Hamilton and A. V.  Ramayya (World Scientific, Singapore, 2013),  
  pp. 597--604 \BibitemShut {NoStop}%
\bibitem [{\citenamefont {Staszczak}\ \emph {et~al.}(2013)\citenamefont
  {Staszczak}, \citenamefont {Baran},\ and\ \citenamefont
  {Nazarewicz}}]{Staszczak2013_PRC87-024320}%
  \BibitemOpen
  \bibfield  {author} {\bibinfo {author} {\bibfnamefont {A.}~\bibnamefont
  {Staszczak}}, \bibinfo {author} {\bibfnamefont {A.}~\bibnamefont {Baran}}, \
  and\ \bibinfo {author} {\bibfnamefont {W.}~\bibnamefont {Nazarewicz}},\
  }\href {\doibase 10.1103/PhysRevC.87.024320} {\bibfield  {journal} {\bibinfo
  {journal} {Phys. Rev. C}\ }\textbf {\bibinfo {volume} {87}},\ \bibinfo
  {pages} {024320} (\bibinfo {year} {2013})}\BibitemShut {NoStop}%
\bibitem [{\citenamefont {Schunck}\ \emph
  {et~al.}(2013{\natexlab{a}})\citenamefont {Schunck}, \citenamefont {Duke},
  \citenamefont {Carr},\ and\ \citenamefont
  {Knoll}}]{Schunck2013_arXiv1311.2616}%
  \BibitemOpen
  \bibfield  {author} {\bibinfo {author} {\bibfnamefont {N.}~\bibnamefont
  {Schunck}}, \bibinfo {author} {\bibfnamefont {D.}~\bibnamefont {Duke}},
  \bibinfo {author} {\bibfnamefont {H.}~\bibnamefont {Carr}}, \ and\ \bibinfo
  {author} {\bibfnamefont {A.}~\bibnamefont {Knoll}},\ }\href
  {http://arxiv.org/abs/1311.2616}
  {arXiv:1311.2616 [nucl-th]}\BibitemShut {NoStop}%
\bibitem [{\citenamefont {Schunck}\ \emph
  {et~al.}(2013{\natexlab{b}})\citenamefont {Schunck}, \citenamefont {Duke},\
  and\ \citenamefont {Carr}}]{Schunck2013_arXiv1311.2620}%
  \BibitemOpen
  \bibfield  {author} {\bibinfo {author} {\bibfnamefont {N.}~\bibnamefont
  {Schunck}}, \bibinfo {author} {\bibfnamefont {D.}~\bibnamefont {Duke}}, \
  and\ \bibinfo {author} {\bibfnamefont {H.}~\bibnamefont {Carr}},\ }\href
  {http://arxiv.org/abs/1311.2620} {arXiv:1311.2620 [nucl-th]}
  \BibitemShut {NoStop}%
\bibitem [{\citenamefont {Egido}\ and\ \citenamefont
  {Robledo}(2000)}]{Egido2000_PRL85-1198}%
  \BibitemOpen
  \bibfield  {author} {\bibinfo {author} {\bibfnamefont {J.~L.}\ \bibnamefont
  {Egido}}\ and\ \bibinfo {author} {\bibfnamefont {L.~M.}\ \bibnamefont
  {Robledo}},\ }\href {\doibase 10.1103/PhysRevLett.85.1198} {\bibfield
  {journal} {\bibinfo  {journal} {Phys. Rev. Lett.}\ }\textbf {\bibinfo
  {volume} {85}},\ \bibinfo {pages} {1198} (\bibinfo {year}
  {2000})}\BibitemShut {NoStop}%
\bibitem [{\citenamefont {Warda}\ and\ \citenamefont
  {Egido}(2012)}]{Warda2012_PRC86-014322}%
  \BibitemOpen
  \bibfield  {author} {\bibinfo {author} {\bibfnamefont {M.}~\bibnamefont
  {Warda}}\ and\ \bibinfo {author} {\bibfnamefont {J.~L.}\ \bibnamefont
  {Egido}},\ }\href {\doibase 10.1103/PhysRevC.86.014322} {\bibfield  {journal}
  {\bibinfo  {journal} {Phys. Rev. C}\ }\textbf {\bibinfo {volume} {86}},\
  \bibinfo {pages} {014322} (\bibinfo {year} {2012})}\BibitemShut {NoStop}%
\bibitem [{\citenamefont {Blum}\ \emph {et~al.}(1994)\citenamefont {Blum},
  \citenamefont {Maruhn}, \citenamefont {Reinhard},\ and\ \citenamefont
  {Greiner}}]{Blum1994_PLB323-262}%
  \BibitemOpen
  \bibfield  {author} {\bibinfo {author} {\bibfnamefont {V.}~\bibnamefont
  {Blum}}, \bibinfo {author} {\bibfnamefont {J.}~\bibnamefont {Maruhn}},
  \bibinfo {author} {\bibfnamefont {P.-G.}\ \bibnamefont {Reinhard}}, \ and\
  \bibinfo {author} {\bibfnamefont {W.}~\bibnamefont {Greiner}},\ }\href
  {\doibase 10.1016/0370-2693(94)91217-3} {\bibfield  {journal} {\bibinfo
  {journal} {Phys. Lett. B}\ }\textbf {\bibinfo {volume} {323}},\ \bibinfo
  {pages} {262} (\bibinfo {year} {1994})}\BibitemShut {NoStop}%
\bibitem [{\citenamefont {Zhang}\ \emph {et~al.}(2003)\citenamefont {Zhang},
  \citenamefont {Zhang}, \citenamefont {Zhang},\ and\ \citenamefont
  {Meng}}]{Zhang2003_CPL20-1694}%
  \BibitemOpen
  \bibfield  {author} {\bibinfo {author} {\bibfnamefont {W.}~\bibnamefont
  {Zhang}}, \bibinfo {author} {\bibfnamefont {S.-S.}\ \bibnamefont {Zhang}},
  \bibinfo {author} {\bibfnamefont {S.-Q.}\ \bibnamefont {Zhang}}, \ and\
  \bibinfo {author} {\bibfnamefont {J.}~\bibnamefont {Meng}},\ }\href {\doibase
  10.1088/0256-307X/20/10/312} {\bibfield  {journal} {\bibinfo  {journal}
  {Chin. Phys. Lett.}\ }\textbf {\bibinfo {volume} {20}},\ \bibinfo {pages}
  {1694} (\bibinfo {year} {2003})}\BibitemShut {NoStop}%
\bibitem [{\citenamefont {Bender}\ \emph {et~al.}(2003)\citenamefont {Bender},
  \citenamefont {Heenen},\ and\ \citenamefont
  {Reinhard}}]{Bender2003_RMP75-121}%
  \BibitemOpen
  \bibfield  {author} {\bibinfo {author} {\bibfnamefont {M.}~\bibnamefont
  {Bender}}, \bibinfo {author} {\bibfnamefont {P.-H.}\ \bibnamefont {Heenen}},
  \ and\ \bibinfo {author} {\bibfnamefont {P.-G.}\ \bibnamefont {Reinhard}},\
  }\href {\doibase 10.1103/RevModPhys.75.121} {\bibfield  {journal} {\bibinfo
  {journal} {Rev. Mod. Phys.}\ }\textbf {\bibinfo {volume} {75}},\ \bibinfo
  {pages} {121} (\bibinfo {year} {2003})}\BibitemShut {NoStop}%
\bibitem [{\citenamefont {L\"u}\ \emph {et~al.}(2006)\citenamefont {L\"u},
  \citenamefont {Geng},\ and\ \citenamefont {Meng}}]{Lu2006_CPL23-2940}%
  \BibitemOpen
  \bibfield  {author} {\bibinfo {author} {\bibfnamefont {H.-F.}\ \bibnamefont
  {L\"u}}, \bibinfo {author} {\bibfnamefont {L.-S.}\ \bibnamefont {Geng}}, \
  and\ \bibinfo {author} {\bibfnamefont {J.}~\bibnamefont {Meng}},\ }\href
  {\doibase 10.1088/0256-307X/23/11/016} {\bibfield  {journal} {\bibinfo
  {journal} {Chin. Phys. Lett.}\ }\textbf {\bibinfo {volume} {23}},\ \bibinfo
  {pages} {2940} (\bibinfo {year} {2006})}\BibitemShut {NoStop}%
\bibitem [{\citenamefont {Li}\ \emph {et~al.}(2010)\citenamefont {Li},
  \citenamefont {Nik\v{s}i\'{c}}, \citenamefont {Vretenar}, \citenamefont
  {Ring},\ and\ \citenamefont {Meng}}]{Li2010_PRC81-064321}%
  \BibitemOpen
  \bibfield  {author} {\bibinfo {author} {\bibfnamefont {Z.~P.}\ \bibnamefont
  {Li}}, \bibinfo {author} {\bibfnamefont {T.}~\bibnamefont {Nik\v{s}i\'{c}}},
  \bibinfo {author} {\bibfnamefont {D.}~\bibnamefont {Vretenar}}, \bibinfo
  {author} {\bibfnamefont {P.}~\bibnamefont {Ring}}, \ and\ \bibinfo {author}
  {\bibfnamefont {J.}~\bibnamefont {Meng}},\ }\href {\doibase
  10.1103/PhysRevC.81.064321} {\bibfield  {journal} {\bibinfo  {journal} {Phys.
  Rev. C}\ }\textbf {\bibinfo {volume} {81}},\ \bibinfo {pages} {064321}
  (\bibinfo {year} {2010})}\BibitemShut {NoStop}%
\bibitem [{\citenamefont {Abusara}\ \emph {et~al.}(2010)\citenamefont
  {Abusara}, \citenamefont {Afanasjev},\ and\ \citenamefont
  {Ring}}]{Abusara2010_PRC82-044303}%
  \BibitemOpen
  \bibfield  {author} {\bibinfo {author} {\bibfnamefont {H.}~\bibnamefont
  {Abusara}}, \bibinfo {author} {\bibfnamefont {A.~V.}\ \bibnamefont
  {Afanasjev}}, \ and\ \bibinfo {author} {\bibfnamefont {P.}~\bibnamefont
  {Ring}},\ }\href {\doibase 10.1103/PhysRevC.82.044303} {\bibfield  {journal}
  {\bibinfo  {journal} {Phys. Rev. C}\ }\textbf {\bibinfo {volume} {82}},\
  \bibinfo {pages} {044303} (\bibinfo {year} {2010})}\BibitemShut {NoStop}%
\bibitem [{\citenamefont {Lu}\ \emph {et~al.}(2012{\natexlab{a}})\citenamefont
  {Lu}, \citenamefont {Zhao},\ and\ \citenamefont
  {Zhou}}]{Lu2012_PRC85-011301R}%
  \BibitemOpen
  \bibfield  {author} {\bibinfo {author} {\bibfnamefont {B.-N.}\ \bibnamefont
  {Lu}}, \bibinfo {author} {\bibfnamefont {E.-G.}\ \bibnamefont {Zhao}}, \ and\
  \bibinfo {author} {\bibfnamefont {S.-G.}\ \bibnamefont {Zhou}},\ }\href
  {\doibase 10.1103/PhysRevC.85.011301} {\bibfield  {journal} {\bibinfo
  {journal} {Phys. Rev. C}\ }\textbf {\bibinfo {volume} {85}},\ \bibinfo
  {pages} {011301(R)} (\bibinfo {year} {2012}{\natexlab{a}})} \BibitemShut
  {NoStop}%
\bibitem [{\citenamefont {Lu}(2012)}]{Lu2012_PhD}%
  \BibitemOpen
  \bibfield  {author} {\bibinfo {author} {\bibfnamefont {B.-N.}\ \bibnamefont
  {Lu}},\ }\emph {\bibinfo {title} {Multi-dimensional constrained relativistic
  mean field theory and the potential energy surfaces and fission barriers of
  actinide nuclei}},\ \href@noop {} {Ph.D. thesis},\ \bibinfo  {school}
  {Institute of Theoretical Physics, Chinese Academy of Sciences} (\bibinfo
  {year} {2012}),\ \bibinfo {note} {{in Chinese}}\BibitemShut {NoStop}%
\bibitem [{\citenamefont {Lu}\ \emph {et~al.}(2012{\natexlab{b}})\citenamefont
  {Lu}, \citenamefont {Zhao}, \citenamefont {Zhao},\ and\ \citenamefont
  {Zhou}}]{Lu2012_EPJWoC38-05003}%
  \BibitemOpen
  \bibfield  {author} {\bibinfo {author} {\bibfnamefont {B.-N.}\ \bibnamefont
  {Lu}}, \bibinfo {author} {\bibfnamefont {J.}~\bibnamefont {Zhao}}, \bibinfo
  {author} {\bibfnamefont {E.-G.}\ \bibnamefont {Zhao}}, \ and\ \bibinfo
  {author} {\bibfnamefont {S.-G.}\ \bibnamefont {Zhou}},\ }\href {\doibase
  10.1051/epjconf/20123805003} {\bibfield  {journal} {\bibinfo  {journal} {EPJ
  Web Conf.}\ }\textbf {\bibinfo {volume} {38}},\ \bibinfo {pages} {05003}
  (\bibinfo {year} {2012}{\natexlab{b}})} \BibitemShut
  {NoStop}%
\bibitem [{\citenamefont {Lu}\ \emph {et~al.}(2013)\citenamefont {Lu},
  \citenamefont {Zhao}, \citenamefont {Zhao},\ and\ \citenamefont
  {Zhou}}]{Lu2013_arXiv1304.6830}%
  \BibitemOpen
  \bibfield  {author} {\bibinfo {author} {\bibfnamefont {B.-N.}\ \bibnamefont
  {Lu}}, \bibinfo {author} {\bibfnamefont {J.}~\bibnamefont {Zhao}}, \bibinfo
  {author} {\bibfnamefont {E.-G.}\ \bibnamefont {Zhao}}, \ and\ \bibinfo
  {author} {\bibfnamefont {S.-G.}\ \bibnamefont {Zhou}},\ }\href
  {http://arxiv.org/abs/1304.6830} {arXiv:1304.6830 [nucl-th]} \BibitemShut
  {NoStop}%
\bibitem [{\citenamefont {Abusara}\ \emph {et~al.}(2012)\citenamefont
  {Abusara}, \citenamefont {Afanasjev},\ and\ \citenamefont
  {Ring}}]{Abusara2012_PRC85-024314}%
  \BibitemOpen
  \bibfield  {author} {\bibinfo {author} {\bibfnamefont {H.}~\bibnamefont
  {Abusara}}, \bibinfo {author} {\bibfnamefont {A.~V.}\ \bibnamefont
  {Afanasjev}}, \ and\ \bibinfo {author} {\bibfnamefont {P.}~\bibnamefont
  {Ring}},\ }\href {\doibase 10.1103/PhysRevC.85.024314} {\bibfield  {journal}
  {\bibinfo  {journal} {Phys. Rev. C}\ }\textbf {\bibinfo {volume} {85}},\
  \bibinfo {pages} {024314} (\bibinfo {year} {2012})}\BibitemShut {NoStop}%
\bibitem [{\citenamefont {Prassa}\ \emph {et~al.}(2012)\citenamefont {Prassa},
  \citenamefont {Nik\v{s}i\'{c}}, \citenamefont {Lalazissis},\ and\
  \citenamefont {Vretenar}}]{Prassa2012_PRC86-024317}%
  \BibitemOpen
  \bibfield  {author} {\bibinfo {author} {\bibfnamefont {V.}~\bibnamefont
  {Prassa}}, \bibinfo {author} {\bibfnamefont {T.}~\bibnamefont
  {Nik\v{s}i\'{c}}}, \bibinfo {author} {\bibfnamefont {G.~A.}\ \bibnamefont
  {Lalazissis}}, \ and\ \bibinfo {author} {\bibfnamefont {D.}~\bibnamefont
  {Vretenar}},\ }\href {\doibase 10.1103/PhysRevC.86.024317} {\bibfield
  {journal} {\bibinfo  {journal} {Phys. Rev. C}\ }\textbf {\bibinfo {volume}
  {86}},\ \bibinfo {pages} {024317} (\bibinfo {year} {2012})}\BibitemShut
  {NoStop}%
\bibitem [{\citenamefont {Prassa}\ \emph {et~al.}(2013)\citenamefont {Prassa},
  \citenamefont {Niksic},\ and\ \citenamefont
  {Vretenar}}]{Prassa2013_PRC88-044324}%
  \BibitemOpen
  \bibfield  {author} {\bibinfo {author} {\bibfnamefont {V.}~\bibnamefont
  {Prassa}}, \bibinfo {author} {\bibfnamefont {T.}~\bibnamefont {Niksic}}, \
  and\ \bibinfo {author} {\bibfnamefont {D.}~\bibnamefont {Vretenar}},\ }\href
  {\doibase 10.1103/PhysRevC.88.044324} {\bibfield  {journal} {\bibinfo
  {journal} {Phys. Rev. C}\ }\textbf {\bibinfo {volume} {88}},\ \bibinfo
  {pages} {044324} (\bibinfo {year} {2013})}\BibitemShut {NoStop}%
\bibitem [{\citenamefont {Afanasjev}(2013)}]{Afanasjev2013_arXiv1303.1206}%
  \BibitemOpen
  \bibfield  {author} {\bibinfo {author} {\bibfnamefont {A.~V.}\ \bibnamefont
  {Afanasjev}},\ }in \textit{Fission and Properties of Neutron-Rich Nuclei: 
  Proceedings of the Fifth International Conference on Fission and Properties 
  of Neutron-Rich Nuclei (ICFN5)}, Nov. 4-10, 2012, Sanibel Island, Florida, 
  USA, Edited by J. H. Hamilton and A. V.  Ramayya (World Scientific, 
  Singapore, 2013),  pp. 303--310 \BibitemShut {NoStop}%
\bibitem [{\citenamefont {Brack}\ \emph {et~al.}(1972)\citenamefont {Brack},
  \citenamefont {Damgaard}, \citenamefont {Jensen}, \citenamefont {Pauli},
  \citenamefont {Strutinsky},\ and\ \citenamefont
  {Wong}}]{Brack1972_RMP44-320}%
  \BibitemOpen
  \bibfield  {author} {\bibinfo {author} {\bibfnamefont {M.}~\bibnamefont
  {Brack}}, \bibinfo {author} {\bibfnamefont {J.}~\bibnamefont {Damgaard}},
  \bibinfo {author} {\bibfnamefont {A.~S.}\ \bibnamefont {Jensen}}, \bibinfo
  {author} {\bibfnamefont {H.~C.}\ \bibnamefont {Pauli}}, \bibinfo {author}
  {\bibfnamefont {V.~M.}\ \bibnamefont {Strutinsky}}, \ and\ \bibinfo {author}
  {\bibfnamefont {C.~Y.}\ \bibnamefont {Wong}},\ }\href {\doibase
  10.1103/RevModPhys.44.320} {\bibfield  {journal} {\bibinfo  {journal} {Rev.
  Mod. Phys.}\ }\textbf {\bibinfo {volume} {44}},\ \bibinfo {pages} {320}
  (\bibinfo {year} {1972})}\BibitemShut {NoStop}%
\bibitem [{\citenamefont {Pashkevich}(1969)}]{Pashkevich1969_NPA133-400}%
  \BibitemOpen
  \bibfield  {author} {\bibinfo {author} {\bibfnamefont {V.~V.}\ \bibnamefont
  {Pashkevich}},\ }\href {\doibase 10.1016/0375-9474(69)90641-1} {\bibfield
  {journal} {\bibinfo  {journal} {Nucl. Phys. A}\ }\textbf {\bibinfo {volume}
  {133}},\ \bibinfo {pages} {400} (\bibinfo {year} {1969})}\BibitemShut
  {NoStop}%
\bibitem [{\citenamefont {M\"oller}\ and\ \citenamefont
  {Nilsson}(1970)}]{Moeller1970_PLB31-283}%
  \BibitemOpen
  \bibfield  {author} {\bibinfo {author} {\bibfnamefont {P.}~\bibnamefont
  {M\"oller}}\ and\ \bibinfo {author} {\bibfnamefont {S.~G.}\ \bibnamefont
  {Nilsson}},\ }\href {\doibase 10.1016/0370-2693(70)90171-1} {\bibfield
  {journal} {\bibinfo  {journal} {Phys. Lett. B}\ }\textbf {\bibinfo {volume}
  {31}},\ \bibinfo {pages} {283} (\bibinfo {year} {1970})}\BibitemShut
  {NoStop}%
\bibitem [{\citenamefont {Randrup}\ \emph {et~al.}(1976)\citenamefont
  {Randrup}, \citenamefont {Larsson}, \citenamefont {Moller}, \citenamefont
  {Nilsson}, \citenamefont {Pomorski},\ and\ \citenamefont
  {Sobiczewski}}]{Randrup1976_PRC13-229}%
  \BibitemOpen
  \bibfield  {author} {\bibinfo {author} {\bibfnamefont {J.}~\bibnamefont
  {Randrup}}, \bibinfo {author} {\bibfnamefont {S.~E.}\ \bibnamefont
  {Larsson}}, \bibinfo {author} {\bibfnamefont {P.}~\bibnamefont {Moller}},
  \bibinfo {author} {\bibfnamefont {S.~G.}\ \bibnamefont {Nilsson}}, \bibinfo
  {author} {\bibfnamefont {K.}~\bibnamefont {Pomorski}}, \ and\ \bibinfo
  {author} {\bibfnamefont {A.}~\bibnamefont {Sobiczewski}},\ }\href {\doibase
  10.1103/PhysRevC.13.229} {\bibfield  {journal} {\bibinfo  {journal} {Phys.
  Rev. C}\ }\textbf {\bibinfo {volume} {13}},\ \bibinfo {pages} {229} (\bibinfo
  {year} {1976})}\BibitemShut {NoStop}%
\bibitem [{\citenamefont {Ledergerber}\ and\ \citenamefont
  {Pauli}(1973)}]{Ledergerber1973_NPA207-1}%
  \BibitemOpen
  \bibfield  {author} {\bibinfo {author} {\bibfnamefont {T.}~\bibnamefont
  {Ledergerber}}\ and\ \bibinfo {author} {\bibfnamefont {H.-C.}\ \bibnamefont
  {Pauli}},\ }\href {\doibase 10.1016/0375-9474(73)90022-5} {\bibfield
  {journal} {\bibinfo  {journal} {Nucl. Phys. A}\ }\textbf {\bibinfo {volume}
  {207}},\ \bibinfo {pages} {1} (\bibinfo {year} {1973})}\BibitemShut {NoStop}%
\bibitem [{\citenamefont {Girod}\ and\ \citenamefont
  {Grammaticos}(1983)}]{Girod1983_PRC27-2317}%
  \BibitemOpen
  \bibfield  {author} {\bibinfo {author} {\bibfnamefont {M.}~\bibnamefont
  {Girod}}\ and\ \bibinfo {author} {\bibfnamefont {B.}~\bibnamefont
  {Grammaticos}},\ }\href {\doibase 10.1103/PhysRevC.27.2317} {\bibfield
  {journal} {\bibinfo  {journal} {Phys. Rev. C}\ }\textbf {\bibinfo {volume}
  {27}},\ \bibinfo {pages} {2317} (\bibinfo {year} {1983})}\BibitemShut
  {NoStop}%
\bibitem [{\citenamefont {Rutz}\ \emph {et~al.}(1995)\citenamefont {Rutz},
  \citenamefont {Maruhn}, \citenamefont {Reinhard},\ and\ \citenamefont
  {Greiner}}]{Rutz1995_NPA590-680}%
  \BibitemOpen
  \bibfield  {author} {\bibinfo {author} {\bibfnamefont {K.}~\bibnamefont
  {Rutz}}, \bibinfo {author} {\bibfnamefont {J.~A.}\ \bibnamefont {Maruhn}},
  \bibinfo {author} {\bibfnamefont {P.~G.}\ \bibnamefont {Reinhard}}, \ and\
  \bibinfo {author} {\bibfnamefont {W.}~\bibnamefont {Greiner}},\ }\href
  {\doibase 10.1016/0375-9474(95)00192-4} {\bibfield  {journal} {\bibinfo
  {journal} {Nucl. Phys. A}\ }\textbf {\bibinfo {volume} {590}},\ \bibinfo
  {pages} {680} (\bibinfo {year} {1995})}\BibitemShut {NoStop}%
\bibitem [{\citenamefont {Bonneau}\ \emph {et~al.}(2004)\citenamefont
  {Bonneau}, \citenamefont {Quentin},\ and\ \citenamefont
  {Samsoen}}]{Bonneau2004_EPJA21-391}%
  \BibitemOpen
  \bibfield  {author} {\bibinfo {author} {\bibfnamefont {L.}~\bibnamefont
  {Bonneau}}, \bibinfo {author} {\bibfnamefont {P.}~\bibnamefont {Quentin}}, \
  and\ \bibinfo {author} {\bibfnamefont {D.}~\bibnamefont {Samsoen}},\ }\href
  {\doibase 10.1140/epja/i2003-10224-x} {\bibfield  {journal} {\bibinfo
  {journal} {Eur. Phys. J. A}\ }\textbf {\bibinfo {volume} {21}},\ \bibinfo
  {pages} {391} (\bibinfo {year} {2004})}\BibitemShut {NoStop}%
\bibitem [{\citenamefont {M\"oller}\ \emph {et~al.}(2009)\citenamefont
  {M\"oller}, \citenamefont {Sierk}, \citenamefont {Ichikawa}, \citenamefont
  {Iwamoto}, \citenamefont {Bengtsson}, \citenamefont {Uhrenholt},\ and\
  \citenamefont {Aberg}}]{Moeller2009_PRC79-064304}%
  \BibitemOpen
  \bibfield  {author} {\bibinfo {author} {\bibfnamefont {P.}~\bibnamefont
  {M\"oller}}, \bibinfo {author} {\bibfnamefont {A.~J.}\ \bibnamefont {Sierk}},
  \bibinfo {author} {\bibfnamefont {T.}~\bibnamefont {Ichikawa}}, \bibinfo
  {author} {\bibfnamefont {A.}~\bibnamefont {Iwamoto}}, \bibinfo {author}
  {\bibfnamefont {R.}~\bibnamefont {Bengtsson}}, \bibinfo {author}
  {\bibfnamefont {H.}~\bibnamefont {Uhrenholt}}, \ and\ \bibinfo {author}
  {\bibfnamefont {S.}~\bibnamefont {Aberg}},\ }\href {\doibase
  10.1103/PhysRevC.79.064304} {\bibfield  {journal} {\bibinfo  {journal} {Phys.
  Rev. C}\ }\textbf {\bibinfo {volume} {79}},\ \bibinfo {pages} {064304}
  (\bibinfo {year} {2009})}\BibitemShut {NoStop}%
\bibitem [{\citenamefont {Skalski}(2007)}]{Skalski2007_PRC76-044603}%
  \BibitemOpen
  \bibfield  {author} {\bibinfo {author} {\bibfnamefont {J.}~\bibnamefont
  {Skalski}},\ }\href {\doibase 10.1103/PhysRevC.76.044603} {\bibfield
  {journal} {\bibinfo  {journal} {Phys. Rev. C}\ }\textbf {\bibinfo {volume}
  {76}},\ \bibinfo {pages} {044603} (\bibinfo {year} {2007})}\BibitemShut
  {NoStop}%
\bibitem [{\citenamefont {Jachimowicz}\ \emph {et~al.}(2011)\citenamefont
  {Jachimowicz}, \citenamefont {Kowal},\ and\ \citenamefont
  {Skalski}}]{Jachimowicz2011_PRC83-054302}%
  \BibitemOpen
  \bibfield  {author} {\bibinfo {author} {\bibfnamefont {P.}~\bibnamefont
  {Jachimowicz}}, \bibinfo {author} {\bibfnamefont {M.}~\bibnamefont {Kowal}},
  \ and\ \bibinfo {author} {\bibfnamefont {J.}~\bibnamefont {Skalski}},\ }\href
  {\doibase 10.1103/PhysRevC.83.054302} {\bibfield  {journal} {\bibinfo
  {journal} {Phys. Rev. C}\ }\textbf {\bibinfo {volume} {83}},\ \bibinfo
  {pages} {054302} (\bibinfo {year} {2011})}\BibitemShut {NoStop}%
\bibitem [{\citenamefont {Serot}\ and\ \citenamefont
  {Walecka}(1986)}]{Serot1986_ANP16-1}%
  \BibitemOpen
  \bibfield  {author} {\bibinfo {author} {\bibfnamefont {B.~D.}\ \bibnamefont
  {Serot}}\ and\ \bibinfo {author} {\bibfnamefont {J.~D.}\ \bibnamefont
  {Walecka}},\ }\href@noop {} {\bibfield  {journal} {\bibinfo  {journal} {Adv.
  Nucl. Phys.}\ }\textbf {\bibinfo {volume} {16}},\ \bibinfo {pages} {1}
  (\bibinfo {year} {1986})}\BibitemShut {NoStop}%
\bibitem [{\citenamefont {Reinhard}(1989)}]{Reinhard1989_RPP52-439}%
  \BibitemOpen
  \bibfield  {author} {\bibinfo {author} {\bibfnamefont {P.~G.}\ \bibnamefont
  {Reinhard}},\ }\href {\doibase 10.1088/0034-4885/52/4/002} {\bibfield
  {journal} {\bibinfo  {journal} {Rep. Prog. Phys.}\ }\textbf {\bibinfo
  {volume} {52}},\ \bibinfo {pages} {439} (\bibinfo {year} {1989})}\BibitemShut
  {NoStop}%
\bibitem [{\citenamefont {Ring}(1996)}]{Ring1996_PPNP37-193}%
  \BibitemOpen
  \bibfield  {author} {\bibinfo {author} {\bibfnamefont {P.}~\bibnamefont
  {Ring}},\ }\href {\doibase 10.1016/0146-6410(96)00054-3} {\bibfield
  {journal} {\bibinfo  {journal} {Prog. Part. Nucl. Phys.}\ }\textbf {\bibinfo
  {volume} {37}},\ \bibinfo {pages} {193} (\bibinfo {year} {1996})}\BibitemShut
  {NoStop}%
\bibitem [{\citenamefont {Vretenar}\ \emph {et~al.}(2005)\citenamefont
  {Vretenar}, \citenamefont {Afanasjev}, \citenamefont {Lalazissis},\ and\
  \citenamefont {Ring}}]{Vretenar2005_PR409-101}%
  \BibitemOpen
  \bibfield  {author} {\bibinfo {author} {\bibfnamefont {D.}~\bibnamefont
  {Vretenar}}, \bibinfo {author} {\bibfnamefont {A. V.}~\bibnamefont {Afanasjev}},
  \bibinfo {author} {\bibfnamefont {G. A.}~\bibnamefont {Lalazissis}}, \ and\
  \bibinfo {author} {\bibfnamefont {P.}~\bibnamefont {Ring}},\ }\href {\doibase
  10.1016/j.physrep.2004.10.001} {\bibfield  {journal} {\bibinfo  {journal}
  {Phys. Rep.}\ }\textbf {\bibinfo {volume} {409}},\ \bibinfo {pages} {101}
  (\bibinfo {year} {2005})}\BibitemShut {NoStop}%
\bibitem [{\citenamefont {Meng}\ \emph
  {et~al.}(2006{\natexlab{a}})\citenamefont {Meng}, \citenamefont {Toki},
  \citenamefont {Zhou}, \citenamefont {Zhang}, \citenamefont {Long},\ and\
  \citenamefont {Geng}}]{Meng2006_PPNP57-470}%
  \BibitemOpen
  \bibfield  {author} {\bibinfo {author} {\bibfnamefont {J.}~\bibnamefont
  {Meng}}, \bibinfo {author} {\bibfnamefont {H.}~\bibnamefont {Toki}}, \bibinfo
  {author} {\bibfnamefont {S.~G.}\ \bibnamefont {Zhou}}, \bibinfo {author}
  {\bibfnamefont {S.~Q.}\ \bibnamefont {Zhang}}, \bibinfo {author}
  {\bibfnamefont {W.~H.}\ \bibnamefont {Long}}, \ and\ \bibinfo {author}
  {\bibfnamefont {L.~S.}\ \bibnamefont {Geng}},\ }\href {\doibase
  10.1016/j.ppnp.2005.06.001} {\bibfield  {journal} {\bibinfo  {journal} {Prog.
  Part. Nucl. Phys.}\ }\textbf {\bibinfo {volume} {57}},\ \bibinfo {pages}
  {470} (\bibinfo {year} {2006}{\natexlab{a}})} \BibitemShut
  {NoStop}%
\bibitem [{\citenamefont {Paar}\ \emph {et~al.}(2007)\citenamefont {Paar},
  \citenamefont {Vretenar},\ and\ \citenamefont {Colo}}]{Paar2007_RPP70-691}%
  \BibitemOpen
  \bibfield  {author} {\bibinfo {author} {\bibfnamefont {N.}~\bibnamefont
  {Paar}}, \bibinfo {author} {\bibfnamefont {D.}~\bibnamefont {Vretenar}}, \
  and\ \bibinfo {author} {\bibfnamefont {G.}~\bibnamefont {Colo}},\ }\href
  {\doibase 10.1088/0034-4885/70/5/R02} {\bibfield  {journal} {\bibinfo
  {journal} {Rep. Prog. Phys.}\ }\textbf {\bibinfo {volume} {70}},\ \bibinfo
  {pages} {691} (\bibinfo {year} {2007})}\BibitemShut {NoStop}%
\bibitem [{\citenamefont {Nik\v{s}i\'{c}}\ \emph {et~al.}(2011)\citenamefont
  {Nik\v{s}i\'{c}}, \citenamefont {Vretenar},\ and\ \citenamefont
  {Ring}}]{Niksic2011_PPNP66-519}%
  \BibitemOpen
  \bibfield  {author} {\bibinfo {author} {\bibfnamefont {T.}~\bibnamefont
  {Nik\v{s}i\'{c}}}, \bibinfo {author} {\bibfnamefont {D.}~\bibnamefont
  {Vretenar}}, \ and\ \bibinfo {author} {\bibfnamefont {P.}~\bibnamefont
  {Ring}},\ }\href {\doibase 10.1016/j.ppnp.2011.01.055} {\bibfield  {journal}
  {\bibinfo  {journal} {Prog. Part. Nucl. Phys.}\ }\textbf {\bibinfo {volume}
  {66}},\ \bibinfo {pages} {519} (\bibinfo {year} {2011})}\BibitemShut
  {NoStop}%
\bibitem [{\citenamefont {Nikolaus}\ \emph {et~al.}(1992)\citenamefont
  {Nikolaus}, \citenamefont {Hoch},\ and\ \citenamefont
  {Madland}}]{Nikolaus1992_PRC46-1757}%
  \BibitemOpen
  \bibfield  {author} {\bibinfo {author} {\bibfnamefont {B.~A.}\ \bibnamefont
  {Nikolaus}}, \bibinfo {author} {\bibfnamefont {T.}~\bibnamefont {Hoch}}, \
  and\ \bibinfo {author} {\bibfnamefont {D.~G.}\ \bibnamefont {Madland}},\
  }\href {\doibase 10.1103/PhysRevC.46.1757} {\bibfield  {journal} {\bibinfo
  {journal} {Phys. Rev. C}\ }\textbf {\bibinfo {volume} {46}},\ \bibinfo
  {pages} {1757} (\bibinfo {year} {1992})}\BibitemShut {NoStop}%
\bibitem [{\citenamefont {Burvenich}\ \emph {et~al.}(2002)\citenamefont
  {Burvenich}, \citenamefont {Madland}, \citenamefont {Maruhn},\ and\
  \citenamefont {Reinhard}}]{Burvenich2002_PRC65-044308}%
  \BibitemOpen
  \bibfield  {author} {\bibinfo {author} {\bibfnamefont {T.}~\bibnamefont
  {Burvenich}}, \bibinfo {author} {\bibfnamefont {D.~G.}\ \bibnamefont
  {Madland}}, \bibinfo {author} {\bibfnamefont {J.~A.}\ \bibnamefont {Maruhn}},
  \ and\ \bibinfo {author} {\bibfnamefont {P.-G.}\ \bibnamefont {Reinhard}},\
  }\href {\doibase 10.1103/PhysRevC.65.044308} {\bibfield  {journal} {\bibinfo
  {journal} {Phys. Rev. C}\ }\textbf {\bibinfo {volume} {65}},\ \bibinfo
  {pages} {044308} (\bibinfo {year} {2002})}\BibitemShut {NoStop}%
\bibitem [{\citenamefont {Boguta}\ and\ \citenamefont
  {Bodmer}(1977)}]{Boguta1977_NPA292-413}%
  \BibitemOpen
  \bibfield  {author} {\bibinfo {author} {\bibfnamefont {J.}~\bibnamefont
  {Boguta}}\ and\ \bibinfo {author} {\bibfnamefont {A.~R.}\ \bibnamefont
  {Bodmer}},\ }\href {\doibase 10.1016/0375-9474(77)90626-1} {\bibfield
  {journal} {\bibinfo  {journal} {Nucl. Phys. A}\ }\textbf {\bibinfo {volume}
  {292}},\ \bibinfo {pages} {413} (\bibinfo {year} {1977})}\BibitemShut
  {NoStop}%
\bibitem [{\citenamefont {Brockmann}\ and\ \citenamefont
  {Toki}(1992)}]{Brockmann1992_PRL68-3408}%
  \BibitemOpen
  \bibfield  {author} {\bibinfo {author} {\bibfnamefont {R.}~\bibnamefont
  {Brockmann}}\ and\ \bibinfo {author} {\bibfnamefont {H.}~\bibnamefont
  {Toki}},\ }\href {\doibase 10.1103/PhysRevLett.68.3408} {\bibfield  {journal}
  {\bibinfo  {journal} {Phys. Rev. Lett.}\ }\textbf {\bibinfo {volume} {68}},\
  \bibinfo {pages} {3408} (\bibinfo {year} {1992})}\BibitemShut {NoStop}%
\bibitem [{\citenamefont {Sugahara}\ and\ \citenamefont
  {Toki}(1994)}]{Sugahara1994_NPA579-557}%
  \BibitemOpen
  \bibfield  {author} {\bibinfo {author} {\bibfnamefont {Y.}~\bibnamefont
  {Sugahara}}\ and\ \bibinfo {author} {\bibfnamefont {H.}~\bibnamefont
  {Toki}},\ }\href {\doibase 10.1016/0375-9474(94)90923-7} {\bibfield
  {journal} {\bibinfo  {journal} {Nucl. Phys. A}\ }\textbf {\bibinfo {volume}
  {579}},\ \bibinfo {pages} {557} (\bibinfo {year} {1994})}\BibitemShut
  {NoStop}%
\bibitem [{\citenamefont {Fuchs}\ \emph {et~al.}(1995)\citenamefont {Fuchs},
  \citenamefont {Lenske},\ and\ \citenamefont {Wolter}}]{Fuchs1995_PRC52-3043}%
  \BibitemOpen
  \bibfield  {author} {\bibinfo {author} {\bibfnamefont {C.}~\bibnamefont
  {Fuchs}}, \bibinfo {author} {\bibfnamefont {H.}~\bibnamefont {Lenske}}, \
  and\ \bibinfo {author} {\bibfnamefont {H.~H.}\ \bibnamefont {Wolter}},\
  }\href {\doibase 10.1103/PhysRevC.52.3043} {\bibfield  {journal} {\bibinfo
  {journal} {Phys. Rev. C}\ }\textbf {\bibinfo {volume} {52}},\ \bibinfo
  {pages} {3043} (\bibinfo {year} {1995})}\BibitemShut {NoStop}%
\bibitem [{\citenamefont {Nik\v{s}i\'{c}}\ \emph {et~al.}(2002)\citenamefont
  {Nik\v{s}i\'{c}}, \citenamefont {Vretenar}, \citenamefont {Finelli},\ and\
  \citenamefont {Ring}}]{Niksic2002_PRC66-024306}%
  \BibitemOpen
  \bibfield  {author} {\bibinfo {author} {\bibfnamefont {T.}~\bibnamefont
  {Nik\v{s}i\'{c}}}, \bibinfo {author} {\bibfnamefont {D.}~\bibnamefont
  {Vretenar}}, \bibinfo {author} {\bibfnamefont {P.}~\bibnamefont {Finelli}}, \
  and\ \bibinfo {author} {\bibfnamefont {P.}~\bibnamefont {Ring}},\ }\href
  {\doibase 10.1103/PhysRevC.66.024306} {\bibfield  {journal} {\bibinfo
  {journal} {Phys. Rev. C}\ }\textbf {\bibinfo {volume} {66}},\ \bibinfo
  {pages} {024306} (\bibinfo {year} {2002})}\BibitemShut {NoStop}%
\bibitem [{\citenamefont {Gambhir}\ \emph {et~al.}(1990)\citenamefont
  {Gambhir}, \citenamefont {Ring},\ and\ \citenamefont
  {Thimet}}]{Gambhir1990_APNY198-132}%
  \BibitemOpen
  \bibfield  {author} {\bibinfo {author} {\bibfnamefont {Y.~K.}\ \bibnamefont
  {Gambhir}}, \bibinfo {author} {\bibfnamefont {P.}~\bibnamefont {Ring}}, \
  and\ \bibinfo {author} {\bibfnamefont {A.}~\bibnamefont {Thimet}},\ }\href
  {\doibase 10.1016/0003-4916(90)90330-Q} {\bibfield  {journal} {\bibinfo
  {journal} {Ann. Phys.}\ }\textbf {\bibinfo {volume} {198}},\ \bibinfo {pages}
  {132} (\bibinfo {year} {1990})}\BibitemShut {NoStop}%
\bibitem [{\citenamefont {Ring}\ \emph {et~al.}(1997)\citenamefont {Ring},
  \citenamefont {Gambhir},\ and\ \citenamefont
  {Lalazissis}}]{Ring1997_CPC105-77}%
  \BibitemOpen
  \bibfield  {author} {\bibinfo {author} {\bibfnamefont {P.}~\bibnamefont
  {Ring}}, \bibinfo {author} {\bibfnamefont {Y.~K.}\ \bibnamefont {Gambhir}}, \
  and\ \bibinfo {author} {\bibfnamefont {G.~A.}\ \bibnamefont {Lalazissis}},\
  }\href {\doibase 10.1016/S0010-4655(97)00022-2} {\bibfield  {journal}
  {\bibinfo  {journal} {Comput. Phys. Commun.}\ }\textbf {\bibinfo {volume}
  {105}},\ \bibinfo {pages} {77} (\bibinfo {year} {1997})}\BibitemShut
  {NoStop}%
\bibitem [{\citenamefont {Zhou}\ \emph {et~al.}(2003)\citenamefont {Zhou},
  \citenamefont {Meng},\ and\ \citenamefont {Ring}}]{Zhou2003_PRC68-034323}%
  \BibitemOpen
  \bibfield  {author} {\bibinfo {author} {\bibfnamefont {S.-G.}\ \bibnamefont
  {Zhou}}, \bibinfo {author} {\bibfnamefont {J.}~\bibnamefont {Meng}}, \ and\
  \bibinfo {author} {\bibfnamefont {P.}~\bibnamefont {Ring}},\ }\href {\doibase
  10.1103/PhysRevC.68.034323} {\bibfield  {journal} {\bibinfo  {journal} {Phys.
  Rev. C}\ }\textbf {\bibinfo {volume} {68}},\ \bibinfo {pages} {034323}
  (\bibinfo {year} {2003})} \BibitemShut {NoStop}%
\bibitem [{\citenamefont {Zhou}\ \emph {et~al.}(2006)\citenamefont {Zhou},
  \citenamefont {Meng},\ and\ \citenamefont {Ring}}]{Zhou2006_AIPCP865-90}%
  \BibitemOpen
  \bibfield  {author} {\bibinfo {author} {\bibfnamefont {S.-G.}\ \bibnamefont
  {Zhou}}, \bibinfo {author} {\bibfnamefont {J.}~\bibnamefont {Meng}}, \ and\
  \bibinfo {author} {\bibfnamefont {P.}~\bibnamefont {Ring}},\ }\href {\doibase
  10.1063/1.2398833} {\bibfield  {journal} {\bibinfo  {journal} {AIP Conf.
  Proc.}\ }\textbf {\bibinfo {volume} {865}},\ \bibinfo {pages} {90} (\bibinfo
  {year} {2006})}\BibitemShut {NoStop}%
\bibitem [{\citenamefont {Zhou}\ \emph {et~al.}(2010)\citenamefont {Zhou},
  \citenamefont {Meng}, \citenamefont {Ring},\ and\ \citenamefont
  {Zhao}}]{Zhou2010_PRC82-011301R}%
  \BibitemOpen
  \bibfield  {author} {\bibinfo {author} {\bibfnamefont {S.-G.}\ \bibnamefont
  {Zhou}}, \bibinfo {author} {\bibfnamefont {J.}~\bibnamefont {Meng}}, \bibinfo
  {author} {\bibfnamefont {P.}~\bibnamefont {Ring}}, \ and\ \bibinfo {author}
  {\bibfnamefont {E.-G.}\ \bibnamefont {Zhao}},\ }\href {\doibase
  10.1103/PhysRevC.82.011301} {\bibfield  {journal} {\bibinfo  {journal} {Phys.
  Rev. C}\ }\textbf {\bibinfo {volume} {82}},\ \bibinfo {pages} {011301(R)}
  (\bibinfo {year} {2010})} \BibitemShut {NoStop}%
\bibitem [{\citenamefont {Li}\ \emph {et~al.}(2012)\citenamefont {Li},
  \citenamefont {Meng}, \citenamefont {Ring}, \citenamefont {Zhao},\ and\
  \citenamefont {Zhou}}]{Li2012_PRC85-024312}%
  \BibitemOpen
  \bibfield  {author} {\bibinfo {author} {\bibfnamefont {L.}~\bibnamefont
  {Li}}, \bibinfo {author} {\bibfnamefont {J.}~\bibnamefont {Meng}}, \bibinfo
  {author} {\bibfnamefont {P.}~\bibnamefont {Ring}}, \bibinfo {author}
  {\bibfnamefont {E.-G.}\ \bibnamefont {Zhao}}, \ and\ \bibinfo {author}
  {\bibfnamefont {S.-G.}\ \bibnamefont {Zhou}},\ }\href {\doibase
  10.1103/PhysRevC.85.024312} {\bibfield  {journal} {\bibinfo  {journal} {Phys.
  Rev. C}\ }\textbf {\bibinfo {volume} {85}},\ \bibinfo {pages} {024312}
  (\bibinfo {year} {2012})} \BibitemShut {NoStop}%
\bibitem [{\citenamefont {Chen}\ \emph {et~al.}(2012)\citenamefont {Chen},
  \citenamefont {Li}, \citenamefont {Liang},\ and\ \citenamefont
  {Meng}}]{Chen2012_PRC85-067301}%
  \BibitemOpen
  \bibfield  {author} {\bibinfo {author} {\bibfnamefont {Y.}~\bibnamefont
  {Chen}}, \bibinfo {author} {\bibfnamefont {L.}~\bibnamefont {Li}}, \bibinfo
  {author} {\bibfnamefont {H.}~\bibnamefont {Liang}}, \ and\ \bibinfo {author}
  {\bibfnamefont {J.}~\bibnamefont {Meng}},\ }\href {\doibase
  10.1103/PhysRevC.85.067301} {\bibfield  {journal} {\bibinfo  {journal} {Phys.
  Rev. C}\ }\textbf {\bibinfo {volume} {85}},\ \bibinfo {pages} {067301}
  (\bibinfo {year} {2012})}\BibitemShut {NoStop}%
\bibitem [{\citenamefont {Geng}\ \emph {et~al.}(2007)\citenamefont {Geng},
  \citenamefont {Meng},\ and\ \citenamefont {Toki}}]{Geng2007_CPL24-1865}%
  \BibitemOpen
  \bibfield  {author} {\bibinfo {author} {\bibfnamefont {L.-S.}\ \bibnamefont
  {Geng}}, \bibinfo {author} {\bibfnamefont {J.}~\bibnamefont {Meng}}, \ and\
  \bibinfo {author} {\bibfnamefont {H.}~\bibnamefont {Toki}},\ }\href {\doibase
  10.1088/0256-307X/24/7/021} {\bibfield  {journal} {\bibinfo  {journal} {Chin.
  Phys. Lett.}\ }\textbf {\bibinfo {volume} {24}},\ \bibinfo {pages} {1865}
  (\bibinfo {year} {2007})}\BibitemShut {NoStop}%
\bibitem [{\citenamefont {Zhang}\ \emph {et~al.}(2010)\citenamefont {Zhang},
  \citenamefont {Li}, \citenamefont {Zhang},\ and\ \citenamefont
  {Meng}}]{Zhang2010_PRC81-034302}%
  \BibitemOpen
  \bibfield  {author} {\bibinfo {author} {\bibfnamefont {W.}~\bibnamefont
  {Zhang}}, \bibinfo {author} {\bibfnamefont {Z.~P.}\ \bibnamefont {Li}},
  \bibinfo {author} {\bibfnamefont {S.~Q.}\ \bibnamefont {Zhang}}, \ and\
  \bibinfo {author} {\bibfnamefont {J.}~\bibnamefont {Meng}},\ }\href {\doibase
  10.1103/PhysRevC.81.034302} {\bibfield  {journal} {\bibinfo  {journal} {Phys.
  Rev. C}\ }\textbf {\bibinfo {volume} {81}},\ \bibinfo {pages} {034302}
  (\bibinfo {year} {2010})}\BibitemShut {NoStop}%
\bibitem [{\citenamefont {Lu}\ \emph {et~al.}(2011)\citenamefont {Lu},
  \citenamefont {Zhao},\ and\ \citenamefont {Zhou}}]{Lu2011_PRC84-014328}%
  \BibitemOpen
  \bibfield  {author} {\bibinfo {author} {\bibfnamefont {B.-N.}\ \bibnamefont
  {Lu}}, \bibinfo {author} {\bibfnamefont {E.-G.}\ \bibnamefont {Zhao}}, \ and\
  \bibinfo {author} {\bibfnamefont {S.-G.}\ \bibnamefont {Zhou}},\ }\href
  {\doibase 10.1103/PhysRevC.84.014328} {\bibfield  {journal} {\bibinfo
  {journal} {Phys. Rev. C}\ }\textbf {\bibinfo {volume} {84}},\ \bibinfo
  {pages} {014328} (\bibinfo {year} {2011})} \BibitemShut
  {NoStop}%
\bibitem [{\citenamefont {Warda}\ \emph {et~al.}(2002)\citenamefont {Warda},
  \citenamefont {Egido}, \citenamefont {Robledo},\ and\ \citenamefont
  {Pomorski}}]{Warda2002_PRC66-014310}%
  \BibitemOpen
  \bibfield  {author} {\bibinfo {author} {\bibfnamefont {M.}~\bibnamefont
  {Warda}}, \bibinfo {author} {\bibfnamefont {J.~L.}\ \bibnamefont {Egido}},
  \bibinfo {author} {\bibfnamefont {L.~M.}\ \bibnamefont {Robledo}}, \ and\
  \bibinfo {author} {\bibfnamefont {K.}~\bibnamefont {Pomorski}},\ }\href
  {\doibase 10.1103/PhysRevC.66.014310} {\bibfield  {journal} {\bibinfo
  {journal} {Phys. Rev. C}\ }\textbf {\bibinfo {volume} {66}},\ \bibinfo
  {pages} {014310} (\bibinfo {year} {2002})}\BibitemShut {NoStop}%
\bibitem [{\citenamefont {Karatzikos}\ \emph {et~al.}(2010)\citenamefont
  {Karatzikos}, \citenamefont {Afanasjev}, \citenamefont {Lalazissis},\ and\
  \citenamefont {Ring}}]{Karatzikos2010_PLB689-72}%
  \BibitemOpen
  \bibfield  {author} {\bibinfo {author} {\bibfnamefont {S.}~\bibnamefont
  {Karatzikos}}, \bibinfo {author} {\bibfnamefont {A.}~\bibnamefont
  {Afanasjev}}, \bibinfo {author} {\bibfnamefont {G.}~\bibnamefont
  {Lalazissis}}, \ and\ \bibinfo {author} {\bibfnamefont {P.}~\bibnamefont
  {Ring}},\ }\href {\doibase 10.1016/j.physletb.2010.04.045} {\bibfield
  {journal} {\bibinfo  {journal} {Phys. Lett. B}\ }\textbf {\bibinfo {volume}
  {689}},\ \bibinfo {pages} {72} (\bibinfo {year} {2010})}\BibitemShut
  {NoStop}%
\bibitem [{\citenamefont {Zeng}\ and\ \citenamefont
  {Cheng}(1983)}]{Zeng1983_NPA405-1}%
  \BibitemOpen
  \bibfield  {author} {\bibinfo {author} {\bibfnamefont {J.~Y.}\ \bibnamefont
  {Zeng}}\ and\ \bibinfo {author} {\bibfnamefont {T.~S.}\ \bibnamefont
  {Cheng}},\ }\href {\doibase 10.1016/0375-9474(83)90320-2} {\bibfield
  {journal} {\bibinfo  {journal} {Nucl. Phys. A}\ }\textbf {\bibinfo {volume}
  {405}},\ \bibinfo {pages} {1} (\bibinfo {year} {1983})}\BibitemShut {NoStop}%
\bibitem [{\citenamefont {Molique}\ and\ \citenamefont
  {Dudek}(1997)}]{Molique1997_PRC56-1795}%
  \BibitemOpen
  \bibfield  {author} {\bibinfo {author} {\bibfnamefont {H.}~\bibnamefont
  {Molique}}\ and\ \bibinfo {author} {\bibfnamefont {J.}~\bibnamefont
  {Dudek}},\ }\href {\doibase 10.1103/PhysRevC.56.1795} {\bibfield  {journal}
  {\bibinfo  {journal} {Phys. Rev. C}\ }\textbf {\bibinfo {volume} {56}},\
  \bibinfo {pages} {1795} (\bibinfo {year} {1997})}\BibitemShut {NoStop}%
\bibitem [{\citenamefont {Meng}\ \emph
  {et~al.}(2006{\natexlab{b}})\citenamefont {Meng}, \citenamefont {Guo},
  \citenamefont {Liu},\ and\ \citenamefont {Zhang}}]{Meng2006_FPC1-38}%
  \BibitemOpen
  \bibfield  {author} {\bibinfo {author} {\bibfnamefont {J.}~\bibnamefont
  {Meng}}, \bibinfo {author} {\bibfnamefont {J.-y.}\ \bibnamefont {Guo}},
  \bibinfo {author} {\bibfnamefont {L.}~\bibnamefont {Liu}}, \ and\ \bibinfo
  {author} {\bibfnamefont {S.-q.}\ \bibnamefont {Zhang}},\ }\href {\doibase
  10.1007/s11467-005-0013-5} {\bibfield  {journal} {\bibinfo  {journal}
  {Frontiers Phys. China}\ }\textbf {\bibinfo {volume} {1}},\ \bibinfo {pages}
  {38} (\bibinfo {year} {2006}{\natexlab{b}})}\BibitemShut {NoStop}%
\bibitem [{\citenamefont {Pillet}\ \emph {et~al.}(2002)\citenamefont {Pillet},
  \citenamefont {Quentin},\ and\ \citenamefont
  {Libert}}]{Pillet2002_NPA697-141}%
  \BibitemOpen
  \bibfield  {author} {\bibinfo {author} {\bibfnamefont {N.}~\bibnamefont
  {Pillet}}, \bibinfo {author} {\bibfnamefont {P.}~\bibnamefont {Quentin}}, \
  and\ \bibinfo {author} {\bibfnamefont {J.}~\bibnamefont {Libert}},\ }\href
  {\doibase 10.1016/S0375-9474(01)01240-4} {\bibfield  {journal} {\bibinfo
  {journal} {Nucl. Phys. A}\ }\textbf {\bibinfo {volume} {697}},\ \bibinfo
  {pages} {141} (\bibinfo {year} {2002})}\BibitemShut {NoStop}%
\bibitem [{\citenamefont {Hao}\ \emph {et~al.}(2012)\citenamefont {Hao},
  \citenamefont {Quentin},\ and\ \citenamefont
  {Bonneau}}]{Hao2012_PRC86-064307}%
  \BibitemOpen
  \bibfield  {author} {\bibinfo {author} {\bibfnamefont {T.~V.~N.}\
  \bibnamefont {Hao}}, \bibinfo {author} {\bibfnamefont {P.}~\bibnamefont
  {Quentin}}, \ and\ \bibinfo {author} {\bibfnamefont {L.}~\bibnamefont
  {Bonneau}},\ }\href {\doibase 10.1103/PhysRevC.86.064307} {\bibfield
  {journal} {\bibinfo  {journal} {Phys. Rev. C}\ }\textbf {\bibinfo {volume}
  {86}},\ \bibinfo {pages} {064307} (\bibinfo {year} {2012})}\BibitemShut
  {NoStop}%
\bibitem [{\citenamefont {Zhang}\ \emph {et~al.}(2011)\citenamefont {Zhang},
  \citenamefont {Zeng}, \citenamefont {Zhao},\ and\ \citenamefont
  {Zhou}}]{Zhang2011_PRC83-011304R}%
  \BibitemOpen
  \bibfield  {author} {\bibinfo {author} {\bibfnamefont {Z.-H.}\ \bibnamefont
  {Zhang}}, \bibinfo {author} {\bibfnamefont {J.-Y.}\ \bibnamefont {Zeng}},
  \bibinfo {author} {\bibfnamefont {E.-G.}\ \bibnamefont {Zhao}}, \ and\
  \bibinfo {author} {\bibfnamefont {S.-G.}\ \bibnamefont {Zhou}},\ }\href
  {\doibase 10.1103/PhysRevC.83.011304} {\bibfield  {journal} {\bibinfo
  {journal} {Phys. Rev. C}\ }\textbf {\bibinfo {volume} {83}},\ \bibinfo
  {pages} {011304(R)} (\bibinfo {year} {2011})} \BibitemShut
  {NoStop}%
\bibitem [{\citenamefont {Zhang}\ \emph {et~al.}(2012)\citenamefont {Zhang},
  \citenamefont {He}, \citenamefont {Zeng}, \citenamefont {Zhao},\ and\
  \citenamefont {Zhou}}]{Zhang2012_PRC85-014324}%
  \BibitemOpen
  \bibfield  {author} {\bibinfo {author} {\bibfnamefont {Z.-H.}\ \bibnamefont
  {Zhang}}, \bibinfo {author} {\bibfnamefont {X.-T.}\ \bibnamefont {He}},
  \bibinfo {author} {\bibfnamefont {J.-Y.}\ \bibnamefont {Zeng}}, \bibinfo
  {author} {\bibfnamefont {E.-G.}\ \bibnamefont {Zhao}}, \ and\ \bibinfo
  {author} {\bibfnamefont {S.-G.}\ \bibnamefont {Zhou}},\ }\href {\doibase
  10.1103/PhysRevC.85.014324} {\bibfield  {journal} {\bibinfo  {journal} {Phys.
  Rev. C}\ }\textbf {\bibinfo {volume} {85}},\ \bibinfo {pages} {014324}
  (\bibinfo {year} {2012})} \BibitemShut {NoStop}%
\bibitem [{\citenamefont {Tian}\ and\ \citenamefont
  {Ma}(2006)}]{Tian2006_CPL23-3226}%
  \BibitemOpen
  \bibfield  {author} {\bibinfo {author} {\bibfnamefont {Y.}~\bibnamefont
  {Tian}}\ and\ \bibinfo {author} {\bibfnamefont {Z.-Y.}\ \bibnamefont {Ma}},\
  }\href {\doibase 10.1088/0256-307X/23/12/029} {\bibfield  {journal} {\bibinfo
   {journal} {Chin. Phys. Lett.}\ }\textbf {\bibinfo {volume} {23}},\ \bibinfo
  {pages} {3226} (\bibinfo {year} {2006})}\BibitemShut {NoStop}%
\bibitem [{\citenamefont {Tian}\ \emph
  {et~al.}(2009{\natexlab{a}})\citenamefont {Tian}, \citenamefont {Ma},\ and\
  \citenamefont {Ring}}]{Tian2009_PLB676-44}%
  \BibitemOpen
  \bibfield  {author} {\bibinfo {author} {\bibfnamefont {Y.}~\bibnamefont
  {Tian}}, \bibinfo {author} {\bibfnamefont {Z.}~\bibnamefont {Ma}}, \ and\
  \bibinfo {author} {\bibfnamefont {P.}~\bibnamefont {Ring}},\ }\href {\doibase
  10.1016/j.physletb.2009.04.067} {\bibfield  {journal} {\bibinfo  {journal}
  {Phys. Lett. B}\ }\textbf {\bibinfo {volume} {676}},\ \bibinfo {pages} {44}
  (\bibinfo {year} {2009}{\natexlab{a}})}\BibitemShut {NoStop}%
\bibitem [{\citenamefont {Tian}\ \emph
  {et~al.}(2009{\natexlab{b}})\citenamefont {Tian}, \citenamefont {Ma},\ and\
  \citenamefont {Ring}}]{Tian2009_PRC79-064301}%
  \BibitemOpen
  \bibfield  {author} {\bibinfo {author} {\bibfnamefont {Y.}~\bibnamefont
  {Tian}}, \bibinfo {author} {\bibfnamefont {Z.-y.}\ \bibnamefont {Ma}}, \ and\
  \bibinfo {author} {\bibfnamefont {P.}~\bibnamefont {Ring}},\ }\href {\doibase
  10.1103/PhysRevC.79.064301} {\bibfield  {journal} {\bibinfo  {journal} {Phys.
  Rev. C}\ }\textbf {\bibinfo {volume} {79}},\ \bibinfo {pages} {064301}
  (\bibinfo {year} {2009}{\natexlab{b}})}\BibitemShut {NoStop}%
\bibitem [{\citenamefont {Bender}\ \emph {et~al.}(2000)\citenamefont {Bender},
  \citenamefont {Rutz}, \citenamefont {Reinhard},\ and\ \citenamefont
  {Maruhn}}]{Bender2000_EPJA8-59}%
  \BibitemOpen
  \bibfield  {author} {\bibinfo {author} {\bibfnamefont {M.}~\bibnamefont
  {Bender}}, \bibinfo {author} {\bibfnamefont {K.}~\bibnamefont {Rutz}},
  \bibinfo {author} {\bibfnamefont {P.-G.}\ \bibnamefont {Reinhard}}, \ and\
  \bibinfo {author} {\bibfnamefont {J.~A.}\ \bibnamefont {Maruhn}},\ }\href
  {\doibase 10.1007/s100530050009} {\bibfield  {journal} {\bibinfo  {journal}
  {Eur. Phys. J. A}\ }\textbf {\bibinfo {volume} {8}},\ \bibinfo {pages} {59}
  (\bibinfo {year} {2000})}\BibitemShut {NoStop}%
\bibitem [{\citenamefont {Zhao}\ \emph {et~al.}(2010)\citenamefont {Zhao},
  \citenamefont {Li}, \citenamefont {Yao},\ and\ \citenamefont
  {Meng}}]{Zhao2010_PRC82-054319}%
  \BibitemOpen
  \bibfield  {author} {\bibinfo {author} {\bibfnamefont {P.~W.}\ \bibnamefont
  {Zhao}}, \bibinfo {author} {\bibfnamefont {Z.~P.}\ \bibnamefont {Li}},
  \bibinfo {author} {\bibfnamefont {J.~M.}\ \bibnamefont {Yao}}, \ and\
  \bibinfo {author} {\bibfnamefont {J.}~\bibnamefont {Meng}},\ }\href {\doibase
  10.1103/PhysRevC.82.054319} {\bibfield  {journal} {\bibinfo  {journal} {Phys.
  Rev. C}\ }\textbf {\bibinfo {volume} {82}},\ \bibinfo {pages} {054319}
  (\bibinfo {year} {2010})}\BibitemShut {NoStop}%
\bibitem [{\citenamefont {Zhao}\ \emph
  {et~al.}(2012{\natexlab{b}})\citenamefont {Zhao}, \citenamefont {Song},
  \citenamefont {Sun}, \citenamefont {Geissel},\ and\ \citenamefont
  {Meng}}]{Zhao2012_PRC86-064324}%
  \BibitemOpen
  \bibfield  {author} {\bibinfo {author} {\bibfnamefont {P.~W.}\ \bibnamefont
  {Zhao}}, \bibinfo {author} {\bibfnamefont {L.~S.}\ \bibnamefont {Song}},
  \bibinfo {author} {\bibfnamefont {B.}~\bibnamefont {Sun}}, \bibinfo {author}
  {\bibfnamefont {H.}~\bibnamefont {Geissel}}, \ and\ \bibinfo {author}
  {\bibfnamefont {J.}~\bibnamefont {Meng}},\ }\href {\doibase
  10.1103/PhysRevC.86.064324} {\bibfield  {journal} {\bibinfo  {journal} {Phys.
  Rev. C}\ }\textbf {\bibinfo {volume} {86}},\ \bibinfo {pages} {064324}
  (\bibinfo {year} {2012}{\natexlab{b}})}\BibitemShut {NoStop}%
\bibitem [{\citenamefont {Kutzelnigg}(1984)}]{Kutzelnigg1984_IJQChem25-107}%
  \BibitemOpen
  \bibfield  {author} {\bibinfo {author} {\bibfnamefont {W.}~\bibnamefont
  {Kutzelnigg}},\ }\href {\doibase 10.1002/qua.560250112} {\bibfield  {journal}
  {\bibinfo  {journal} {Int. J. Quantum Chem.}\ }\textbf {\bibinfo {volume}
  {25}},\ \bibinfo {pages} {107} (\bibinfo {year} {1984})}\BibitemShut
  {NoStop}%
\bibitem [{\citenamefont {Fillion-Gourdeau}\ \emph {et~al.}(2012)\citenamefont
  {Fillion-Gourdeau}, \citenamefont {Lorin},\ and\ \citenamefont
  {Bandrauk}}]{Fillion-Gourdeau2012_PRA85-022506}%
  \BibitemOpen
  \bibfield  {author} {\bibinfo {author} {\bibfnamefont {F.}~\bibnamefont
  {Fillion-Gourdeau}}, \bibinfo {author} {\bibfnamefont {E.}~\bibnamefont
  {Lorin}}, \ and\ \bibinfo {author} {\bibfnamefont {A.~D.}\ \bibnamefont
  {Bandrauk}},\ }\href {\doibase 10.1103/PhysRevA.85.022506} {\bibfield
  {journal} {\bibinfo  {journal} {Phys. Rev. A}\ }\textbf {\bibinfo {volume}
  {85}},\ \bibinfo {pages} {022506} (\bibinfo {year} {2012})}\BibitemShut
  {NoStop}%
\bibitem [{\citenamefont {Reichstein}\ and\ \citenamefont
  {Bary~Malik}(1976)}]{Reichstein1976_AoP98-322}%
  \BibitemOpen
  \bibfield  {author} {\bibinfo {author} {\bibfnamefont {I.}~\bibnamefont
  {Reichstein}}\ and\ \bibinfo {author} {\bibfnamefont {F.}~\bibnamefont
  {Bary~Malik}},\ }\href {\doibase 10.1016/0003-4916(76)90157-3} {\bibfield
  {journal} {\bibinfo  {journal} {Ann. Phys.}\ }\textbf {\bibinfo {volume}
  {98}},\ \bibinfo {pages} {322} (\bibinfo {year} {1976})}\BibitemShut
  {NoStop}%
\bibitem [{\citenamefont {Pomorska}(1979)}]{Pomorska1979_NPA327-1}%
  \BibitemOpen
  \bibfield  {author} {\bibinfo {author} {\bibfnamefont {B.~N.}\ \bibnamefont
  {Pomorska}},\ }\href {\doibase 10.1016/0375-9474(79)90314-2} {\bibfield
  {journal} {\bibinfo  {journal} {Nucl. Phys. A}\ }\textbf {\bibinfo {volume}
  {327}},\ \bibinfo {pages} {1} (\bibinfo {year} {1979})}\BibitemShut {NoStop}%
\bibitem [{\citenamefont {Tondeur}(1985)}]{Tondeur1985_NPA442-460}%
  \BibitemOpen
  \bibfield  {author} {\bibinfo {author} {\bibfnamefont {F.}~\bibnamefont
  {Tondeur}},\ }\href {\doibase 10.1016/S0375-9474(85)80026-9} {\bibfield
  {journal} {\bibinfo  {journal} {Nucl. Phys. A}\ }\textbf {\bibinfo {volume}
  {442}},\ \bibinfo {pages} {460} (\bibinfo {year} {1985})}\BibitemShut
  {NoStop}%
\bibitem [{\citenamefont {Bender}\ \emph {et~al.}(2004)\citenamefont {Bender},
  \citenamefont {Heenen},\ and\ \citenamefont
  {Bonche}}]{Bender2004_PRC70-054304}%
  \BibitemOpen
  \bibfield  {author} {\bibinfo {author} {\bibfnamefont {M.}~\bibnamefont
  {Bender}}, \bibinfo {author} {\bibfnamefont {P.-H.}\ \bibnamefont {Heenen}},
  \ and\ \bibinfo {author} {\bibfnamefont {P.}~\bibnamefont {Bonche}},\ }\href
  {\doibase 10.1103/PhysRevC.70.054304} {\bibfield  {journal} {\bibinfo
  {journal} {Phys. Rev. C}\ }\textbf {\bibinfo {volume} {70}},\ \bibinfo
  {pages} {054304} (\bibinfo {year} {2004})}\BibitemShut {NoStop}%
\bibitem [{\citenamefont {Andreev}\ \emph {et~al.}(2005)\citenamefont
  {Andreev}, \citenamefont {Adamian}, \citenamefont {Antonenko},\ and\
  \citenamefont {Ivanova}}]{Andreev2005_EPJA26-327}%
  \BibitemOpen
  \bibfield  {author} {\bibinfo {author} {\bibfnamefont {A.}~\bibnamefont
  {Andreev}}, \bibinfo {author} {\bibfnamefont {G.}~\bibnamefont {Adamian}},
  \bibinfo {author} {\bibfnamefont {N.}~\bibnamefont {Antonenko}}, \ and\
  \bibinfo {author} {\bibfnamefont {S.}~\bibnamefont {Ivanova}},\ }\href
  {\doibase 10.1140/epja/i2005-10179-x} {\bibfield  {journal} {\bibinfo
  {journal} {Eur. Phys. J. A}\ }\textbf {\bibinfo {volume} {26}},\ \bibinfo
  {pages} {327} (\bibinfo {year} {2005})}\BibitemShut {NoStop}%
\bibitem [{\citenamefont {Robledo}\ \emph {et~al.}(2008)\citenamefont
  {Robledo}, \citenamefont {Baldo}, \citenamefont {Schuck},\ and\ \citenamefont
  {Vinas}}]{Robledo2008_PRC77-051301}%
  \BibitemOpen
  \bibfield  {author} {\bibinfo {author} {\bibfnamefont {L.~M.}\ \bibnamefont
  {Robledo}}, \bibinfo {author} {\bibfnamefont {M.}~\bibnamefont {Baldo}},
  \bibinfo {author} {\bibfnamefont {P.}~\bibnamefont {Schuck}}, \ and\ \bibinfo
  {author} {\bibfnamefont {X.}~\bibnamefont {Vinas}},\ }\href {\doibase
  10.1103/PhysRevC.77.051301} {\bibfield  {journal} {\bibinfo  {journal} {Phys.
  Rev. C}\ }\textbf {\bibinfo {volume} {77}},\ \bibinfo {pages} {051301}
  (\bibinfo {year} {2008})}\BibitemShut {NoStop}%
\bibitem [{\citenamefont {Younes}\ and\ \citenamefont
  {Gogny}(2009)}]{Younes2009_PRC80-054313}%
  \BibitemOpen
  \bibfield  {author} {\bibinfo {author} {\bibfnamefont {W.}~\bibnamefont
  {Younes}}\ and\ \bibinfo {author} {\bibfnamefont {D.}~\bibnamefont {Gogny}},\
  }\href {\doibase 10.1103/PhysRevC.80.054313} {\bibfield  {journal} {\bibinfo
  {journal} {Phys. Rev. C}\ }\textbf {\bibinfo {volume} {80}},\ \bibinfo
  {pages} {054313} (\bibinfo {year} {2009})}\BibitemShut {NoStop}%
\bibitem [{\citenamefont {Capote}\ \emph {et~al.}(2009)\citenamefont {Capote},
  \citenamefont {Herman}, \citenamefont {Oblozinsky}, \citenamefont {Young},
  \citenamefont {Goriely}, \citenamefont {Belgya}, \citenamefont {Ignatyuk},
  \citenamefont {Koning}, \citenamefont {Hilaire}, \citenamefont {Plujko},
  \citenamefont {Avrigeanu}, \citenamefont {Bersillon}, \citenamefont
  {Chadwick}, \citenamefont {Fukahori}, \citenamefont {Ge}, \citenamefont
  {Han}, \citenamefont {Kailas}, \citenamefont {Kopecky}, \citenamefont
  {Maslov}, \citenamefont {Reffo}, \citenamefont {Sin}, \citenamefont
  {Soukhovitskii},\ and\ \citenamefont {Talou}}]{Capote2009_NDS110-3107}%
  \BibitemOpen
  \bibfield  {author} {\bibinfo {author} {\bibfnamefont {R.}~\bibnamefont
  {Capote}}, \bibinfo {author} {\bibfnamefont {M.}~\bibnamefont {Herman}},
  \bibinfo {author} {\bibfnamefont {P.}~\bibnamefont {Oblozinsky}}, \bibinfo
  {author} {\bibfnamefont {P.}~\bibnamefont {Young}}, \bibinfo {author}
  {\bibfnamefont {S.}~\bibnamefont {Goriely}}, \bibinfo {author} {\bibfnamefont
  {T.}~\bibnamefont {Belgya}}, \bibinfo {author} {\bibfnamefont
  {A.}~\bibnamefont {Ignatyuk}}, \bibinfo {author} {\bibfnamefont
  {A.}~\bibnamefont {Koning}}, \bibinfo {author} {\bibfnamefont
  {S.}~\bibnamefont {Hilaire}}, \bibinfo {author} {\bibfnamefont
  {V.}~\bibnamefont {Plujko}}, \bibinfo {author} {\bibfnamefont
  {M.}~\bibnamefont {Avrigeanu}}, \bibinfo {author} {\bibfnamefont
  {O.}~\bibnamefont {Bersillon}}, \bibinfo {author} {\bibfnamefont
  {M.}~\bibnamefont {Chadwick}}, \bibinfo {author} {\bibfnamefont
  {T.}~\bibnamefont {Fukahori}}, \bibinfo {author} {\bibfnamefont
  {Z.}~\bibnamefont {Ge}}, \bibinfo {author} {\bibfnamefont {Y.}~\bibnamefont
  {Han}}, \bibinfo {author} {\bibfnamefont {S.}~\bibnamefont {Kailas}},
  \bibinfo {author} {\bibfnamefont {J.}~\bibnamefont {Kopecky}}, \bibinfo
  {author} {\bibfnamefont {V.}~\bibnamefont {Maslov}}, \bibinfo {author}
  {\bibfnamefont {G.}~\bibnamefont {Reffo}}, \bibinfo {author} {\bibfnamefont
  {M.}~\bibnamefont {Sin}}, \bibinfo {author} {\bibfnamefont {E.}~\bibnamefont
  {Soukhovitskii}}, \ and\ \bibinfo {author} {\bibfnamefont {P.}~\bibnamefont
  {Talou}},\ }\bibfield  {booktitle} {\emph {\bibinfo {booktitle} {Special
  Issue on Nuclear Reaction Data}},\ }\href {\doibase
  10.1016/j.nds.2009.10.004} {\bibfield  {journal} {\bibinfo  {journal} {Nucl.
  Data Sheets}\ }\textbf {\bibinfo {volume} {110}},\ \bibinfo {pages} {3107}
  (\bibinfo {year} {2009})}\BibitemShut {NoStop}%
\bibitem [{\citenamefont {Singh}\ \emph {et~al.}(2002)\citenamefont {Singh},
  \citenamefont {Zywina},\ and\ \citenamefont
  {Firestone}}]{Singh2002_NDS97-241}%
  \BibitemOpen
  \bibfield  {author} {\bibinfo {author} {\bibfnamefont {B.}~\bibnamefont
  {Singh}}, \bibinfo {author} {\bibfnamefont {R.}~\bibnamefont {Zywina}}, \
  and\ \bibinfo {author} {\bibfnamefont {R.~B.}\ \bibnamefont {Firestone}},\
  }\href {\doibase 10.1006/ndsh.2002.0018} {\bibfield  {journal} {\bibinfo
  {journal} {Nucl. Data Sheets}\ }\textbf {\bibinfo {volume} {97}},\ \bibinfo
  {pages} {241} (\bibinfo {year} {2002})}\BibitemShut {NoStop}%
\bibitem [{\citenamefont {Lalazissis}\ \emph {et~al.}(1997)\citenamefont
  {Lalazissis}, \citenamefont {Konig},\ and\ \citenamefont
  {Ring}}]{Lalazissis1997_PRC55-540}%
  \BibitemOpen
  \bibfield  {author} {\bibinfo {author} {\bibfnamefont {G.~A.}\ \bibnamefont
  {Lalazissis}}, \bibinfo {author} {\bibfnamefont {J.}~\bibnamefont {Konig}}, \
  and\ \bibinfo {author} {\bibfnamefont {P.}~\bibnamefont {Ring}},\ }\href
  {\doibase 10.1103/PhysRevC.55.540} {\bibfield  {journal} {\bibinfo  {journal}
  {Phys. Rev. C}\ }\textbf {\bibinfo {volume} {55}},\ \bibinfo {pages} {540}
  (\bibinfo {year} {1997})}\BibitemShut {NoStop}%
\bibitem [{\citenamefont {Lalazissis}\ \emph {et~al.}(2009)\citenamefont
  {Lalazissis}, \citenamefont {Karatzikos}, \citenamefont {Fossion},
  \citenamefont {Arteaga}, \citenamefont {Afanasjev},\ and\ \citenamefont
  {Ring}}]{Lalazissis2009_PLB671-36}%
  \BibitemOpen
  \bibfield  {author} {\bibinfo {author} {\bibfnamefont {G.~A.}\ \bibnamefont
  {Lalazissis}}, \bibinfo {author} {\bibfnamefont {S.}~\bibnamefont
  {Karatzikos}}, \bibinfo {author} {\bibfnamefont {R.}~\bibnamefont {Fossion}},
  \bibinfo {author} {\bibfnamefont {D.~P.}\ \bibnamefont {Arteaga}}, \bibinfo
  {author} {\bibfnamefont {A.~V.}\ \bibnamefont {Afanasjev}}, \ and\ \bibinfo
  {author} {\bibfnamefont {P.}~\bibnamefont {Ring}},\ }\href {\doibase
  10.1016/j.physletb.2008.11.070} {\bibfield  {journal} {\bibinfo  {journal}
  {Phys. Lett. B}\ }\textbf {\bibinfo {volume} {671}},\ \bibinfo {pages} {36}
  (\bibinfo {year} {2009})}\BibitemShut {NoStop}%
\bibitem [{\citenamefont {Bender}\ \emph {et~al.}(1999)\citenamefont {Bender},
  \citenamefont {Rutz}, \citenamefont {Reinhard}, \citenamefont {Maruhn},\ and\
  \citenamefont {Greiner}}]{Bender1999_PRC60-034304}%
  \BibitemOpen
  \bibfield  {author} {\bibinfo {author} {\bibfnamefont {M.}~\bibnamefont
  {Bender}}, \bibinfo {author} {\bibfnamefont {K.}~\bibnamefont {Rutz}},
  \bibinfo {author} {\bibfnamefont {P.-G.}\ \bibnamefont {Reinhard}}, \bibinfo
  {author} {\bibfnamefont {J.~A.}\ \bibnamefont {Maruhn}}, \ and\ \bibinfo
  {author} {\bibfnamefont {W.}~\bibnamefont {Greiner}},\ }\href {\doibase
  10.1103/PhysRevC.60.034304} {\bibfield  {journal} {\bibinfo  {journal} {Phys.
  Rev. C}\ }\textbf {\bibinfo {volume} {60}},\ \bibinfo {pages} {034304}
  (\bibinfo {year} {1999})}\BibitemShut {NoStop}%
\bibitem [{\citenamefont {Lalazissis}\ \emph {et~al.}(2005)\citenamefont
  {Lalazissis}, \citenamefont {Nik\v{s}i\'{c}}, \citenamefont {Vretenar},\ and\
  \citenamefont {Ring}}]{Lalazissis2005_PRC71-024312}%
  \BibitemOpen
  \bibfield  {author} {\bibinfo {author} {\bibfnamefont {G.~A.}\ \bibnamefont
  {Lalazissis}}, \bibinfo {author} {\bibfnamefont {T.}~\bibnamefont
  {Nik\v{s}i\'{c}}}, \bibinfo {author} {\bibfnamefont {D.}~\bibnamefont
  {Vretenar}}, \ and\ \bibinfo {author} {\bibfnamefont {P.}~\bibnamefont
  {Ring}},\ }\href {\doibase 10.1103/PhysRevC.71.024312} {\bibfield  {journal}
  {\bibinfo  {journal} {Phys. Rev. C}\ }\textbf {\bibinfo {volume} {71}},\
  \bibinfo {pages} {024312} (\bibinfo {year} {2005})}\BibitemShut {NoStop}%
\bibitem [{\citenamefont {Nik\v{s}i\'{c}}\ \emph {et~al.}(2008)\citenamefont
  {Nik\v{s}i\'{c}}, \citenamefont {Vretenar},\ and\ \citenamefont
  {Ring}}]{Niksic2008_PRC78-034318}%
  \BibitemOpen
  \bibfield  {author} {\bibinfo {author} {\bibfnamefont {T.}~\bibnamefont
  {Nik\v{s}i\'{c}}}, \bibinfo {author} {\bibfnamefont {D.}~\bibnamefont
  {Vretenar}}, \ and\ \bibinfo {author} {\bibfnamefont {P.}~\bibnamefont
  {Ring}},\ }\href {\doibase 10.1103/PhysRevC.78.034318} {\bibfield  {journal}
  {\bibinfo  {journal} {Phys. Rev. C}\ }\textbf {\bibinfo {volume} {78}},\
  \bibinfo {pages} {034318} (\bibinfo {year} {2008})}\BibitemShut {NoStop}%
\bibitem [{\citenamefont {Jachimowicz}\ \emph {et~al.}(2012)\citenamefont
  {Jachimowicz}, \citenamefont {Kowal},\ and\ \citenamefont
  {Skalski}}]{Jachimowicz2012_PRC85-034305}%
  \BibitemOpen
  \bibfield  {author} {\bibinfo {author} {\bibfnamefont {P.}~\bibnamefont
  {Jachimowicz}}, \bibinfo {author} {\bibfnamefont {M.}~\bibnamefont {Kowal}},
  \ and\ \bibinfo {author} {\bibfnamefont {J.}~\bibnamefont {Skalski}},\ }\href
  {\doibase 10.1103/PhysRevC.85.034305} {\bibfield  {journal} {\bibinfo
  {journal} {Phys. Rev. C}\ }\textbf {\bibinfo {volume} {85}},\ \bibinfo
  {pages} {034305} (\bibinfo {year} {2012})}\BibitemShut {NoStop}%
\bibitem [{\citenamefont {Dubray}\ and\ \citenamefont
  {Regnier}(2012)}]{Dubray2012_CPC183-2035}%
  \BibitemOpen
  \bibfield  {author} {\bibinfo {author} {\bibfnamefont {N.}~\bibnamefont
  {Dubray}}\ and\ \bibinfo {author} {\bibfnamefont {D.}~\bibnamefont
  {Regnier}},\ }\href {\doibase 10.1016/j.cpc.2012.05.001} {\bibfield
  {journal} {\bibinfo  {journal} {Comput. Phys. Commun.}\ }\textbf {\bibinfo
  {volume} {183}},\ \bibinfo {pages} {2035} (\bibinfo {year}
  {2012})}\BibitemShut {NoStop}%
\bibitem [{\citenamefont {Yao}\ \emph {et~al.}(2010)\citenamefont {Yao},
  \citenamefont {Meng}, \citenamefont {Ring},\ and\ \citenamefont
  {Vretenar}}]{Yao2010_PRC81-044311}%
  \BibitemOpen
  \bibfield  {author} {\bibinfo {author} {\bibfnamefont {J.~M.}\ \bibnamefont
  {Yao}}, \bibinfo {author} {\bibfnamefont {J.}~\bibnamefont {Meng}}, \bibinfo
  {author} {\bibfnamefont {P.}~\bibnamefont {Ring}}, \ and\ \bibinfo {author}
  {\bibfnamefont {D.}~\bibnamefont {Vretenar}},\ }\href {\doibase
  10.1103/PhysRevC.81.044311} {\bibfield  {journal} {\bibinfo  {journal} {Phys.
  Rev. C}\ }\textbf {\bibinfo {volume} {81}},\ \bibinfo {pages} {044311}
  (\bibinfo {year} {2010})}\BibitemShut {NoStop}%
\bibitem [{\citenamefont {Yao}\ \emph {et~al.}(2011{\natexlab{a}})\citenamefont
  {Yao}, \citenamefont {Mei}, \citenamefont {Chen}, \citenamefont {Meng},
  \citenamefont {Ring},\ and\ \citenamefont {Vretenar}}]{Yao2011_PRC83-014308}%
  \BibitemOpen
  \bibfield  {author} {\bibinfo {author} {\bibfnamefont {J.~M.}\ \bibnamefont
  {Yao}}, \bibinfo {author} {\bibfnamefont {H.}~\bibnamefont {Mei}}, \bibinfo
  {author} {\bibfnamefont {H.}~\bibnamefont {Chen}}, \bibinfo {author}
  {\bibfnamefont {J.}~\bibnamefont {Meng}}, \bibinfo {author} {\bibfnamefont
  {P.}~\bibnamefont {Ring}}, \ and\ \bibinfo {author} {\bibfnamefont
  {D.}~\bibnamefont {Vretenar}},\ }\href {\doibase 10.1103/PhysRevC.83.014308}
  {\bibfield  {journal} {\bibinfo  {journal} {Phys. Rev. C}\ }\textbf {\bibinfo
  {volume} {83}},\ \bibinfo {pages} {014308} (\bibinfo {year}
  {2011}{\natexlab{a}})}\BibitemShut {NoStop}%
\bibitem [{\citenamefont {Yao}\ \emph {et~al.}(2011{\natexlab{b}})\citenamefont
  {Yao}, \citenamefont {Peng}, \citenamefont {Meng},\ and\ \citenamefont
  {Ring}}]{Yao2011_SCG54-198}%
  \BibitemOpen
  \bibfield  {author} {\bibinfo {author} {\bibfnamefont {J.-M.}\ \bibnamefont
  {Yao}}, \bibinfo {author} {\bibfnamefont {J.}~\bibnamefont {Peng}}, \bibinfo
  {author} {\bibfnamefont {J.}~\bibnamefont {Meng}}, \ and\ \bibinfo {author}
  {\bibfnamefont {P.}~\bibnamefont {Ring}},\ }\href {\doibase
  10.1007/s11433-010-4214-8} {\bibfield  {journal} {\bibinfo  {journal} {Sci.
  China Phys. Mech. Astron.}\ }\textbf {\bibinfo {volume} {54}},\ \bibinfo
  {pages} {198} (\bibinfo {year} {2011}{\natexlab{b}})}\BibitemShut {NoStop}%
\bibitem [{\citenamefont {Liu}\ \emph {et~al.}(2011)\citenamefont {Liu},
  \citenamefont {Xu}, \citenamefont {Sun}, \citenamefont {Walker},\ and\
  \citenamefont {Wyss}}]{Liu2011_EPJA47-135}%
  \BibitemOpen
  \bibfield  {author} {\bibinfo {author} {\bibfnamefont {H.~L.}\ \bibnamefont
  {Liu}}, \bibinfo {author} {\bibfnamefont {F.~R.}\ \bibnamefont {Xu}},
  \bibinfo {author} {\bibfnamefont {Y.}~\bibnamefont {Sun}}, \bibinfo {author}
  {\bibfnamefont {P.~M.}\ \bibnamefont {Walker}}, \ and\ \bibinfo {author}
  {\bibfnamefont {R.}~\bibnamefont {Wyss}},\ }\href {\doibase
  10.1140/epja/i2011-11135-y} {\bibfield  {journal} {\bibinfo  {journal} {Eur.
  Phys. J. A}\ }\textbf {\bibinfo {volume} {47}},\ \bibinfo {pages} {135}
  (\bibinfo {year} {2011})}\BibitemShut {NoStop}%
\bibitem [{\citenamefont {Nhan~Hao}\ \emph {et~al.}(2012)\citenamefont
  {Nhan~Hao}, \citenamefont {Le~Bloas}, \citenamefont {Koh}, \citenamefont
  {Bonneau},\ and\ \citenamefont {Quentin}}]{Hao2012_IJMPE21-1250051}%
  \BibitemOpen
  \bibfield  {author} {\bibinfo {author} {\bibfnamefont {T.~V.}\ \bibnamefont
  {Nhan~Hao}}, \bibinfo {author} {\bibfnamefont {J.}~\bibnamefont {Le~Bloas}},
  \bibinfo {author} {\bibfnamefont {M.-H.}\ \bibnamefont {Koh}}, \bibinfo
  {author} {\bibfnamefont {L.}~\bibnamefont {Bonneau}}, \ and\ \bibinfo
  {author} {\bibfnamefont {P.}~\bibnamefont {Quentin}},\ }\href {\doibase
  10.1142/S0218301312500516} {\bibfield  {journal} {\bibinfo  {journal} {Int.
  J. Mod. Phys. E}\ }\textbf {\bibinfo {volume} {21}},\ \bibinfo {pages}
  {1250051} (\bibinfo {year} {2012})}\BibitemShut {NoStop}%
\end{thebibliography}


\end{document}